\documentclass[apsrev4-2,aps,physrev,superscriptaddress,twocolumn,floatfix,longbibliography,]{revtex4-2}
\usepackage{colortbl}
\usepackage{color}
\usepackage{amsfonts,amsmath,amssymb}
\usepackage[dvipsnames]{xcolor}
\usepackage{tabularx,hhline,tikz,multirow,array}
\usepackage{bm,enumitem}
\usepackage{dcolumn,stmaryrd}
\usepackage[colorlinks=true,pdfborder={0 0 0},linkcolor=red,citecolor = blue]{hyperref}
\usepackage{comment}
\usepackage[caption=false]{subfig} 

\def\beq{\begin{equation}}
\def\eeq{\end{equation}}
\def\bea{\begin{eqnarray}}
\def\eea{\end{eqnarray}}

\newcommand{\Sp}{\mathbf{S}}
\newcommand{\Dp}{\mathbf{D}}
\def\rv {{\bf r}}
\def\kv {{\bf k}}
\def\Rv {{\bf R}}
\def\rv {{\bf r}}
\def\kv {{\bf k}}

\begin{document}

\title{From chiral spin liquids to skyrmion fluids and crystals, and their interplay with itinerant electrons}

\author{F.A. G\'omez Albarrac\'in}
\email[]{albarrac@fisica.unlp.edu.ar}
\affiliation{Instituto de F\'isica de L\'iquidos y Sistemas Biol\'ogicos (IFLYSIB), UNLP-CONICET, La Plata, Argentina and Departamento de F\'isica, Facultad de Ciencias Exactas,
Universidad Nacional de La Plata, c.c. 16, suc. 4, 1900 La Plata, Argentina.}
\affiliation{Departamento de Ciencas B\'asicas, Facultad de Ingenier\'ia, UNLP, La Plata, Argentina}
\author{H. Diego Rosales}
\email[]{rosales@fisica.unlp.edu.ar}
\affiliation{Instituto de F\'isica de L\'iquidos y Sistemas Biol\'ogicos (IFLYSIB), UNLP-CONICET, La Plata, Argentina and Departamento de F\'isica, Facultad de Ciencias Exactas,
Universidad Nacional de La Plata, c.c. 16, suc. 4, 1900 La Plata, Argentina.}
\affiliation{Departamento de Ciencas B\'asicas, Facultad de Ingenier\'ia, UNLP, La Plata, Argentina}
\author{Masafumi Udagawa}
\email[]{Masafumi.Udagawa@gakushuin.ac.jp}
\affiliation{Department of Physics, Gakushuin University, Mejiro, Toshima-ku, 171-8588, Japan}

\author{P. Pujol}
\email[]{pierre.pujol@irsamc.ups-tlse.fr}
\affiliation{Laboratoire de Physique Theorique, CNRS and Universit\'e de Toulouse, UPS, Toulouse, F-31062, France.}
\author{Ludovic D. C Jaubert}
\email[]{ludovic.jaubert@cnrs.fr}
\affiliation{CNRS, Universit\'e de Bordeaux, LOMA, UMR 5798, 33400 Talence, France}

\begin{abstract}

The physics of skyrmions, and in particular the issue of how to isolate and manipulate them individually, is a subject of major importance nowadays in the community of magnetism. In this article we present an in-depth extension of a study on this issue that was recently proposed by some of the authors [H. D. Rosales, et al. Phys. Rev. Lett. \textbf{130}, 106703 (2023)]. More precisely, we analyse the competition between skyrmions and a chiral spin liquid in a model on the kagome lattice. We first present an analytical overview of the low-energy states using the Luttinger-Tisza approximation. We then study the effect of thermal fluctuations thanks to large-scale Monte-Carlo simulations, and explore the entire parameter space with a magnetic field $B$, in-plane $D^{xy}$ and out-of-plane $D^z$ Dzyaloshinskii-Moriya interactions. While skyrmions and the chiral spin liquid live in different regions of the parameter space, we show how to bring them together, stabilizing a skyrmion fluid in between; a region where the density of well-defined skyrmions can be tuned  before obtaining an ordered phase. We investigate in particular the melting of the skyrmion solid. Our analysis also brings to light a long-range ordered phase with Z$_3$ symmetry. At last, we initiate the study of this rich magnetic background on conduction electrons that are coupled to the local spins. We study how the different chiral magnetic textures stabilized in this model (skyrmion solid, liquid and gas and chiral spin liquid) induce a topological Quantum Hall effect. We observe in the ordered skyrmion phase the appearance of Landau levels which persist even in the skyrmion-liquid regime and gradually disappear as the skyrmion density decreases to form a gas.

\end{abstract}

\maketitle

\section{Introduction}

Magnetic skyrmions have attracted great interest owing to their unique topological spin texture \cite{bogdanov1989ther,bogdanov1994ther,roessler2006spontaneous,muhlbauer2009sk,yu2010re,nagaosa2013topo}, especially for potential applications to next-generation magnetic memory and logic computing devices in spintronics \cite{fert2017magnetic,everschor2018perspective}. In general, the formation of skyrmion lattices arises from the interplay of competing interactions. A variety of stabilizing mechanisms have been well established over the years, starting from the competition between the ferromagnetic exchange interaction and the Dzyaloshinskii-Moriya (DM) interaction \cite{DM1,DM2}, to exchange frustration \cite{Okubo2012,mohylna2022spontaneous}, bond-dependent exchange anisotropy \cite{Gao2020Nat,Amoroso2020,Wang2021,Hayami2021-a,Utesov2021,Yambe2021,Amoroso2021,Rosales22,Hayami2022-a}, the Ruderman-Kittel-Kasuya-Yosida (RKKY) interaction in itinerant magnets\cite{PhysRevLett.108.096401,WangRKKY2020,hayami21,10.21468/SciPostPhys.15.4.161} or higher-order exchange interactions\cite{Paul2020}. Generally at zero field, $B=0$, the low-temperature physics favors the development of a helical (H) phase characterized by one-dimensional magnetic stripes. As the magnetic field $B$ increases from a low but finite value, these stripes, through the superposition of multiple stripes along different directions, develop into a periodic arrangement of skyrmions (SkX). During the transition between the H and SkX phases, the emergence of elongated skyrmions known as bimerons is often observed \cite{ezawa2011c}. Ultimately, at high magnetic fields, the SkX phase undergoes a transition into a field-polarized (FP) regime.

The intermediate region between the SkX and FP phases holds significant importance due to the intriguing presence of a dilute fluid of skyrmions. Typically, the density of skyrmions can be manipulated at low temperatures by adjusting the field strength $B$ which acts as an effective chemical potential \cite{huang2020m,balavz2021m,nishikawa2019s}. However, attempting to control the number of skyrmions with temperature (by heating) is more challenging. Thermal fluctuations do melt the SkX phase, but they eventually destroy the skyrmions. In that sense, temperature does not really tune the density of skyrmions but rather disintegrates them into paramagnetic fluctuations. In a recent publication, we introduced a frustrated microscopic model that effectively separates the SkX order from the paramagnetic regime by introducing an intervening chiral spin liquid (CSL) in between \cite{rosales2023skyrmion}. This CSL bears a finite magnetization that couples to the magnetic field, stabilizing the FP regime for lower fields at intermediate temperatures. As a result, the FP regime circles around the SkX phase, and the skyrmion density can now be continuously tuned from high to low upon heating.

Our objective in the present manuscript is to conduct an in-depth study of the model proposed in Ref.~[\onlinecite{rosales2023skyrmion}]. Firstly, we analyze the low-energy physics using a Luttinger-Tisza approximation. Then, through Monte-Carlo simulations, we explore the entire phase diagram in multi-dimensional parameter space, varying the in-plane $D^{xy}$ and out-of-plane $D^z$ Dzyaloshinskii-Moriya interactions as well as the magnetic field $B$, as a function of temperature $T$. The chiral spin liquid and skyrmions live in different regions of parameter space, and we explain under which conditions these regions can be brought together. As a side benefit, we discover the presence of a long-range order with Z$_3$ symmetry. Then, focusing on parameters where \textit{four} distinct chiral magnetic textures are successively stabilized upon heating (skyrmion solid, liquid, and gas \& chiral spin liquid), we characterize the two-dimensional solid/liquid transition of skyrmions  \cite{huang2020m,balavz2021m,nishikawa2019s}.  Last but not least, we analyze the topological Hall response of itinerant electrons coupled to the local magnetic moments to probe these different chiral phases. We find the traditional topological quantum Hall effect in the skyrmion-solid phase \cite{yi09_skyrm_anomal_hall_effec_dzyal,hamamoto2015quantized} but also, and more surprisingly, in the skyrmion-liquid regime despite the absence of broken translational symmetry.

\section{The model}
\label{sec:model}

\begin{figure}[thb]
\includegraphics[width=0.7\columnwidth]{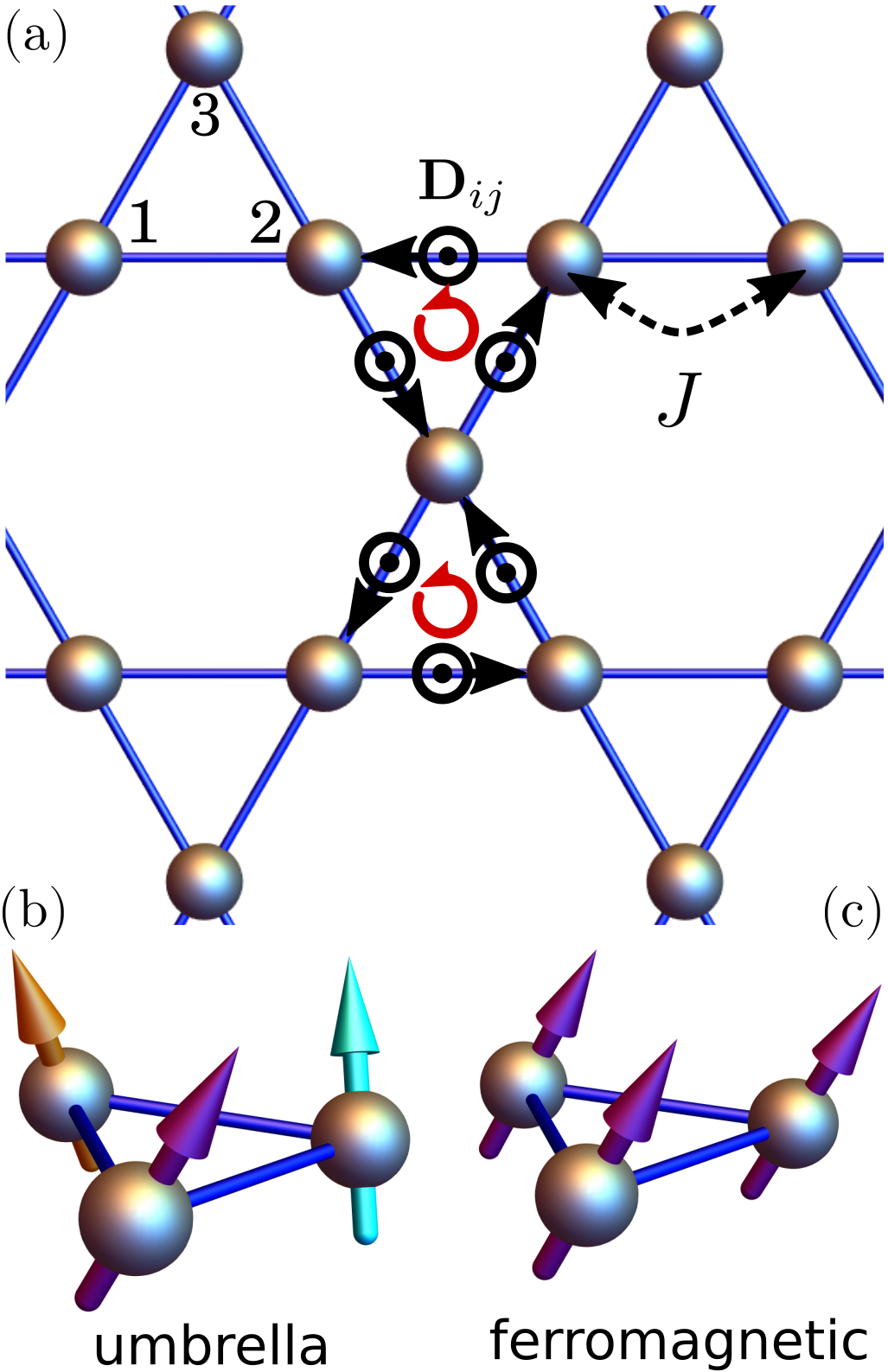}
\caption{(a) kagome lattice and DM vectors. Labels $1,2,3$ represent the three sublattices, (b) umbrella configuration in a triangular plaquette (c) ferromagnetic configuration. Both (b) and (c) states have the same out-of-plane magnetization.}
\label{fig:latt}
\end{figure}

We consider the following spin Hamiltonian on a kagome lattice
\bea
\mathcal{H}&=&-\sum_{\langle ij\rangle}J\;\Sp_i\cdot\Sp_j- \sum_{\langle ij\rangle}\Dp_{ij}\cdot(\Sp_i\times\Sp_j)-B\sum_i S^{z}_i
\label{eq:HamSpin}
\eea
where $\mathbf{S}_i$ represents classical Heisenberg spins of unit length ($|\mathbf{S}_i|=1$) at site $i$, $J>0$ (for simplicity, we will fix $J=1$ for the rest of the manuscript), the DM interaction includes in-plane and perpendicular (to the lattice) contribution as $\Dp_{ij}=D^z_{ij}\,\hat{z}+\Dp^{xy}_{ij}$ (see Fig.~\ref{fig:latt}(a)), being $\Dp^{xy}_{ij}=D^{xy}{\bf \hat{r}}_{ij}=D^{xy}(\rv_j-\rv_i)/|\rv_j-\rv_i|$,  and $B$ is the external magnetic field perpendicular to the lattice plane.

In the kagome lattice, at $B=0$, the model defined by Eq.~(\ref{eq:HamSpin}) exhibits a particular point in the parameter space. When $D^{xy}=0$ and $D^{z}/J=\pm\sqrt{3}$, the ferromagnetic (FM) coupling perfectly balances the $D^{z}$ interaction. As a result, the chiral ``umbrella'' and FM ground-state configurations illustrated in Fig.~\ref{fig:latt}(b,c) minimize the classical energy for each triangle \cite{essafi2016k}. This leads to the spins' $xy$ components displaying chiral-spin-liquid (CSL) behavior, characterized by extensive degeneracy and algebraic correlations, while the $z$ components assume a finite uniform value.

However, in the case of pure in-plane DM interaction where $D^{xy}>0$ and $D^{z}=0$, the scenario undergoes a significant change. A non-zero $D^{xy}$ value promotes the emergence of typical spin helical configurations, commonly observed in ferromagnetic materials, as well as the formation of skyrmion phases for a finite external magnetic field. Thus, the interplay between these two types of DM interactions is expected to give rise to exotic phenomena. In our previous work \cite{rosales2023skyrmion}, we initiated this study by focusing on a combination of these cases, specifically considering $D^{z} = \sqrt{3}$ and $D^{xy} = 0.5$. We demonstrated the possibility of utilizing a CSL as an entropic buffer to induce a quasi-vacuum of skyrmions. Building upon these findings, our present study aims to further explore this model and investigate the phenomenology across a broader range of parameters.

\begin{figure*}[thb]
\includegraphics[width=0.99\textwidth]{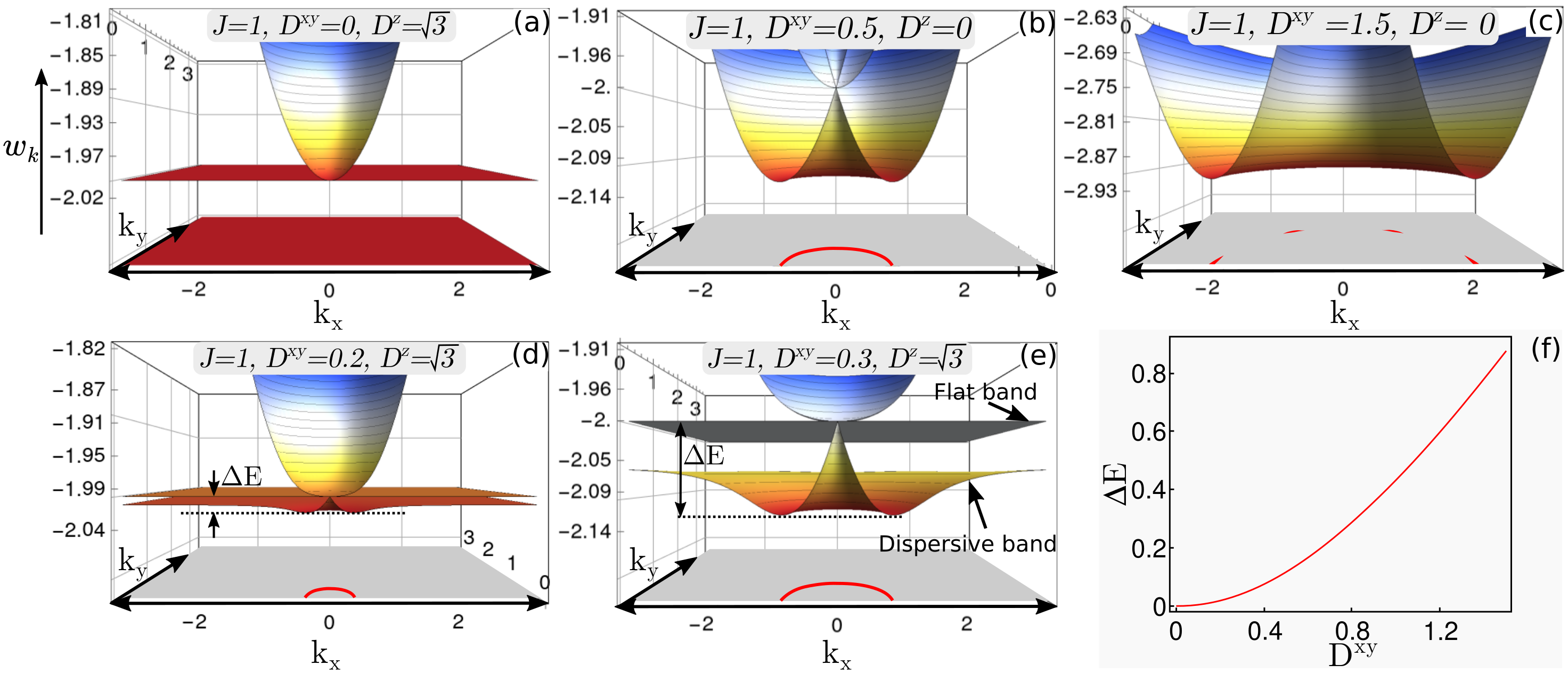}
\caption{(a-e) Cut of the band structure obtained from the Luttinger-Tisza approximation (LTA) in momentum space for $(D^z,D^{xy})=$ (a) $(\sqrt{3},0)$ (pure chiral spin liquid) (b)   $(0,0.5)$ (c)  $(0,1.5)$ (pure skyrmion models), and (d)  $(\sqrt{3},0.2)$ (e)  $(\sqrt{3},0.3)$, combination of CSL and pure skyrmion models. The bottom of each panel shows the position of the energy minima at $E_{gs}$. Panel (f) shows the energy difference $\Delta E$ between the energy minima and the degenerate flat bands as a function of $D^{xy}$, fixing $D^z=\sqrt{3}$.}
\label{fig:LTA1}
\end{figure*}
%

\section{The Luttinger-Tisza approximation}
\label{sec:LTA}

To explore the low energy configurations in the absence of a magnetic field at zero temperature, we resort to the LTA \cite{Luttinger1946,Luttinger1951}. In this approximation, the local fixed spin-length constraint $|\mathbf{S}_i|=1$ is replaced by a global constraint $\sum_j \mathbf{S}_j^2=N$, where $N$ represents the number of lattice sites. By introducing this softer constraint, the spin Hamiltonian (\ref{eq:HamSpin}) can be diagonalized using the Fourier transformation $S^{\alpha}_j=\sum_{\mathbf{k}}e^{i\mathbf{k}\cdot \mathbf{r}_j}S^{\alpha}_{\mathbf{k}}$. Here, $\alpha = x, y, z$, while $\mathbf{r}_j$ and $\mathbf{k}$ denote the position and pseudo-momentum, respectively. Following the diagonalization process, the resulting spectrum within this method consists of nine bands $w_{\mathbf{k},a}$ ($a=1,...9$, representing the three spin components and three sublattices) and the lowest energy configuration is associated with the bottom of the lowest band which defines the ordering wave-vector ${\bf k}^*$. 

In Fig.~\ref{fig:LTA1} we summarize several cases of the model defined in Eq.~(\ref{eq:HamSpin}). Panel (a) shows the pure chiral spin liquid case, $(D^z,D^{yx})=(\sqrt{3},0)$, which has the characteristic lowest energy flat bands. In the ``pure skyrmion'' model ($D^z=0$), at lower values of the in-plane DM interaction the lowest energy band has ring-like minima for smaller $D^{xy}$, and a triple-k set appears as $D^{xy}$ is increased (panels (b) and (c)). The inclusion of $D^z=\sqrt{3}$ does not significantly change the bands' minima (panels (d) and (e)), but it introduces a flat band with energy $-2J$. The energy difference $\Delta E$ between the minima and the flat bands starts from zero (gapless) and smoothly increases with $D^{xy}$ (panel (f)). At low $D^{xy}$ this difference is fairly small compared to the ground state energy $E_{gs}$, and thus we may expect a strong influence of the chiral spin liquid physics at finite temperature. On the other hand for higher $D^{xy}$, $\Delta E$  is significantly larger, so we expect the chiral spin liquid effects to be erased. At the particular value $D^{xy}=0.5$, $\Delta E \approx 0.2 \approx 10\% E_{gs}$, which would be consistent with the evidence of chiral spin liquid behavior at intermediate temperatures.  Thus, from this analysis we are able to predict the range of $D^{xy}$ where chiral spin liquid effects are visible in Monte Carlo simulations: they would be dominant for low $D^{xy}$ and negligible for larger $D^{xy} \approx 1$. At intermediate values, $D^{xy}\approx 0.3-0.6$, an interesting interplay between skyrmion  and spin liquid physics can be expected. But the LTA remains an approximation, which is why we shall now turn our attention to Monte Carlo simulations in order to confirm this analytical intuition.

\section{Phase diagrams by classical Monte Carlo simulations }

Here we thoroughly investigate the proposed model in Eq.~(\ref{eq:HamSpin}) by exploring different parameter values of the DM interactions and magnetic field, using extensive Monte-Carlo simulations, with a combination of the Metropolis algorithm and the overrelaxation method (microcanonical updates), lowering the temperature in an annealing scheme. We performed our simulations in $N=3\times L^2$ site clusters, with $L=48-192$, with periodic boundary conditions.  $10^5-10^6$ Monte Carlo steps (MCS) were used for initial relaxation, and measurements were taken in twice as much MCS.

 \begin{figure*}[ht!]
\includegraphics[width=0.99\textwidth]{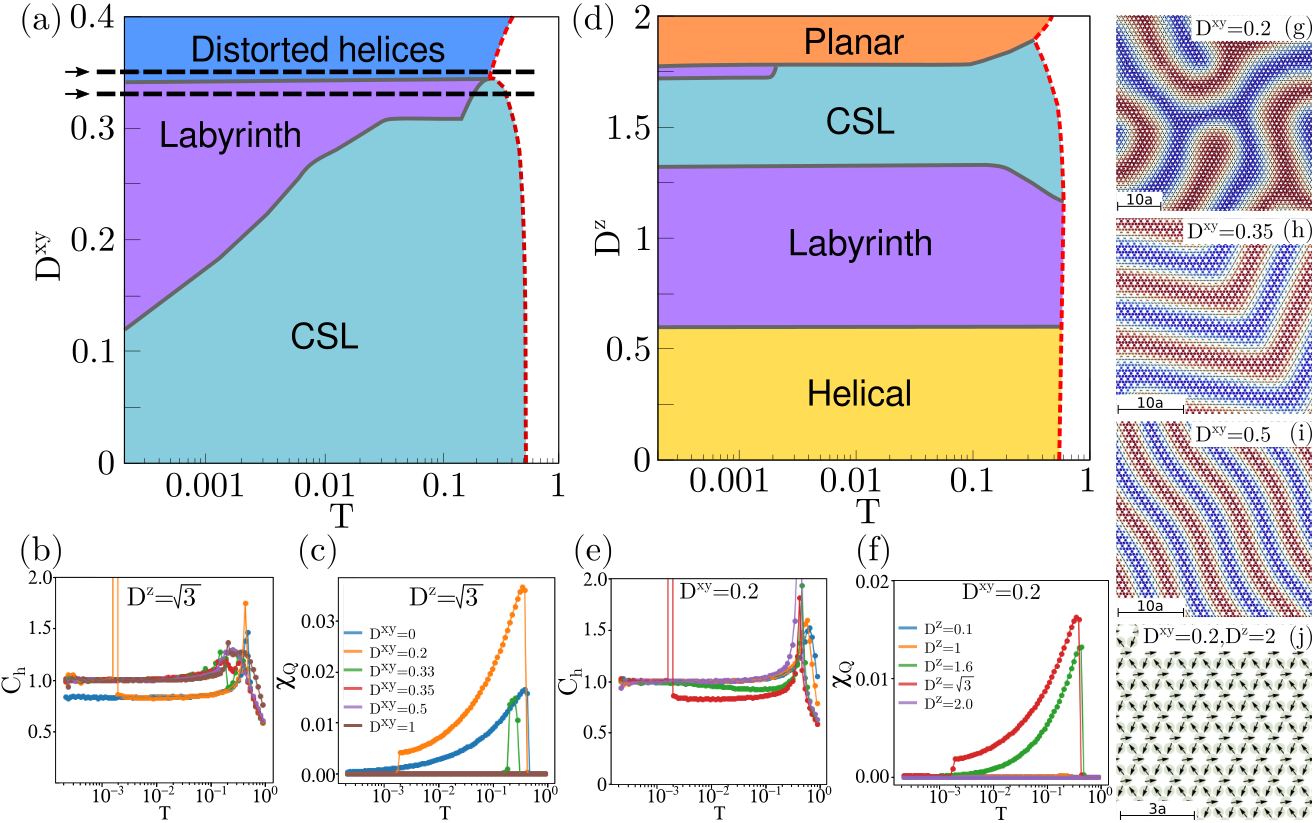}
\caption{(a) $D^{xy}$ vs $T$ phase diagram at $B=0$, $D^z=\sqrt{3}$ (b,c) $C_h$ and $\chi_Q$ vs $T$ for different values of $D^{xy}$, $D^z=\sqrt{3}$ (d) $D^{z}$ vs $T$ phase diagram at $B=0$, fixing $D^{xy}=0.2$ (e,f) $C_h$ and $\chi_Q$ vs $T$ for different values of $D^z$, $D^{xy}=0.2$. Representative snapshots for different phases at $T=2\times 10^{-4}$ for $D^{xy}=0.2,0.35,0.5$ (panels g-i), $D^z=\sqrt{3}$, and for $D^{xy}=0.2, D^z=2$ (panel j). In panels (g-j) the scale in terms of lattice spacing (a) is indicated.} 
\label{fig:PDB0}
\end{figure*}

We first give a general overview of the competition between chiral spin liquid (CSL) and skyrmion physics, varying $D^{xy}$ and  $D^z$ at zero field $B=0$. Then, we describe in detail the emergent physics in the $B-T$ phase diagram that was presented in Ref.~[\onlinecite{rosales2023skyrmion}] for $D^{xy}=0.5$. The different magnetic phases are characterized via the magnetization $M=\langle \frac{1}{N}\sum_i S^z_i\rangle$  and the specific heat $C_h=(\langle E^2\rangle-\langle E\rangle^2)/T^2$.  Another convenient quantities to identify various types of chiral states are the scalar chirality $\chi_{ijk}=\Sp_i\cdot(\Sp_j\times\Sp_k)$ defined for each triangular plaquette and, what is more appropriate to characterize skyrmions states, the discretized scalar chirality defined as \cite{rosales2015three}

\bea
\chi_Q=\frac{1}{4\pi N}\langle \sum_{i}A_{ijk}\,\text{sgn}\left[\chi_{ijk}\right]+A_{ij'k'}\,\text{sgn}\left[\chi_{ij'k'}\right] \rangle
\eea
where $i,j,k$ ($i,j',k'$) are the sites involved in the calculation and $A_{ijk}$ is the local area spanned by three spins $i,j,k$ (and $i,j',k'$) on every elementary triangle. In addition to these quantities computed from the local spins $\{\Sp_i\}$, we introduce the plaquette spin, calculated as the average spin in each elementary triangle (or plaquette), $\Sp^T_j=\sum_{i\in  \bigtriangleup_j,\bigtriangledown_j}\Sp_i/3$, with plaquette index $j$.  Taking into account that each type of plaquette defines a triangular lattice, we also define the plaquette chirality $\chi^T_Q$, calculated by the triple product between plaquette spins $\{\Sp^T_j\}$.

 Finally, it is well known that the structure factor will have distinct characteristics for the CSL phase (the presence of pinch-points \cite{henley10a,essafi2016k}) and for a skyrmion lattice (six bright peaks or triple-k phase \cite{muhlbauer2009sk}). Therefore, we have also calculated he static spin structure factor,  with components $S_{\perp}(\kv)$ and $S_{||}(\kv)$, perpendicular and parallel to the external field respectively, defined as $S_{\perp}(\kv)=\frac{1}{N}\sum_{a=x,y}\langle|\sum_j S^{a}_{j}e^{i\kv\cdot\rv}|^2\rangle$ and $S_{||}(\kv)=\frac{1}{N}\langle|\sum_j S^z_{j}e^{i\kv\cdot\rv}|^2\rangle$, and the related plaquette variables, $S^T_{\perp}(\kv)$ and $S^T_{||}(\kv)$. As seen in the following section, we will focus on the use of the plaquette variables at $D^{xy}=0.5$, to evidence the effect of the CSL physics.

\subsection{In zero field \texorpdfstring{$B=0$}{B=0}}

As we described in Sec.~\ref{sec:model}, there are two disconnected limits of the model presented in Eq.~(\ref{eq:HamSpin}): (a) $D^{xy}=0$ and $D^{z}=\pm\sqrt{3}$, with the spins $xy$ components displaying chiral-spin-liquid (CSL) behavior and (b) $D^{xy}>0$ and $D^{z}=0$ that promotes the emergence of typical spin helical configurations, necessary for the formation of skyrmion phases in a field. Let us see what happens when moving from one limit to the other in zero field.  In Fig.~\ref{fig:PDB0} we present our simulation results, for lattice size $L=48$. The phase diagrams where constructed combining the specific heat, the nearest neighbor chirality and inspection of the snapshots.

We first fix $D^z=\sqrt{3}$ and $B=0$ (Fig.~\ref{fig:PDB0}(a)), inducing the CSL phase for $D^{xy}=0$ with two clear signatures: (i) the presence of the well-known ``pinch points'' in the spin structure factor $S_{\perp}(\kv)$  at low-temperatures  \cite{henley10a,essafi2016k} and (ii) a specific heat $C_h\approx 5/6<1$ (see Fig.~\ref{fig:PDB0}(b)) which reflects the presence of soft-modes \cite{Chalker92a}. In the context of the LTA, as presented in section \ref{sec:LTA}, the initial scenario occurs when the dispersive band intersects with the lowest energy flat band, leading to the observation of pinch points \cite{davier2023combined,yan2023classification}. As the CSL is chiral, time-reversal symmetry in this phase is broken. In the absence of any external magnetic field, it is then a spontaneously broken symmetry, indicating a phase transition between the paramagnetic and CSL phases. As explained in the introduction, this broken symmetry only applies to the $z-$component of the spins (out-of-plane magnetization); the spin-liquid degrees of freedom are entirely in the ($S^x,S^y$) plane \cite{Essafi2017}. For $0< D^{xy} \lesssim 0.35$,  from the LTA analysis we expect strong CSL effects. Indeed,   the CSL phase persists at intermediate temperatures ($C_h<1$), until a sharp peak in specific heat arises at lower temperatures, indicating a transition to a ``labyrinth'' phase (see Fig.~\ref{fig:PDB0}.(a,b,g)). This phase is similar to that found in other chiral magnets \cite{GARANIN2020165724, OharaNANO, Wang-Wenbin, Du-Wenhui}, lacking a defining wavevector but clearly showing a scale in the ``thickness'' of the labyrinth, which varies with $D^{xy}$. This competition between the CSL and labyrinth phases is consistent with the LTA results presented in Sec.~\ref{sec:model}; for low enough $D^{xy}$ the gap to the flat bands remains fairly small (see Fig.~\ref{fig:LTA1}.(f)), which means that the chiral spin liquid state may be accessed through thermal fluctuations. For $D^{xy} \gtrsim 0.35$, the  CSL region is suppressed and the system stabilizes in a distorted helix (DH) phase, which gets smoother as $D^{xy}$ is increased (compare snapshots for $D^{xy}=0.35,0.5$ in panels (h,i)).

Starting from the other limit, we now fix $D^{xy}=0.2$ and vary $D^z$. The resulting phase diagram is presented in Fig.~\ref{fig:PDB0}.(d). The helical order persists up to $D^z\approx 0.6$. For intermediate values, $0.6 \lesssim D^z < 1.35$, the competition between in-plane and out-of-plane DM interactions distorts the helical order and induces a labyrinth phase, while CSL  appears at higher temperatures for $1.35 < D^z \lesssim \sqrt{3}$. Since the phase diagram of panel (d) at $D^z=\sqrt{3}$ overlaps with the one of panel (a), we know that the labyrinth phase must re-emerge at low temperature below the CSL (see the upper violet region). Finally, for $D^z \gtrsim \sqrt{3}$ the system goes into a $q=0$ planar order (see for example panel (j) in Fig.~\ref{fig:PDB0}), due to the dominant positive $D^z$ term \cite{Essafi2017}.

\begin{figure}[ht]
\includegraphics[width=0.99\columnwidth]{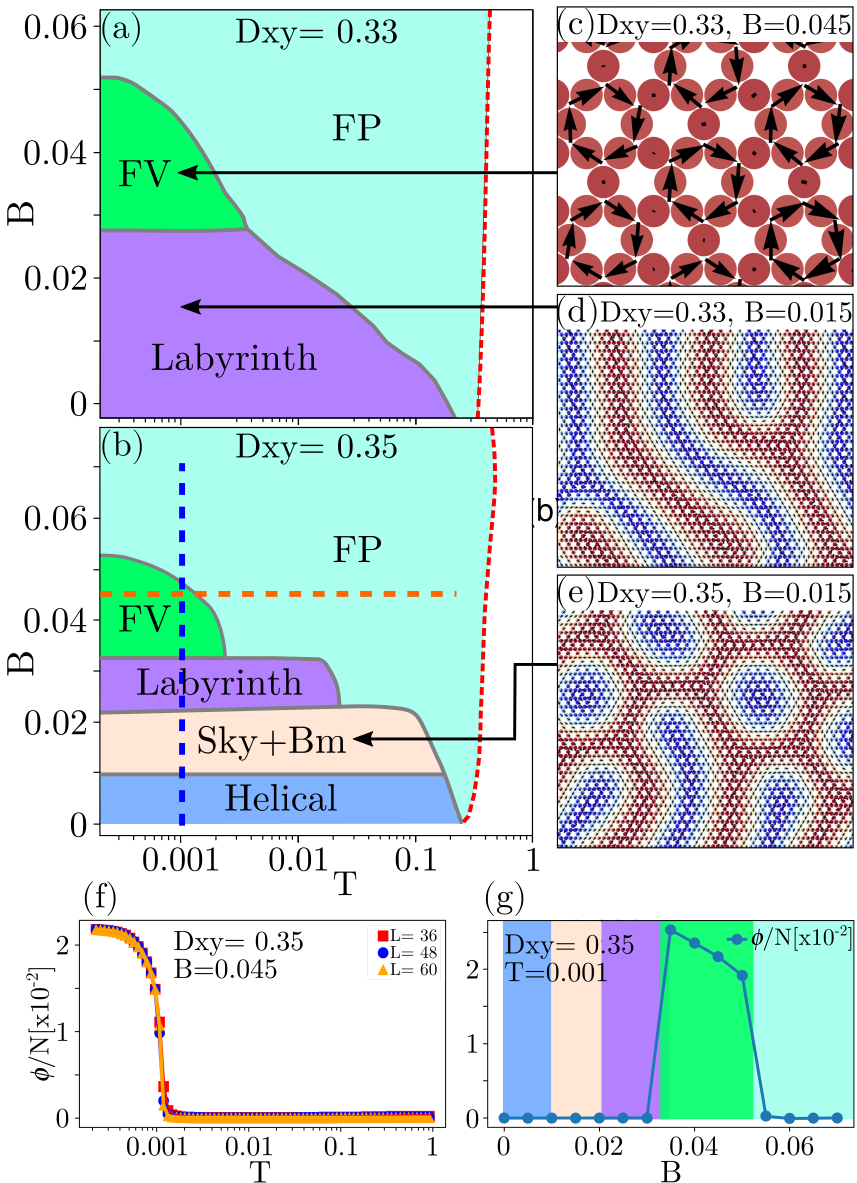}
\caption{$B$ vs $T$ phase diagrams for $D^z=\sqrt{3}$  and $D^{xy}=0.33$ (panel (a)) and $D^{xy}=0.35$ (panel (b)) with representative spin configurations (c-e)  (lattice size $L=48$).
In panel (f)  we present the order parameter $\phi$ (scaled by system size $N$) as a function of temperature for three different system sizes $L=36, 48, 60$ for $D^{xy}=0.35$ and magnetic fields $B=0.045$.}  Panel (g) presents the evolution of $\phi$ as a function of the magnetic field $B$ showing a well-defined region where the FV phase emerges. In panels (f) and (g) the error bars are the size of the markers.  \label{fig:BTDxy033035}
\end{figure}

In a nutshell, it is possible to tune the system from the chiral spin liquid to the helical phase, passing by the labyrinth phase. In particular, the helical phase is accessible, albeit distorted, for $D^z=\sqrt{3}$, which is the closest in parameter space to the CSL. From now on, we will only consider $D^z=\sqrt{3}$ (except briefly in section \ref{sec:BvsT_009}). Now that the zero-field physics is under control, let us turn on the magnetic field to determine in which parameter space we can find skyrmions.

\subsection{The search of skyrmions in a magnetic field}

To explore the emergence of skyrmions in a finite magnetic field, we investigate the cases of  $D^{xy}=0.33$ and $D^{xy}=0.35$ (dashed black lines in Fig.~\ref{fig:PDB0}(a)). As seen, at zero magnetic field, for $D^{xy}=0.33$ there is a small CSL region encroaching upon the labyrinth region at higher temperature, which is not seen for $D^{xy}=0.35$. We thus explore the behavior of the model for these two values of $D^{xy}$ in the presence of a magnetic field, to see the effect of the zero-field CSL region in the formation of skyrmions; in particular, if they can appear out of the labyrinth phase, or if they require the more traditional helical one \cite{han2010}. In Fig.~ \ref{fig:BTDxy033035}(a,b) we present the resulting $B$ vs $T$ phase diagrams.

For $D^{xy}=0.33$, the labyrinth phase persists at low field (a representative snapshot is shown in Fig.~\ref{fig:BTDxy033035}(d)), and then turns into a ``frustrated vortex'' (FV) phase, which we describe in detail below, and at higher field it goes into the field-polarized (FP) phase. This FP phase is actually the magnetized evolution of the CSL that was present in a small temperature window at $B=0$, but where the time-reversal symmetry is broken by the field instead of spontaneously by a transition. The out-of-plane spin components are magnetized when increasing $B$, but the in-plane spin components continue to support a noticeable chirality $|\chi_{Q}|$, signature of the underlying CSL. As seen in the LTA analysis, for these values of $D^{xy}$ the flat band, related to the CSL, is not far from the minima, and thus is accessible through thermal fluctuations and magnetic field.  For $D^{xy}=0.35$, however, the labyrinth phase and the CSL are absent at zero field. When increasing the magnetic field, bimerons and skyrmions appear above the distorted-helical phase,  (Fig.~\ref{fig:BTDxy033035}(e)) before the labyrinth phase, which then evolves into the FV phase. 

\begin{figure*}[ht!]
\includegraphics[width=0.97\textwidth]{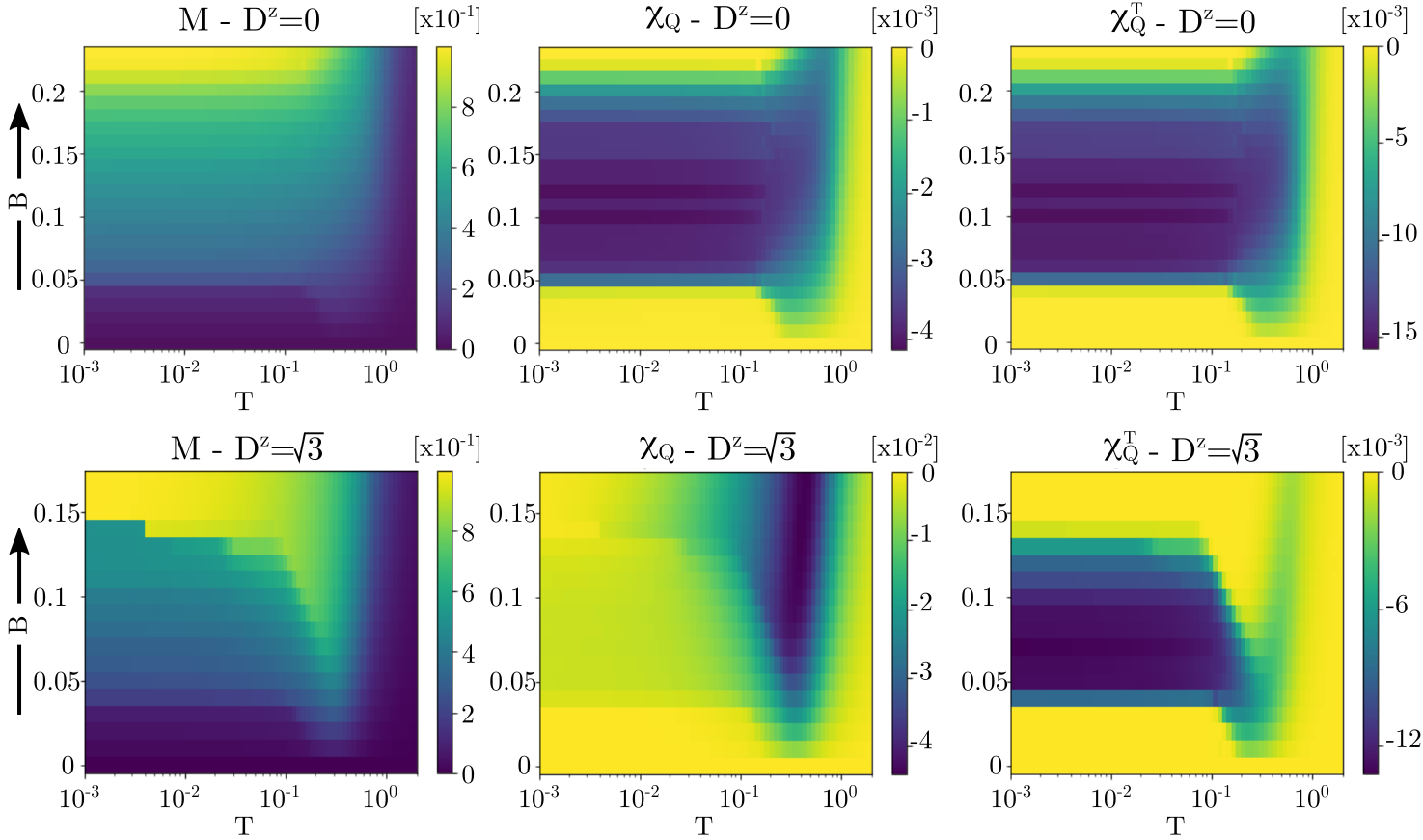}
\caption{Magnetization $M$, first nearest neighbors chirality $\chi_Q$ and plaquette chirality $\chi_Q^T$ presented as density plots in $B-T$ phase diagrams for $D^z=0$ (top) and $D^z=\sqrt{3}$ (bottom), fixing $D^{xy}=0.5$ (lattice size $L=48$). Please note the sharp increase of chirality $\chi_Q$ when $D^z=\sqrt{3}$; the color scale is an order of magnitude higher than for $D^z=0$. }
\label{fig:PDcompasironNoDz}
\end{figure*}

The frustrated vortex phase is favored at higher fields, but it gets smaller as $D^{xy}$ is increased until it disapears at higher $D^{xy}$. As depicted in the snapshot in Fig.~\ref{fig:BTDxy033035}(c), the $xy$ components of the spin in the hexagons of the kagome lattice form vortexes with alternating circulation. It is easily seen that there is a  $Z_3$ symmetry breaking due to the three possible tilings, which have a $3\times 3 $ hexagons magnetic unit cell. These structures are similar to those proposed in \cite{Tchernyshyov2012}, but here the sense of the circulation is fixed by the sign of the DM interaction. To identify this phase, we calculate as an order parameter $\phi$, the planar structure factor $\phi=S_{\perp}(\kv^*)$ at $\kv^*=(4\pi/3,0)$ \cite{Tchernyshyov2012}. In Fig.~\ref{fig:BTDxy033035}(f) we illustrate MC results for $\phi/N$  as a function of $T$ for three different system sizes ($L=36,48,60$)  at $D^{xy}=0.35$ and  $B=0.045$. The order parameter $\phi$ jumps to a finite value below a critical temperature, where the FV phase emerges. Furthermore, in panel (g) we present $\phi$ as a function of $B$ for $D^{xy}=0.35$ at the lowest temperature $T=0.0002$, averaged over independent copies, showing that $\phi$  distinguishes the FV phase from the other ones.

The labyrinth phase at zero field thus seems to prevent the apparition of skyrmions at finite field. This is probably because the skyrmion crystal comes from multi-$q$ order, or in other words, from the interference pattern of the stripes of spins oriented in multiple directions \cite{han2010}. In the labyrinth phase, these different orientations can naturally co-exist in the same spin configuration (see Fig.~\ref{fig:PDB0}.(g,h)). More intuitively speaking, there is no need for the interference pattern of the skyrmion crystal to minimize the energy.\bigskip

As a conclusion of this analysis, and combining the results from simulations and the Luttinger-Tisza approximation, we understand the necessary balance on the value of $D^{xy}$ in order to support both the CSL and skyrmion physics. On one hand, the LTA shows that we need small values of $D^{xy}$ to keep the energy gap of the flat bands reasonable (see Fig.~\ref{fig:LTA1}.(f)), and thus the physics of the chiral spin liquid accessible to thermal fluctuations. On the other hand, we also need $D^{xy}$ to be large enough to support a (distorted) helical phase at zero field; otherwise, the presence of the labyrinth phase at low temperature would prevent the apparition of skyrmions in a field. Hence, we expect the competition between CSL and skyrmion physics to be stronger at intermediate $D^{xy}$. Through the rest of this work, we fix $D^{xy}=0.5$. Now, we shall describe in detail the phase diagram of this model in a field $B$.

\begin{figure*}[htb!]
\includegraphics[width=0.95\textwidth]{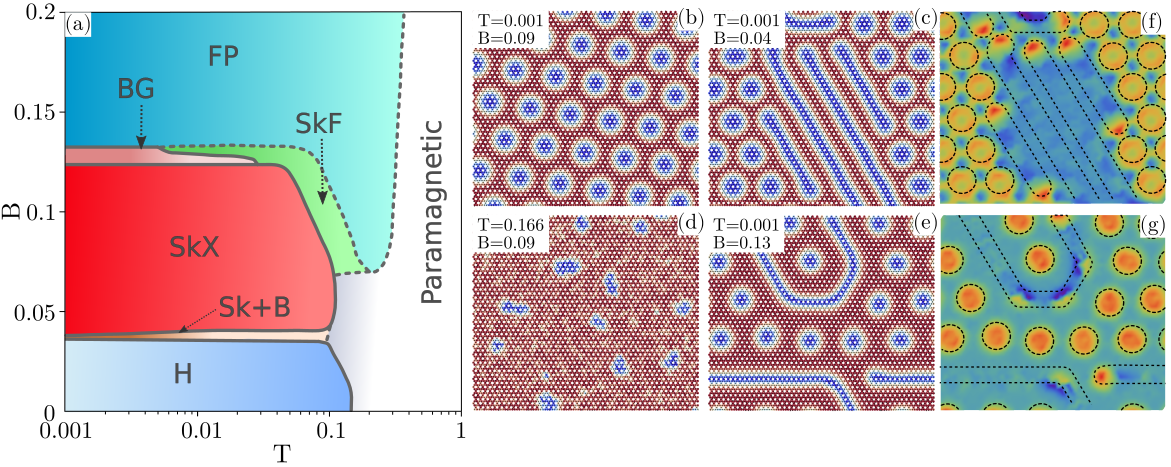}
\caption{
(a) $B-T$ phase diagram obtained from MC simulations for $D^{xy}=0.5, D^z=\sqrt{3}$, where there are helical (H), mixed skyrmion and bimeron (Sk+B), skyrmion crystal (SkX), skyrmion fluid (SkF), bimeron glass (BG) and field polarized (FP) phases, taken from Ref.[\onlinecite{rosales2023skyrmion}]. (b-e) Typical real-space spin configurations obtained by Monte Carlo simulations on a $N=3 \times 48^2$ system size at different temperatures and magnetic field, illustrating various phases of panel (a). (f,g) Comparison of the two phases where bimerons are found, i.e. the low-field Sk+B phase (f) and the high-field BG phase (g), as seen from their plaquette chirality density per plaquette position. Bimerons are surrounded by dashed curves, showing that the chirality is stronger in the extremities.
}
\label{fig:PDsnaps_LowT}
\end{figure*}
\begin{figure*}[tb!]
\includegraphics[width=0.98\textwidth]{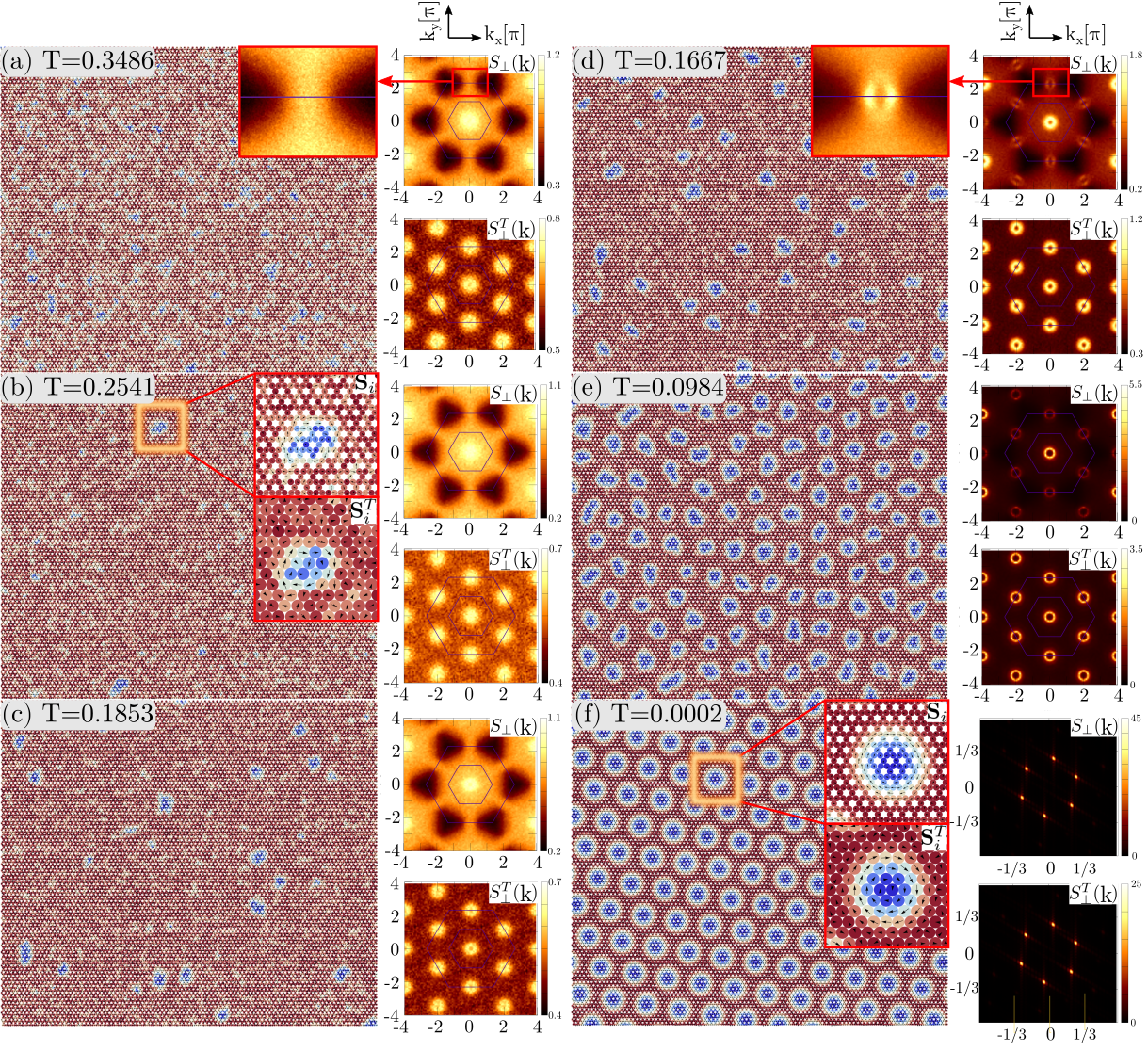}
\caption{
Snapshots of spin configurations taken at different temperatures for our model at $D^{xy}=0.5$ and $D^z=\sqrt{3}$ at fixed magnetic field $B=0.09$, for large system sizes ($L=96$). Each panel is accompanied by the corresponding structure factors $S_{\perp}(\kv)$ (top right) and $S^T_{\perp}(\kv)$ (bottom right), averaged over Monte Carlo time. $S_{\perp}(\kv)$ is the traditional magnetic structure factor for the transverse $(S^x,S^y)$ spin component. $S^T_{\perp}(\kv)$ is the same but the degrees of freedom are the total spin of each triangular plaquette. In panels (b) and (f), we zoom in on a skyrmion (inset, above) and show how it looks like from the point of view of the total spins per plaquette (inset, below). Please note that the Fourier space in panel (f) is smaller than the others; it is zoomed in to emphasize the 6 Bragg peaks characteristic of the skyrmion lattice.
}
\label{fig:texturesCSL2SkL}
\end{figure*}
%

\subsection{Competition between skyrmions and a chiral spin liquid}
\label{sec:BvsT_009}

In order to quantify the influence of the spin liquid, it is useful to know what happens in absence of it. To do so, we compare our model (where $D^z=\sqrt{3}$) to a standard model for skyrmions (where $D^z=0$). In Fig.~\ref{fig:PDcompasironNoDz} we present $B-T$ density plots for the magnetization, nearest neighbor chirality, and plaquette chirality for both Hamiltonians.

The $D^z=0$ case presents the features of a typical skyrmion model. At low field is the helical phase, with quasi-zero magnetization and chirality because the stripes in the spin configuration bear alternatively positive and negative magnetization. As we increase the field $B$, there is a jump in the spin and plaquette chirality and a gradual increase of the magnetization; this is the region of the skyrmion crystal (SkX). The chirality vanishes again when the magnetization saturates at high field; this is the field-polarized (FP) phase.

The picture becomes noticeably different when $D^z=\sqrt{3}$. The FP phase goes down at high temperature, circling around the skyrmion crystal, and accompanied by a sharp increase of the spin chirality; $\chi_Q$ is 10 times bigger in this high-temperature FP regime than in the SkX ! This high-temperature FP regime is clearly not due to skyrmions because the plaquette chirality $\chi_Q^T$ now behaves differently from the spin chirality $\chi_Q$. This is due to the CSL, where under a magnetic field the out-of-plane spin components tend to be aligned with the field. As the planar spins components in each plaquette are either chiral or ferromagnetic (see Fig.~\ref{fig:LTA1}), this structure is washed out when adding the three spins in each plaquette to calculate $\chi_Q^T$. Therefore, we may identify an extended chiral spin liquid region in the density plots by noticing there is a large region at higher temperatures, starting at intermediate fields, where $\chi_Q$ is quite large but $\chi_Q^T$ is significantly smaller.

The phase diagram for $D^z=\sqrt{3}$ and $D^{xy}=0.5$ is given in Fig.~\ref{fig:PDsnaps_LowT}(a), as determined in Ref.~[\onlinecite{rosales2023skyrmion}]. We summarize the main results of Ref.~[\onlinecite{rosales2023skyrmion}] below:\bigskip

\noindent (i) {\it Helical} (H): in Fig.~\ref{fig:PDB0}(i) we show an example of a helical phase, present at low temperatures and low fields.\\
\noindent (ii) {\it Skyrmions + Bimerons} (Sk+B):  a very thin metastable region characterized by a mixture of  skyrmions + bimerons \cite{ezawa2011c,mohanta2020} (Fig.~\ref{fig:PDsnaps_LowT}(c)) emerges between the helical and the skyrmion lattice phases.\\
\noindent (iii) {\it Skyrmion Lattice} (SkX): a typical skyrmion lattice (Fig.~\ref{fig:PDsnaps_LowT}(b)) is found at intermediate fields and low temperature, matching the region where $\chi_Q^T$ is higher in Fig.~\ref{fig:PDcompasironNoDz} (bottom right panel). This phase is a superposition of three spiral orders that preserves the C3 symmetry of the lattice.\\
\noindent (iv) {\it Field Polarized} (FP): in this phase, as the temperature is lowered, magnetic moments are further aligned with  the field, but the $xy$ components retain the extended degeneracy of the chiral spin liquid, as explained in section \ref{sec:model}.\\
\noindent (v) {\it Skyrmion-Fluid} (SkF): this phase can be qualitatively separated in a dense, or skyrmion liquid phase (SkL) and a dilute phase, dubbed skyrmion gas (SkG). Unlike the SkF usually present at higher fields and low temperatures in a typical skyrmion model ($D^z=0$), this SkF phase is found at higher temperatures, and skyrmions are found on a chiral spin liquid background (Fig.~\ref{fig:PDsnaps_LowT}(d)). \\
\noindent (vi) {\it Bimeron Glass} (BG): an intermediate phase between the SkX and field polarized region emerges at lower temperatures (Fig.~\ref{fig:PDsnaps_LowT}(e)), which is not present when $D^z=0$. In Fig.~\ref{fig:PDsnaps_LowT}(f,g), we present the plaquette chirality density of the Sk+B and  BG phases, illustrating that the chirality in bimerons is located in their extremities and that bimerons in the BG phase are more extended.\\
%
\begin{figure*}[htb!]
\includegraphics[width=1.0\textwidth]{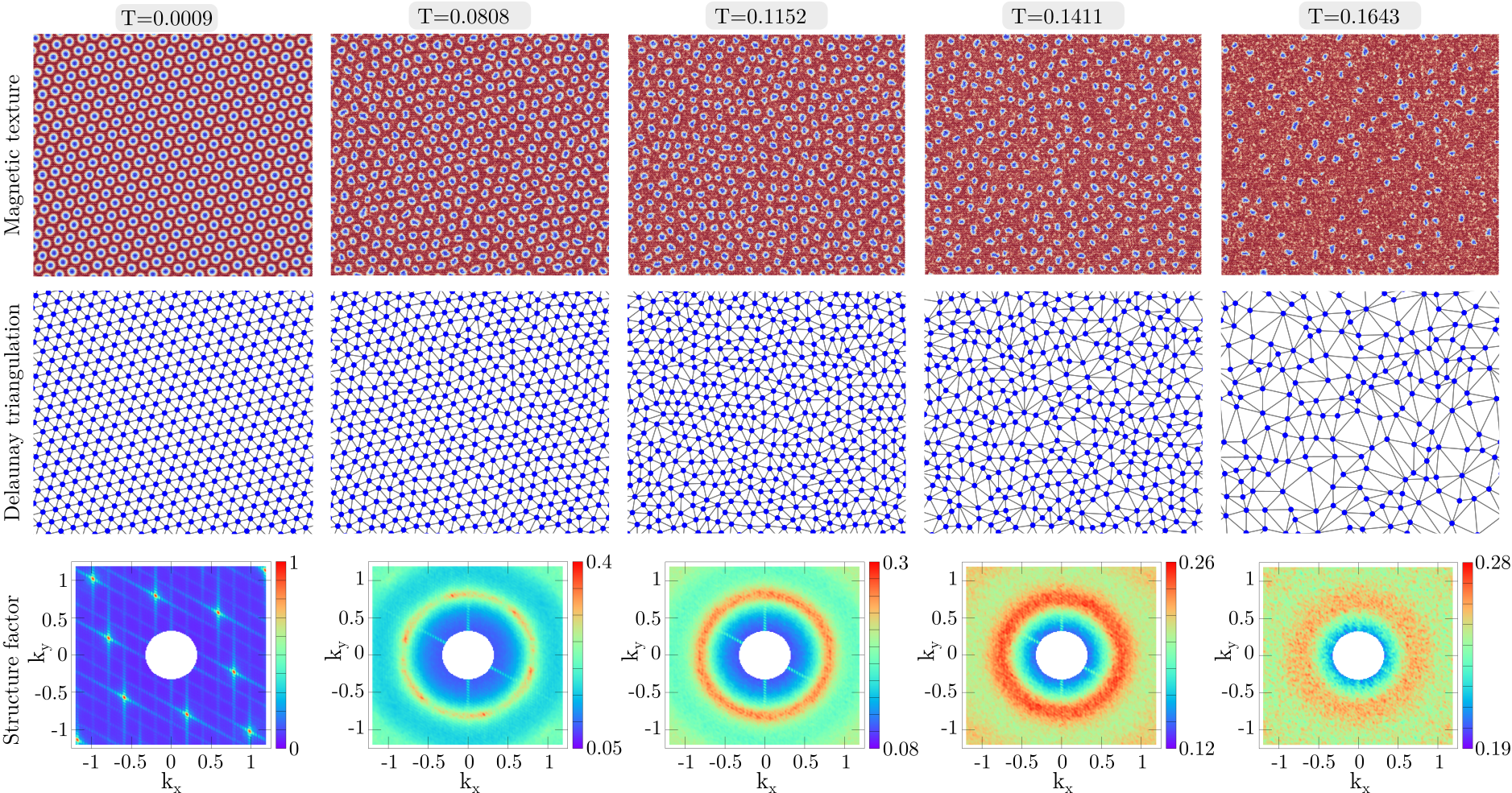}
\caption{ Spin configuration (top row),  Delaunay triangulation (middle row) and  structure factor $S_q({\bf q})$ (bottom row) for a representative set of temperatures of the different regimes for $D^{\perp}=0.5$, $D^z =\sqrt{3}$ and $B=0.09$. Here it is the structure factor of the position of skyrmions, given by $S_q ({\bf q})= \frac{1}{N_{\text{Sk}}}\sum_{{\bf r}}e^{-{\bf q}\cdot{\bf r}}\sum_{i,j} \delta({\bf r}-{\bf r}_{ij})$ where the summation in r is over all skyrmion positions, the summation in $i,j$ is over all skyrmion pairs, and ${\bf r}_{ij}$ is the vector between skyrmions $i$ and $j$.  Figure taken from Ref.[\onlinecite{rosales2023skyrmion}], Supp. Material. (lattice size $L=192$)}
\label{fig:delauney}
\end{figure*}

A remarkable feature of our model is the formation of a skyrmion gas at high temperature, emerging from a CSL background. The difference with traditional skyrmion models (e.g. when $D^z=0$) is that the CSL provides a large entropy to the field-polarized phase (FP). The chirality $\chi_Q$ of Fig.~\ref{fig:PDcompasironNoDz} clearly illustrates how this enhanced entropy of the FP phase separates the paramagnetic regime to the SkX order, imposing an intervening region with quasi-zero skyrmions. This is how the density of skyrmions becomes controllable with temperature, even forming a diluted gas, before skyrmions are destroyed by paramagnetic fluctuations.

In Fig.~\ref{fig:texturesCSL2SkL} we present several snapshots and their corresponding spin-spin structure factor $S_{\perp}(\kv)$ and plaquette-plaquette structure factor $S_{\perp}^T(\kv)$ (calculated with the total spin per plaquette), at different temperatures for $B=0.09$. From panels (a) to (c) we see how well-defined skyrmions start to emerge as $T$ is lowered. The underlying CSL background is signaled by the characteristic pinch points in $S_{\perp}(\kv)$, albeit broadened by thermal fluctuations and the presence of skyrmions. As $T$ is lowered further, more skyrmions appear with no particular pattern, but with distinctive changes in the structure factors (panel (d)).  Pinch points are replaced by ``half moons'' \cite{robert08a,mizoguchi2018m,yan2018half}, indicating the disappearance of the CSL. And the circles of intensity in $S_{\perp}^T(\kv)$ become hollow circles, whose radius indicates the apparition of a length scale, typically the mean distance between skyrmions. These bright circles, which strongly suggest a liquid-like behavior, are still present as the lattice is more densely populated by skyrmions (panel (e)), and disappear at the lowest temperatures in favor of Bragg peaks when the skyrmion lattice is formed (panel (f)). Given the rich variety of emergent phases and their unique characteristics, in the following section, we focus on carefully characterizing this transition from fluid to solid.

\section{Fluid to solid phase transition}

In this section we focus on the transition from the CSL phase to the SkX phase, going through a skyrmion fluid, and how the skyrmions ``crystallize'' at low temperatures. For this purpose, we choose a region in parameter space where several different phases emerged as the temperature was lowered, $D^{xy}=0.5, D^z=\sqrt{3}$ and $B=0.09$, as discussed in Sec.~\ref{sec:BvsT_009}.
An important fact to carry on this analysis is that the individual skyrmions persist throughout the whole melting process \cite{timm1998sk}. 
Apart from a crystal/solid phase at low temperature and a fluid at high temperature,  a third intermediate hexatic phase may emerge, characterized by short-range translational order and quasi-long-range orientational order. However, this is still an open problem with several scenarios depending on the nature of model \cite{huang2020m,balavz2021m,nishikawa2019s}.

\begin{figure}[thb]
\includegraphics[width=0.97\columnwidth]{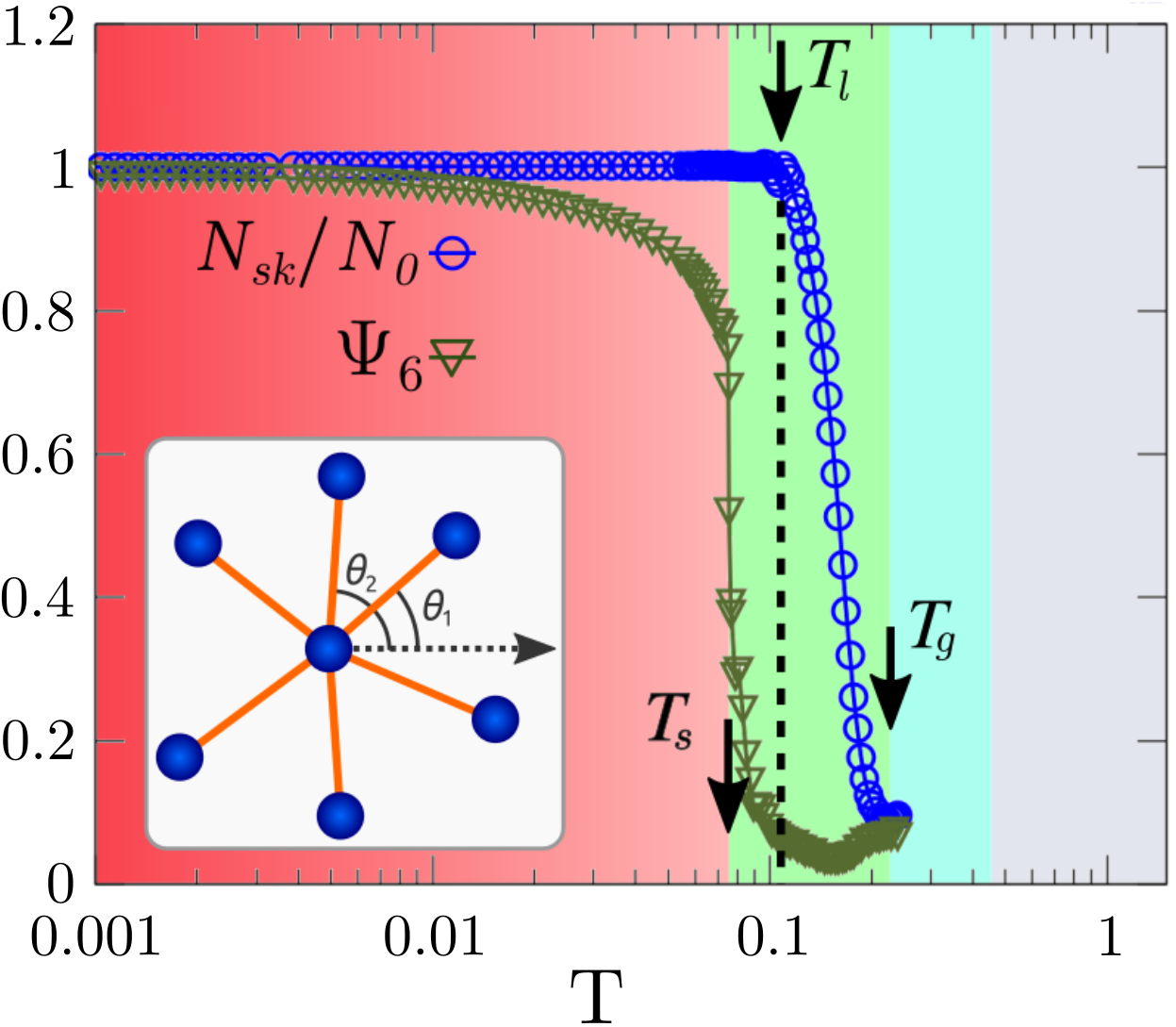}
\caption{Competition between the chiral spin liquid and skyrmion solid at $B=0.09$ as measured from: \textbf{(a)} the specific heat $C_{h}$ and magnetisation $M_{z}$, \textbf{(b)} the normalised number of skyrmion $N_{S}$ and orientational order parameter $\Psi_{6}$ of Eq.~(\ref{eq:psi6}) whose angle $\theta_{ij}$ is defined in the inset. $N_{0}$ is is the saturated number of skyrmions for $B=0.09$. (Figure taken from Ref.[\onlinecite{rosales2023skyrmion}])}
\label{fig:CvNPsi6vsT}
\end{figure}

Here, by means of a combination of Monte Carlo algorithms (Metropolis and Heat-Bath), using large system sizes ($L=96,192$), real-space identification of the skyrmion positions, and a Delaunay triangulation, we study the emergence of SkX tuning the temperature $T$. To do so, we first assigned a real-value position to each skyrmion in the lattice, computing it as the center of mass of one individual skyrmion $\Rv_{i}=\frac{1}{n}\sum_{a\in Sk}\rv_a$ where $\rv_a$ denotes the position of the $n$ spins in skyrmion $i$, defined as a contiguous cluster of spins with $S^z<0$. Each skyrmion $i$ is surrounded by $n_i$ nearest neighbors determined by a standard Delaunay triangulation. Each nearest neighbor $j\in \{1,...,n_i\}$ sits at position $\Rv_i + \vec r_{ij}$. If $\theta_{ij}$ is the angle formed between the $x-$axis and $\vec r_{ij}$ (see inset in Fig.~\ref{fig:CvNPsi6vsT}(b)), then we can define the local orientational order parameter $\psi_6(\Rv_i)$ [\onlinecite{nelson1979d}],
\begin{eqnarray}
\psi_6(\Rv_i)&=&\frac{1}{n_i}\sum_{j=1}^{n_i}e^{i\,\theta_{ij}},
\end{eqnarray}
which is a standard measure to quantify the emergence of local hexagonal order in a 2D ensemble of particles. We can also define the global orientational order parameter as
\begin{eqnarray}
\Psi_6&=&\frac{1}{N}\sum_{i=1}^{N}\psi_6(R_i)
\label{eq:psi6}
\end{eqnarray}
Then, the orientational correlation function $G_6(R)$ is defined as:
\begin{eqnarray}
G_6(R)&=&\frac{1}{n_R}\sum_{|\Rv_i-\Rv_j|=R}\psi_6(R_i)\,\psi^{*}_6(R_j)
\end{eqnarray}
where the sum is over all $n_R$ particle pairs at distance $R$. In addition, we introduce the  translational correlation function $G_k(R)$ defined as

\begin{eqnarray}
G_K(R)&=&\frac{1}{6}\sum_{a=1}^{6}\frac{1}{n_R}\sum_{|\Rv_i-\Rv_j|=R}\psi_{k_a}(R_i)\,\psi^{*}_{k_a}(R_j)
\end{eqnarray}
where $\psi_{k_a}(\Rv_i)=e^{i\,\kv\cdot\Rv_i}$ is the translational order parameter and $\kv_a$ $(a = 1, ..., 6)$ are the reciprocal lattice vectors determined by the positions of the first-order Bragg peaks in the structure factor of the position of the skyrmions $\rv$, given by $S_k(\kv)=\frac{1}{N_S}\sum_{\Rv}\exp^{i\kv\cdot\Rv}\sum_{i,j}\delta(\Rv-\Rv_{ij})$, where $N_S$ is the total number of skyrmions in the lattice. The nature of the orientational and translational correlation functions, whether they are short-ranged or quasi-long ranged, together with the values of the exponents, will be the key to classify the various phases.

 Before exploring these parameters, we briefly recall some relevant results from our previous work \cite{rosales2023skyrmion}, to show how the structure factor from the Delauney triangulation (Fig.~\ref{fig:delauney}) and variables such as the specific heat and the chirality (Fig.~\ref{fig:CvNPsi6vsT}) behave in the different phases. In Fig.~\ref{fig:delauney} we show, on the first row, the real-space skyrmion structures at five different temperatures and, on the second row, the filtered particle configurations and corresponding Delaunay triangulation. At  low temperature ($T=0.0009$), the skyrmion crystal is stabilized in the form of a triangular lattice, characterized by the well-known six bright peaks in the structure factor. This lattice gets distorted at higher temperatures, and the Bragg peaks turn into a bright circle ($T=0.1152$), suggesting a fluid-type behavior, which broadens upon heating until the lattice structure essentially vanishes. This disorder in the positions of skyrmions is accompanied by a reduction of their density, granting them more freedom to occupy non-regular positions.

In Fig.~\ref{fig:CvNPsi6vsT}, we compare the behavior of the skyrmion density $N_{S}$ (normalized with $N_0$, the number of skyrmions at $T\to 0$) with the thermally averaged $\langle|\Psi_6|\rangle$ parameter. We can  observe that $\langle|\Psi_6|\rangle$ is saturated at low temperature (crystal phase), and sharply decreases at $T_s$. In this window of temperature ($T_s<T<T_l$), $N_{S}/N_0$ remains saturated to 1, and thus corresponds to a dense fluid of skyrmions, or skyrmion liquid (SkL).  $N_{S}/N_0$ then drops from $T_l$ to $T_g$, defining the region for the skyrmion gas (SkG).

\begin{figure}[ht]
\includegraphics[width=1.0\columnwidth]{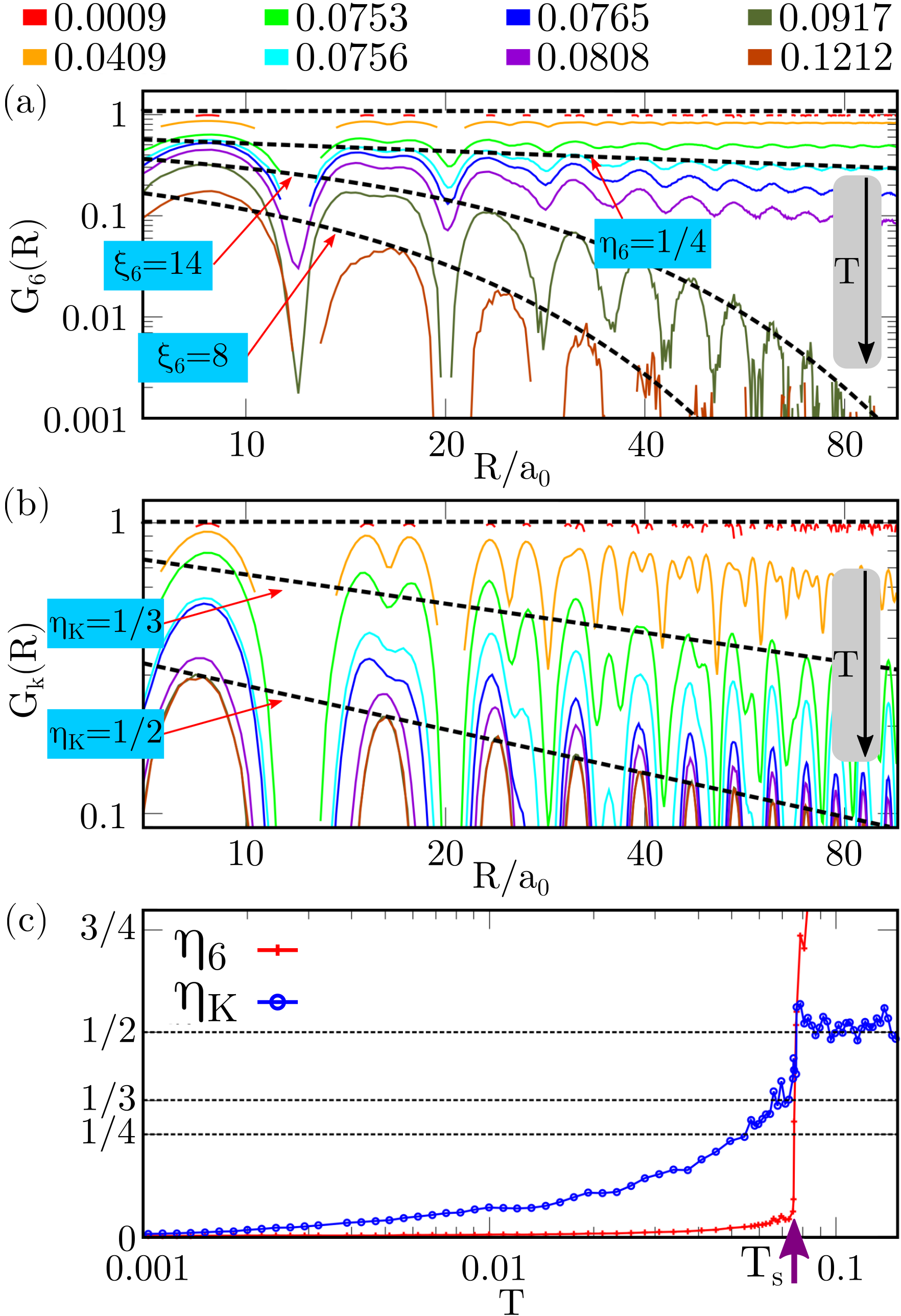}
\caption{ Orientational $G_6(R)$ (a) and positional $G_K(R)$ (b) correlation functions as a function of the distance between skyrmions ($R$), calculated for a fixed magnetic field $B/J=0.09$ at different temperatures $T$. The dashed black curves are typical fits of their local maxima to power-law  or exponential decay. (c) Temperature dependence of the power-law exponents $\eta_6$ and $\eta_K$.}
\label{fig:G6andGKvsR}
\end{figure}

 In order to clarify and characterize the different phases that emerge as temperature varies, in Fig.~\ref{fig:G6andGKvsR} we show the calculated $G_6(R)$ and $G_K(R)$ and the corresponding
fits of its upper envelopes at different temperatures in log-log scale. Following the theory proposed by the Kosterlitz, Thouless, Halperin, Nelson and Young (KTHNY) \cite{Kosterlitz1973,Halperin1978,nelson1979d,Young1979}, we  recall the expected behaviour of the correlations in each type of phase:

\begin{itemize}
\item crystal phase: $G_6(R)\to 1$ and $G_K(R)\to 1$
\item solid phase:  $G_6(R)\to 1$ and $G_K(R)\propto R^{-\eta_K}$, with $\eta_K<1/3$. 
\item hexatic phase: $G_6(R)\propto R^{-\eta_6}$, with $\eta_6<1/4$  and $G_K(R)\propto e^{-R/\xi_K}$.
\item liquid phase: $G_6(R)\propto e^{-R/\xi_6}$  and $G_K(R)\propto e^{-R/\xi_K}$.
\end{itemize}

Figs.~\ref{fig:G6andGKvsR}(a) and (b) show the $G_6(R)$ and $G_K(R)$ as a function of distance $R$ between skyrmions for different temperatures. It can be seen that  at the lowest temperature  ($T/J\sim 9\times 10^{-4}$), both the orientational and positional correlation functions are constant with skyrmion distance and remain close $G_6(R)\sim 1$ and $G_K(R)\sim 1$, indicating a crystal phase. By increasing the temperature, $G_6(R)\sim 1$ while $G_K(R)$ follows an algebraic behavior with $\eta_K<1/3$, consistent with the characteristic quasi-long-range positional ordering of a solid-like phase.

In Fig.~\ref{fig:G6andGKvsR}(c)  we show the evolution of the power-law exponents $\eta_6$ and $\eta_K$ obtained from the fits of the local maxima of the correlation functions at a given temperature, in the range where an algebraic fit is adequate. We observe that $\eta_6$ increases rapidly when the temperature approaches $T\sim T_{s}$. This abrupt change is also reflected in the parameter $\langle|\Psi_6|\rangle$ (Fig.~\ref{fig:CvNPsi6vsT}). At the temperature  $T=T_{s}\sim 0.0769\,J$, $\eta_6$ crosses the specific value $\eta_6= 1/4$, the upper bound for $\eta_6$ if the system were in a hexatic phase, which means we have entered the liquid phase. However, for $T<T_{s}$, $G_K(R)$ decay as a power-law with $\eta_K<1/3$, as expected for a solid phase. This situation differs from the hexatic state in 2D systems, where $G_K(R)$ is expected to decay exponentially at a large distance. A similar situation has earlier been studied in the context of a vortex lattice in a Type II superconductor \cite{ganguli2015d}. This state could be an orientational glass (OG) with a slowly decaying orientational order. At $T\approx T_{s}$, the structure factor presents a reduction in the peak amplitude while the sharp peaks are widened, which is similar to that characteristic of the hexatic phase  (see Fig.~\ref{fig:delauney}, bottom row). Nonetheless, the present results suggest that the potential hexatic phase in our model is either in a very narrow temperature range or non-existent.

\begin{figure}[htb]
\includegraphics[width=0.9\columnwidth]{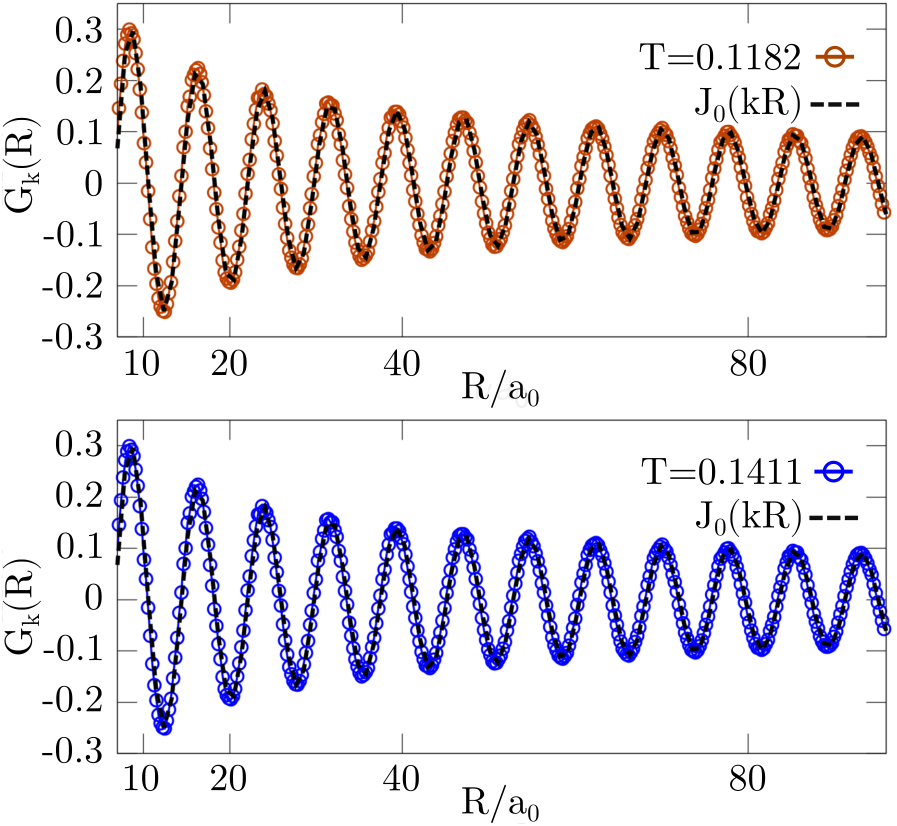}
\caption{Positional correlation function calculated for $\langle G_{|K|}(R)\rangle$ (dashed black line) and $G_{K}(R)$ from MC simulations  at fixed magnetic field $B/J=0.09$. For both temperatures the system is in the SkG phase.} 
\label{fig:Gk_largeT}
\end{figure}

For temperatures $T_{s}<T<T_{g}$, $G_6(R)$ presents  short-range correlations (Fig.~\ref{fig:G6andGKvsR}(a)), while $G_K(R)\propto R^{-\eta_K}$ ($\eta_K \sim 1/2$). This phase corresponds to the fluid phase with a characteristic widened ring distribution in the structure factor (Fig.~\ref{fig:delauney}). Although for a fluid phase short-range order for the translational correlations is expected, the present behavior for $G_K(R)$  can be understood as follows: the crystal/solid phase presents well-defined sharp peaks in the structure factor, while the fluid phase supports a ring-like distribution of peaks with homogeneous amplitude. Then, we must calculate the average of $G_K(R)$ in the whole $\kv-$ring. If we write $\kv\cdot\Rv_{ij}=k\,R\,\cos\theta^{k}_{ij}$ in the definition of the translational order parameter $\psi_{k_a}(\Rv_i)$, then

\begin{eqnarray}
\langle G_{|K|}(R)\rangle&=&\frac{1}{2\pi k}\int_{0}^{2\pi} d\theta^{k}_{ij}\,k\,G_{K}(R)\nonumber\\
&=&\frac{1}{2\pi}\int d\theta_k \left(\frac{1}{6}\sum_{a=1}^{6}\frac{1}{n_R}\sum_{R}\cos(k\,R\,\cos\theta^{k}_{ij})\right)\nonumber\\
&=&\frac{1}{2\pi}\int_{0}^{2\pi} d\theta_k \left(\frac{2}{n_R}\sum_{R}\right)\cos(kR\,\cos\theta^{k}_{ij})\nonumber\\
&\propto&J_{0}(k\,R)
\label{fig:GkAvq}
\end{eqnarray}

\noindent where $J_{0}(k\,R)$ is Bessel function of the first kind ($n=0$). This description is confirmed in  Fig.~\ref{fig:Gk_largeT} where we compare the average value $\langle G_{|K|}(R)\rangle$ in Eq.~(\ref{fig:GkAvq}) to $G_K(R)$ from  MC simulations.

Finally, for temperatures $T>T_{g}$ we have the chiral spin liquid behavior discussed in the previous sections.

\section{Rotation of the Hamiltonian}

When $D^{xy}=0$, the ground state of Hamiltonian (\ref{eq:HamSpin}) is the chiral spin liquid, where each triangle is either ferromagnetic or an umbrella state with a fixed magnetic chirality imposed by the sign of $D^z$ \cite{essafi2016k} (see Fig.~\ref{fig:latt}(b,c)). It has been shown that this Hamiltonian can be mapped exactly onto a specific point of the XXZ model on the kagome lattice \cite{essafi2016k}, without any DM interaction. Since the mapping is exact, the ground state of this XXZ Hamiltonian is an equivalent spin liquid, but where all triangles are in an umbrella state, with either positive or negative magnetic chirality; our chiral spin liquid has thus lost its magnetic chirality. And since the mapping is a rotation of in-plane spin components (different for each kagome sublattice), the out-of-plane magnetic field is invariant. When applying this mapping to our present model with finite $D^{xy}$, we thus obtain an XXZ model in a magnetic field, with rotated in-plane DM interactions,

\begin{figure}[h!]
    \centering
    \includegraphics[width=0.9\columnwidth]{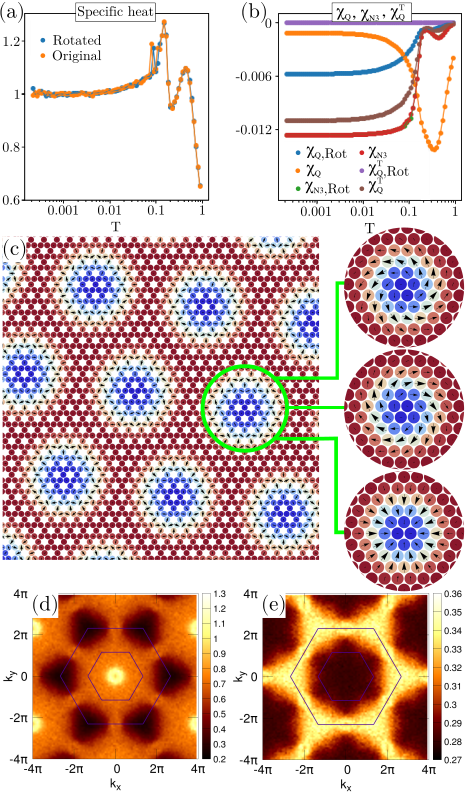}
    \caption{Top: Comparison of specific heat (a), nearest neighbor chirality $\chi_Q$, third nearest neighbor chirality $\chi_Q^{N3}$ and plaquette chirality $\chi_Q^T$ (b) as a function of temperature for the original (Eq.~(\ref{eq:HamSpin})) and rotated (Eq.~\ref{eq:hrot}) Hamiltonians. (c) Low-temperature spin configuration of the SkX lattice (left) and for each sublattice (right). Comparison of the magnetic structure factor $S^\perp$ in the spin-liquid regime at $T=0.1853$ between the (d) original and (e) rotated Hamiltonians. All data have been measured for $D^{xy}=0.5$ and $B=0.09$  (lattice size $L=48$).}
    \label{fig:rot}
\end{figure}
\bea
\noindent \mathcal{H}_{rot}&=&2|J|\sum_{i,j}\left(\Sp_i^{\perp}\cdot\Sp_j^{\perp} -\frac{1}{2}S^z_iS^z_j\right) \\ \nonumber
\noindent && + \frac{D^{xy}}{2}\left(-S^x_iS^z_j + S_i^zS_j^x\right) \\
&& - \frac{\sqrt{3}}{2}D^{xy}\left(S_i^yS_j^z + S_i^zS_j^y\right)  - B\sum_iS^z_i. \nonumber
\label{eq:hrot}
\eea

Here it is important to understand that since the transformation between Hamiltonians (\ref{eq:HamSpin}) and (\ref{eq:hrot}) is exact and conserves the spin length, the energy spectrum of these two models is the same; it is only the corresponding eigenstates (i.e. spin configurations) that are transformed by the mapping. It means that the phase diagram of Hamiltonian (\ref{eq:hrot}) is the same as in Fig.~\ref{fig:PDsnaps_LowT}, albeit with different magnetic textures. For example, the specific heat is the same between the two models in Fig.~\ref{fig:rot}(a), but the magnetic chiralities in Fig.~\ref{fig:rot}(b) are different. The spin chirality $\chi_Q$ is much lower at high temperature for the rotated Hamiltonian (\ref{eq:hrot}) because the spin liquid is not chiral anymore; there are as many umbrella states with positive than with negative chirality ion average. And the in-plane spin components of the umbrella states always have zero magnetization; the plaquette magnetization is thus only along the $z-$axis and the plaquette chirality $\chi_Q^T$ is necessarily zero.

This mapping nonetheless raises the question of skyrmions. Thermodynamically, there must be phases equivalent to the skyrmion solid and fluid in the phase diagram, but the form of their magnetic texture after transformation is less straightforward. The presence of skyrmions can be measured by the third-nearest-neighbor chirality $\chi_{N3}$, where we see that both models give the same result in Fig.~\ref{fig:rot}(b). This is confirmed by looking at snapshots of spin configurations in Fig.~\ref{fig:rot}(c), where we show that the transformation simply changes the helicity in each sublattice. Finally, the magnetic structure factor $S^\perp$ of the spin liquid changes from ferromagnetic to antiferromagnetic pinch points in Fig.~\ref{fig:rot}(d,e), as was measured in Ref.~[\onlinecite{essafi2016k}].

In summary, we have shown here that all the physics developed in this paper does not necessarily requires an out-of-plane Dzyaloshinskii-Moriya term, but can also be obtained in a simple XXZ model on kagome.

\section{Hall conductivity of chiral magnets}

Itinerant electrons are especially useful probes for chiral magnetic textures  and have a long history with frustrated magnetism \cite{batista16a,doi:10.1126/science.1058161,PhysRevLett.98.057203,Uehara22}, especially on the kagome lattice, due to the natural occurrence of scalar spin chirality in this geometry \cite{ohgushi2000spin,Taillefumier06a,xu15a,rosales2019frustrated}, even if virtual \cite{Mertig2020} or attached to antiferromagnetic order \cite{Chen14a,Chen2022}, spin liquids \cite{Udagawa13a,Ishizuka2013,Chern2014} and more recently to Fe$_3$Sn$_2$ \cite{ye2018}. The unconventional electron conduction properties of skyrmion systems have also attracted a rather intense effort over the past years, especially related to anomalous and topological Hall effects (THE) \cite{doi:10.1143/JPSJ.73.2624,yi09_skyrm_anomal_hall_effec_dzyal,hamamoto2015quantized,gobel2017unconventional,gobel2017signatures,gobel2018family,tome2021topo,Wang2021,Kolincio23,Kathyat2021,Mukherjee2023,terasawa2023anomalous}. In this section, we address how the competition between the chiral spin liquid and skyrmion phases (gas, liquid and solid) leads to different types of THE, focusing on the parameter set of $D^{xy}=0.5, D^z=\sqrt{3}$ and $B=0.09$ where all these chiral phases are stable.

\subsection{The Kondo lattice}

To describe the coupling between the itinerant electrons and magnetic textures, we consider the classical Kondo lattice Hamiltonian, which has been widely used to address the transport properties of the electron-spin coupled systems\cite{furukawa1994transport,Udagawa2012,Udagawa2015,Wang2016a,Chern2013a}:

\begin{eqnarray}
\mathcal{H}_K = \mathcal{H}_{\rm kin} + \mathcal{H}_{\rm int} + \mathcal{H},
\label{eq:Ham_cl_Kondo}
\end{eqnarray}
where $\mathcal{H}$ is the Hamiltonian of Eq.~\ref{eq:HamSpin} between localized moments, and 
\begin{eqnarray}
\begin{cases}
\mathcal{H}_{\rm kin} = -t\sum_{\langle ij\rangle,s}(c^{\dag}_{is}c_{js} + c^{\dag}_{js}c_{is}),\\
\mathcal{H}_{\rm int} = -J_{\rm K}\sum_i {\mathbf s}_i\cdot{\mathbf S}_i.
\end{cases}
\label{eq:Ham_cl_Kondo_parts}
\end{eqnarray}
Here, $\mathcal{H}_{\rm kin}$ is the kinetic energy of itinerant electrons: $c^{\dag}_{is} (c_{is})$ is a creation (annihilation) operator of an electron at a site $i$ and spin $s$. The summation over $\langle ij\rangle$ is taken over the nearest-neighbor pairs of sites. The electrons interact with the localized moments through the Kondo coupling represented by $\mathcal{H}_{\rm int}$, where ${\mathbf S}_i = (S_{ix}, S_{iy}, S_{iz})$ is a classical localized moment defined at a site $i$ on the kagome lattice, and ${\mathbf s}_i\equiv\frac{1}{2}c^{\dag}_{is}{\bm\sigma}_{ss'}c_{is'}$ is an electron spin, with ${\bm\sigma}_{ss'}$, the vector notation of Pauli matrices. Note that we do not consider the direct coupling of itinerant electrons with the external magnetic field $B$.
As a transport property of the system, we focus on the Hall conductivity, which is obtained from the Kubo formula, as
\begin{align}
\sigma_{xy}=\frac{2\pi}{V}\sum\limits_{m,m'}&(f(E_{m}) - f(E_{m'}))\nonumber\\
&\times\frac{{\rm Im}(\langle m|J_x|m'\rangle\langle m'|J_y|m\rangle)}{(E_{m} - E_{m'})^2 + 1/\tau^2}.
\label{eq:KuboFormula}
\end{align}
Here, $\sigma_{xy}$ is given in unit of $e^2/h$. $V$ is the volume (area) of the system.
In Eq.~(\ref{eq:KuboFormula}), $|m\rangle$ is the one-particle eigenstate of the Hamiltonian (\ref{eq:Ham_cl_Kondo}), and $E_m$ is the corresponding eigenenergy labeled in ascending order, namely $E_m\leq E_{m+1}$. $J_{\nu}$ is the current operator in $\nu$ direction, and $1/\tau$ is a dumping rate due to non-magnetic impurities, which are implicitly assumed.
We consider a clean system, and set the damping rate $1/\tau$ to be small, $1/\tau=0.001$.
As we are generally interested in the low-temperature transport compared with the electric energy scale, we assume a low temperature $\sim0.01t$, which allows us to connect the chemical potential $\mu$ and the total number of electrons $N_{\rm el}$ as $\mu\simeq\frac{1}{2}(E_{N_{\rm el}-1} + E_{N_{\rm el}})$, in the Fermi distribution function $f(\varepsilon)$.

As to the coupling constant $J_{\rm K}$, we set $J_{\rm K}=100t$, which is almost the strong-coupling limit, $J_{\rm K}\to\infty$, where we expect the influence of magnetic texture on the transport to be the clearest. We mainly focus on the region of small electron density $n$, even though we present results for all range of $n$.  In general, itinerant electrons mediate the effective interactions between localized spins. However, in the strong-coupling limit, the energy scale of effective interaction is small in the low-density region $J_{\rm eff}\propto nt$. Accordingly, we can exclude the possibility that the presence of electrons alters the magnetic structure due to the localized part $\mathcal{H}$.
With this assumption, we obtain the Hall conductivity as a sample average of $\sigma_{xy}$ over 120 configurations of $\{{\mathbf S}_i\}$, obtained from Monte Carlo simulations of Hamiltonian $\mathcal{H}$ at different temperature, $T$.

\begin{figure}[htb]
\includegraphics[width=\columnwidth]{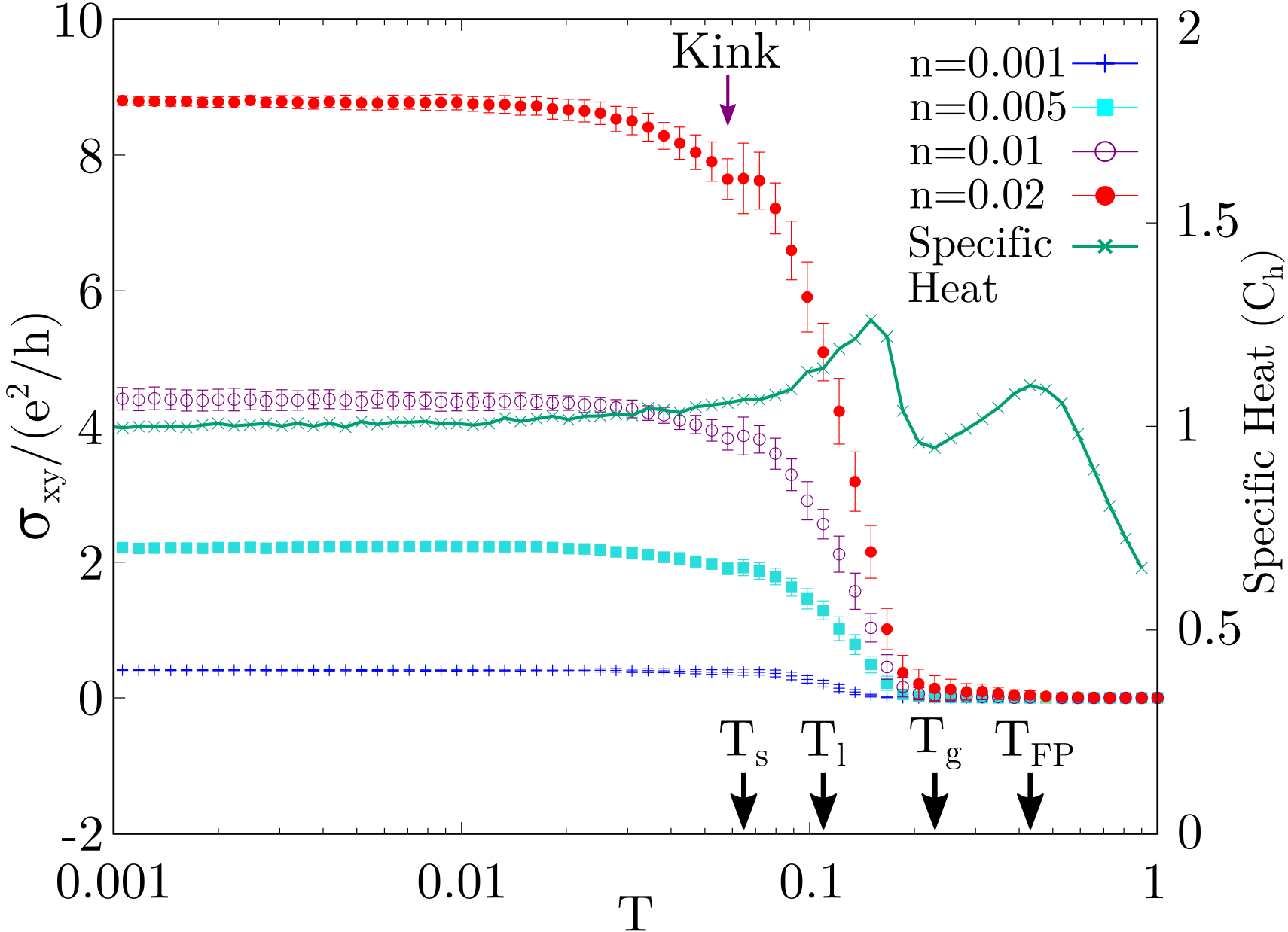}
\caption{Temperature dependence of Hall conductivity for $n=0.001, 0.005, 0.01$, and $0.02$, compared to the evolution of the specific heat. Characteristic temperatures are indicated by arrows.} 
\label{fig:Hall_smalln}
\end{figure}
%

\subsection{Hall conductivity in each phase}

As we discussed in Sec.~\ref{sec:BvsT_009}, the competition between skyrmions and a chiral spin liquid results in a rich phase diagram for the parameter set of $D^{xy}=0.5, D^z=\sqrt{3}$ and $B=0.09$, as is evident in Fig.~\ref{fig:PDsnaps_LowT} and Fig.~\ref{fig:CvNPsi6vsT}. We here address the behavior of Hall conductivity in each phase. To this aim, from the high-temperature side, we define the characteristic temperatures, $T_{\rm FP}=0.430$, $T_{\rm g}=0.229$, $T_{\rm l}=0.109$, and $T_{\rm s}=0.065$ below which the field polarized state containing the chiral spin liquid (FP), skyrmion gas state (SkG), skyrmion liquid state (SkL), and skyrmion crystal state (SkX) are respectively realized. 

In Fig.~\ref{fig:Hall_smalln}, we show the overall temperature dependence of Hall conductivity $\sigma_{xy}$ at several small electron densities, $n=0.001, 0.005, 0.01$ and $0.02$, together with the specific heat.
As shown here, $\sigma_{xy}$ shows larger values for larger $n$ in the whole temperature range,  and at low temperatures, $\sigma_{xy}$ is approximately proportional to $n$, as we discuss in more details later.
The temperature variation of $\sigma_{xy}$ shows several characteristic features.
Upon cooling from high temperature, $\sigma_{xy}$ starts to show a gradual increase around $T_{\rm FP}$, as seen more clearly for $n=0.02$. Then $\sigma_{xy}$ turns to a rapid increase when entering the skyrmion fluid below $T_{g}$, while no conspicuous change is found at the boundary between the skyrmion gas and liquid. Finally, at the transition to the SkX phase at $T_{\rm s}$, a small kink appears in $\sigma_{xy}$.

\begin{figure}[htb]
\includegraphics[width=0.9\columnwidth]{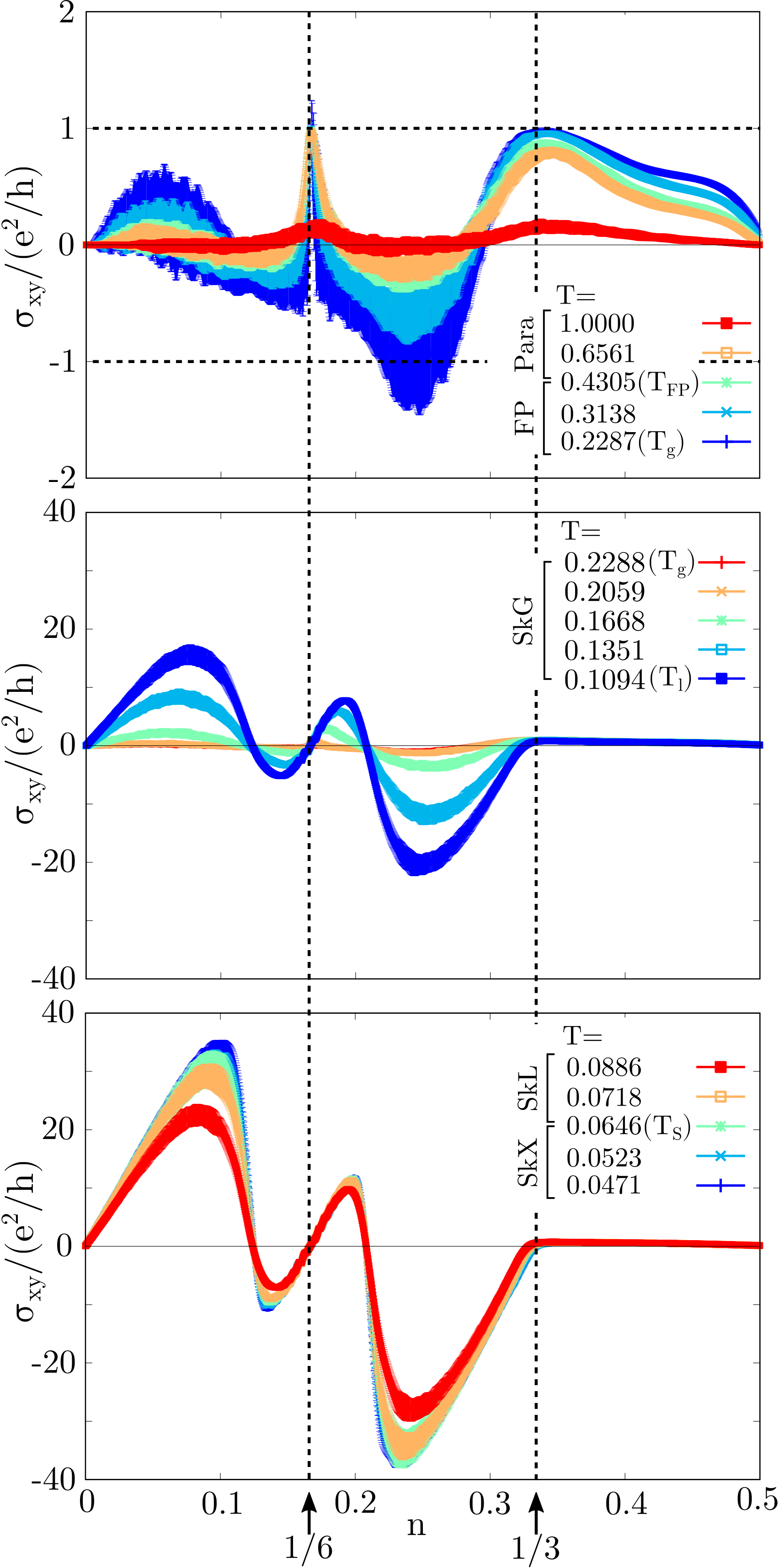}
\caption{
Hall conductivity $\sigma_{xy}$ as a function of doping $n$ (top) around $T_{\rm FP}$, (middle) between $T_{\rm g}$ and $T_{\rm l}$, and (bottom) around $T_{\rm s}$.
Please note that the different scale of longitudinal axis in the top panel. For the guide to eye, we draw vertical dashed lines to indicate $n=1/6$ and $n=1/3$, and horizontal dashed lines to indicate the quantized Hall conductivity.
} 
\label{fig:Hall_ndep}
\end{figure}

To understand the behavior of the Hall conductivity in each region, we now plot the doping dependence of $\sigma_{xy}$ at several temperatures in Figs.~\ref{fig:Hall_ndep}. Please note that in the limit of $J_{\rm K}\to\infty$, $\sigma_{xy}$ satisfies the symmetry relation: $\sigma_{xy}\to-\sigma_{xy}$ when changing $n\to n+\frac{1}{2}$. We will thus only consider doping values, $0<n<1/2$.

Fig.~\ref{fig:Hall_ndep}(a) shows the evolution of $\sigma_{xy}$ at high temperature for $T > T_g$. As mentioned previously, $\sigma_{xy}$ gradually starts to grow below $T_{\rm FP}$. Even though its magnitude remains small, the Hall conductivity displays large error bars in this region. Since these curves are obtained after sample averaging over 120 spin configurations obtained from Monte Carlo simulations, these large error bars indicate a strong sample dependence in the chiral spin liquid. This might be because umbrella states (with scalar chirality) and ferromagnetic states (without scalar chirality) are equiprobable in the spin-liquid ground state \cite{essafi2016k} (see Fig.~\ref{fig:latt}); their relative ratio may thus vary a lot from one sample to the other. This sample dependence vanishes at $n=1/6$ where $\sigma_{xy}$ exhibits a quantized value of $e^/h$. Since $n=1/6$ corresponds to band touching in the tight-binding spectrum on the kagome lattice at $J_{\rm K}=0$, this quantum Hall effect is a consequence of the degeneracy lift and gap opening when the Kondo coupling to the spin texture $J_K$ is turned on. More precisely, the quantization of the transverse conductivity at $n=1/6$ is consistent with the spin scalar chirality of the umbrella states in the chiral spin liquid, generating a  Berry curvature responsible for the topological Hall effect \cite{ohgushi2000spin,Taillefumier06a}.

Fig.~\ref{fig:Hall_ndep}(b) shows $\sigma_{xy}$ inside the skyrmion fluid phase, $T_l<T<T_g$, where the error bar is suppressed, indicating little sample dependence. $\sigma_{xy}$ shows a quick and monotonic evolution upon cooling, approximately 50-fold between $T_g$ and $T_l$.

\begin{figure}[htb]
\includegraphics[width=0.99\columnwidth]{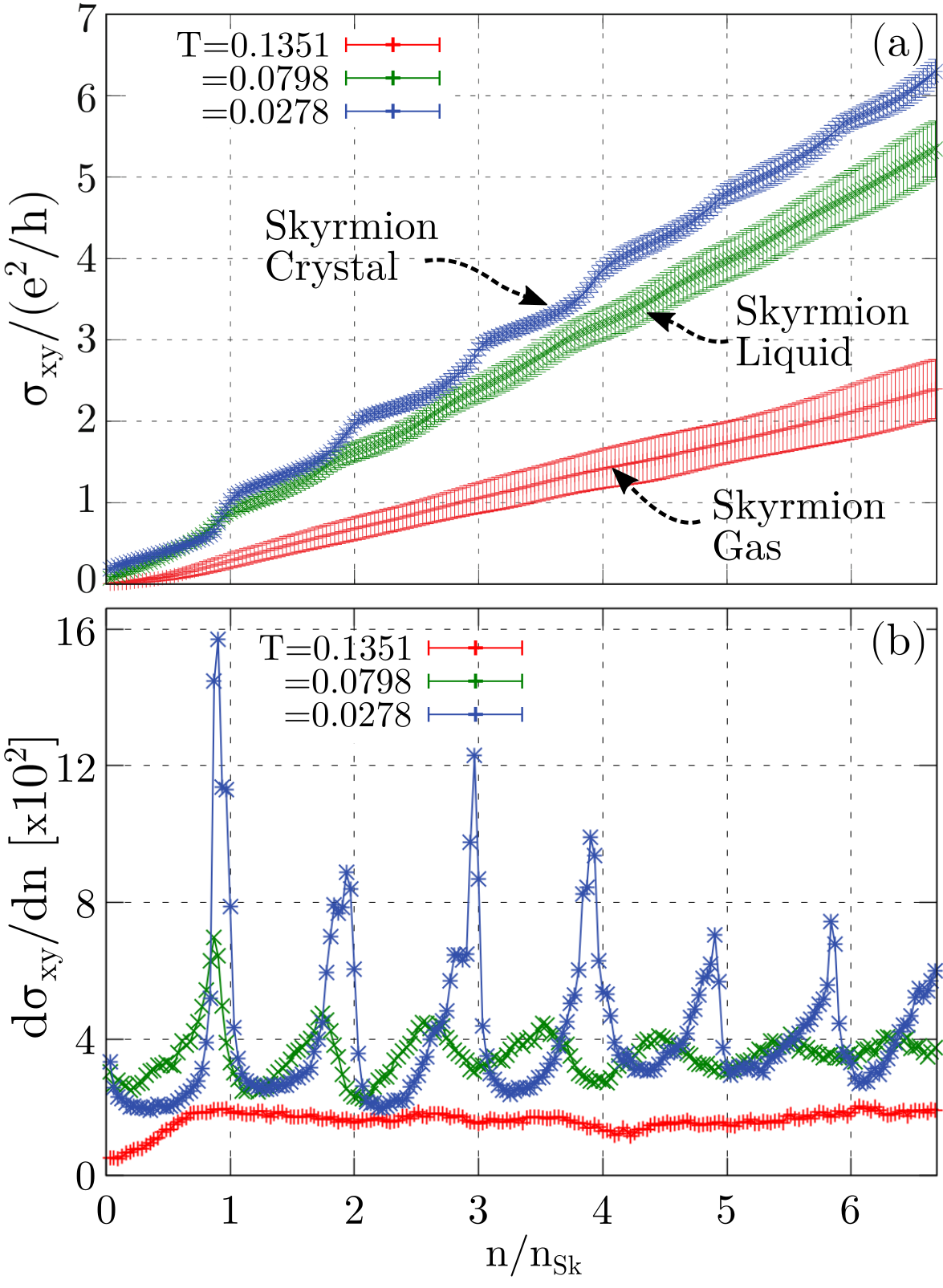}
\caption{
(a) Hall conductivity $\sigma_{xy}$ as a function of doping $n$ for $T/t=0.1351$, $0.0798$, and $0.0278$, which correspond to SkG, SkL, and SkX regions, respectively. The electron density $n$ is scaled with the skyrmion density $n_{\rm Sk}\equiv 2.2425\times10^{-3}$ of the SkX phase. Namely, $n/n_{\rm Sk}=1$ means an electron density just filling up the lowest skyrmion band.
(b) Derivative of $\sigma_{xy}$ with respect to doping $n$. The Hall conductivity quantization appears as peaks in the SkX and SkL regimes, but disappears in the SkG region (except for a weak signal at very low doping). The change of periodicity of the peaks between the SkX and SkL is due to the decrease of skyrmions with temperature.
}
\label{fig:Hall_step}
\end{figure}

Finally, Fig.~\ref{fig:Hall_ndep}(c) shows $\sigma_{xy}$ around the SkX phase. Across the transition point of the skyrmion ordering, $\sigma_{xy}$ seems to exhibit a rather gradual increase. Inside the SkX phase, $\sigma_{xy}$ continues to grow slowly as decreasing temperature and saturates at low temperature. From the smooth evolution of Fig.~\ref{fig:Hall_ndep}(c), it is unclear how the kink of Fig.~\ref{fig:Hall_step} appears at $T_{\rm s}$. This is because we need to zoom in at the low doping behavior of $\sigma_{xy}$.

\subsection{Landau level formation without skyrmion crystallization}

Indeed, Fig.~\ref{fig:Hall_step} shows that at low doping, $\sigma_{xy}$ increases smoothly in the skyrmion gas ($T=0.135$) but develops a step-like behavior upon cooling, which is especially clear in the SkX phase ($T=0.027$). The edge of the plateau corresponds to quantized value of $\sigma_{xy}=n\frac{e^2}{h}$ where integer $n$ changes by $1$ between neighboring plateaus. Additionally, the jump between plateaux occurs when $n$ is the integer multiple of skyrmion density $n_{\rm Sk}\equiv2.2425\times10^{-3}$. In the SkX phase for our Hamiltonian parameters, the number of skyrmions is 31 in the systems of $48\times48\times3$ sites. This feature naturally explains the origin of the kink at $T_s$ where each skyrmion is the source of an emergent magnetic field \cite{nagaosa2013topo}.

\begin{figure}[ht!]
\includegraphics[width=0.99\columnwidth]{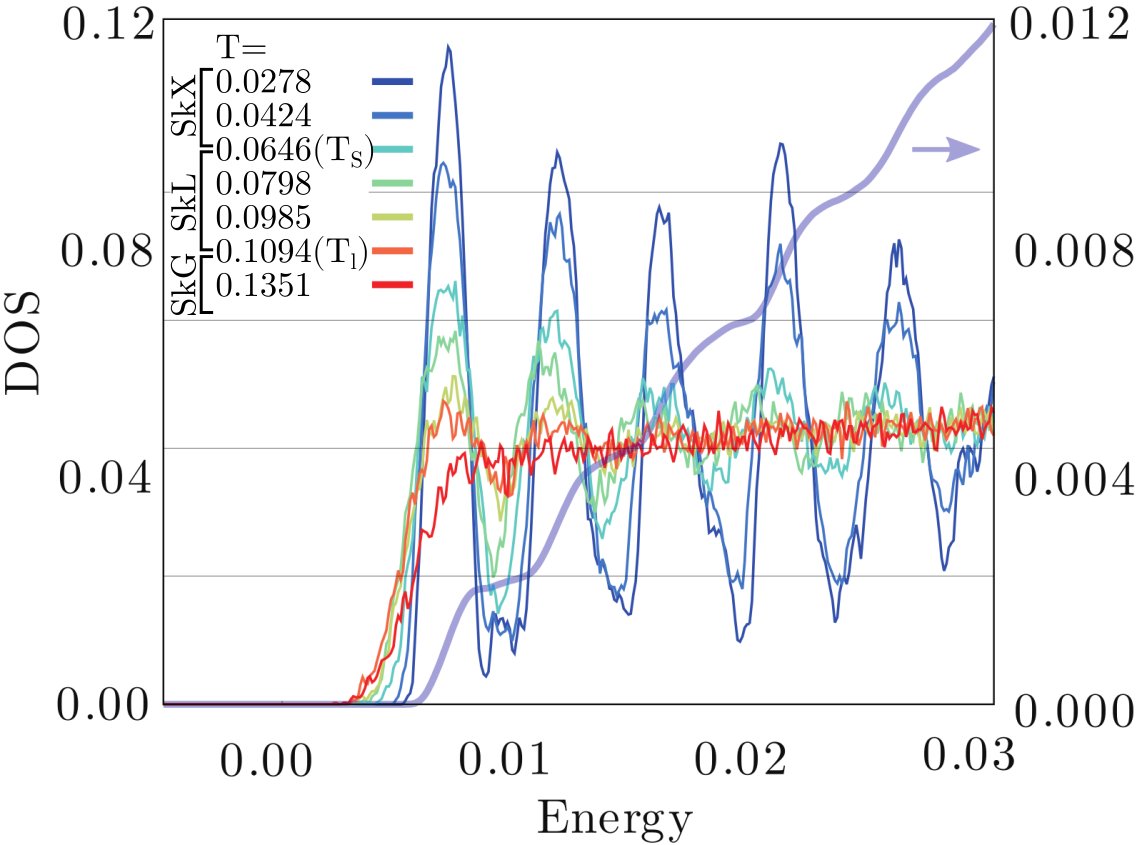}
\caption{A set of one particle electron density of states (DOS) is shown for several temperatures across $T_{\rm s}$ and $T_{\rm l}$.  To facilitate the comparison of oscillation amplitudes, we shift energies to each curve to place the lowest peak at the same position. After this shift, we set the origin of the energy near the band bottom. We also show the accumulated density of states at $T=0.0278$. The thin horizontal lines are integer multiples of $n_{\rm Sk}$ to compare with the accumulated DOS.} 
\label{fig:density_of_states}
\end{figure}

To understand the microscopic origin of this step-like behavior, we plot the averaged density of states (DOS) in the lower band in Fig.~\ref{fig:density_of_states}, with properly shifted energies depending on $n$. The DOS shows oscillatory behavior as a function of the energy, especially in the SkX phase. As a result of the DOS oscillations, the integrated DOS shows a step-like behavior whose jumps are consistent with the width of the plateaux in Fig.~\ref{fig:Hall_step}, $n_{\rm Sk}$. This implies that the step-like behaviour of $\sigma_{xy}$ can be attributed to the oscillations in the DOS.

Presumably, the oscillation of DOS in the SkX phase may not be surprising \cite{yi09_skyrm_anomal_hall_effec_dzyal,hamamoto2015quantized}, since the skyrmions exhibit periodic ordering, which explicitly breaks the translational symmetry of the kagome lattice. As a result of this translational symmetry breaking, the electric states are decomposed into skyrmion subbands. If each skyrmion subbands has the Chern number $\nu=1$, $\sigma_{xy}$ would show a step-like behavior, as shown in Fig.~\ref{fig:Hall_step}, and each plateau should be quantized as an integer multiple of $\frac{e^2}{h}$. Moreover, the jump of the integrated DOS corresponds to the number of the states contained in one subband, which precisely corresponds to the electron density of $n_{\rm Sk}$.

What is remarkable in Fig.~\ref{fig:density_of_states} is that the DOS oscillations persist even above $T_s=0.065$, in the skyrmion liquid. Reflecting this oscillatory character of DOS, $\sigma_{xy}$ also shows rounded but nonetheless distinctive steps at $T=0.080$. In other words, there is a signature of subband formation in the electric state without translational symmetry breaking in the magnetic state! 

As an origin of the subband formation, it is interesting to point out the role of Berry phase arising from the coupling of itinerant electrons to the disordered magnetic texture of the skyrmion fluid. In the strong coupling limit: $J_{\rm}\to\infty$, when an itinerant electron moves around a triplet of spins, it acquires the Berry phase $\Phi=\frac{\Omega}{2}$, where $\Omega$ is the solid angle spanned by the triplet of the spins. It means the presence of one skyrmion generates the effective magnetic field equivalent to one magnetic flux quantum.

Accordingly, if the effective magnetic field is uniformly distributed, one Landau level is formed per one skyrmion. We expect it is actually what happens in the SkL phase. The assumption of uniform effective magnetic field is, however, far from trivial. The translational symmetry in the skyrmion fluid phase is obtained after the sample average. Each spin configuration suffers a strong spatial variation of the effective magnetic field, which may destroy the equally-spaced Landau level structure. Indeed, in the skyrmion gas phase, Landau levels do not appear to exist, as DOS lacks the oscillatory components (See $T=0.135$ of  Fig.~\ref{fig:density_of_states}).

Presumably, in the SkL phase, the highly packed nature of skyrmions, as is implied by the saturation of skyrmion density [Fig.~\ref{fig:CvNPsi6vsT}], leads to the suppression of the fluctuation of the effective magnetic field. In contrast, in the SkG phase, loosely placed skyrmions may result in a random distribution of effective magnetic field, and makes it difficult to form Landau levels.

So far, electric response of skyrmion fluid states has not been well explored, in contrast to abundant studies on the skyrmion crystal case. Our study reveals two important aspects of electric transport in the skyrmion fluid phase. Firstly, the Berry phase from the magnetic texture induces the Landau levels. And secondly, packed configurations of skyrmions might be necessary for the equally-space Landau levels to be actually observable. 
These findings will give insights into this nontrivial spin-electron coupled state and will be useful to further studies.

\section{Conclusion and perspectives}

In this work we have extensively studied a magnetic model with an extremely rich variety of relatively unconventional phenomena. Starting from a model defined in the kagome lattice with isotropic couplings and an unidirectional DM interaction, which is known to host a chiral spin liquid phase, we investigated the effect of a perpendicular DM term and a magnetic field. Depending on the values of each parameter, we found an extremely rich behavior of the system which can be summarized in figures \ref{fig:PDB0}, \ref{fig:BTDxy033035} and \ref{fig:PDsnaps_LowT}.

 We first analyze the purely magnetic system with the LT approximation.  The main result with this approach is that for small in-plane DM interaction ($D^{xy}$), the energy gap between the minima and the flat band associated with the chiral spin liquid remains relatively modest in comparison to the ground state energy, suggesting a substantial impact of chiral spin liquid physics at finite temperatures.  Thus, we resort to large-scale Monte Carlo simulations to incorporate thermal fluctuations and broaden the exploration in parameter space. The zero field behavior is already quite rich, with a dubbed labyrinth phase, distorted helices order, and a chiral spin liquid. In the presence of an external magnetic field,  a large variety of additional phases unfolds. When the in-plane $D^{xy}$ is limited to a narrower range, we identify a region where a frustrated vortex phase emerges exhibiting long-range order and spontaneous breaking of $Z_3$ symmetry. Regarding skyrmion textures, for large enough $D^{xy}$ we have identified three different phases,  liquid, solid, and crystalline, according to the long-range behavior of positional and orientational order parameters.   On top of the skyrmion crystalline phase lies a bimeron glass phase whose study goes beyond the scope of the present work but certainly deserves further investigation as another horizon for a very rich phenomenology, this time in the subject of slow dynamics and out-of-equilibrium physics.

As magnetic chiral structures induce interesting behaviors in itinerant electrons, the rich variety of chiral phases found in this model motivated our investigation in this direction. We studied a Kondo-lattice Hamiltonian, where the non-zero but relatively low measured Hall conductance in the chiral spin liquid phase sees a dramatic rise once the skyrmions start to form until reaching a very high level in the dense skyrmion phase. At the crystalline phase, the Hall conductance shows a steplike behavior indicating the formation of band with non-zero Chern number which originates from the larger periodic structure of the skymion Crystal. This can be easily seen in the plot of the DOS as the function of the filling where distinctive oscillation of quasi-band structure shows up. What is remarkable is that the DOS oscillations persist even in the skyrmion liquid phase, indicating that there is a signature of subband formation in the electric state without translational symmetry breaking in the magnetic state.

It is quite surprising that a relatively simple model, such as the one studied here, can give rise to such a spectacular variety of phenomena, some of which would deserve further investigation. The issues of the dynamics in the bimeron glass as well as the behavior of the Hall conductivity of itinerant electrons in all these phases certainly open promising perspectives for future studies.

\section*{Acknowledgments}
The authors thank the CNRS International Research Project COQSYS for their support. H.D.R. and F.A.G.A. acknowledge financial support from CONICET (PIP 2021-11220200101480CO), Agencia I+D+i (PICT 2020 Serie A 03205) and SECyT-UNLP (I+D X893),  F.A.G.A. from PIBAA 28720210100698CO (CONICET), M. U. from JSPS KAKENHI (Grant No.JP20H05655 and JP22H01147), and L.D.C.J. from ANR-18-CE30-0011-01. L.D.C.J would like to thank IFLYSIB (CONICET-UNLP) for their hospitality during completion of this project.


\begin{thebibliography}{87}%
\makeatletter
\providecommand \@ifxundefined [1]{%
 \@ifx{#1\undefined}
}%
\providecommand \@ifnum [1]{%
 \ifnum #1\expandafter \@firstoftwo
 \else \expandafter \@secondoftwo
 \fi
}%
\providecommand \@ifx [1]{%
 \ifx #1\expandafter \@firstoftwo
 \else \expandafter \@secondoftwo
 \fi
}%
\providecommand \natexlab [1]{#1}%
\providecommand \enquote  [1]{``#1''}%
\providecommand \bibnamefont  [1]{#1}%
\providecommand \bibfnamefont [1]{#1}%
\providecommand \citenamefont [1]{#1}%
\providecommand \href@noop [0]{\@secondoftwo}%
\providecommand \href [0]{\begingroup \@sanitize@url \@href}%
\providecommand \@href[1]{\@@startlink{#1}\@@href}%
\providecommand \@@href[1]{\endgroup#1\@@endlink}%
\providecommand \@sanitize@url [0]{\catcode `\\12\catcode `\$12\catcode
  `\&12\catcode `\#12\catcode `\^12\catcode `\_12\catcode `\%12\relax}%
\providecommand \@@startlink[1]{}%
\providecommand \@@endlink[0]{}%
\providecommand \url  [0]{\begingroup\@sanitize@url \@url }%
\providecommand \@url [1]{\endgroup\@href {#1}{\urlprefix }}%
\providecommand \urlprefix  [0]{URL }%
\providecommand \Eprint [0]{\href }%
\providecommand \doibase [0]{https://doi.org/}%
\providecommand \selectlanguage [0]{\@gobble}%
\providecommand \bibinfo  [0]{\@secondoftwo}%
\providecommand \bibfield  [0]{\@secondoftwo}%
\providecommand \translation [1]{[#1]}%
\providecommand \BibitemOpen [0]{}%
\providecommand \bibitemStop [0]{}%
\providecommand \bibitemNoStop [0]{.\EOS\space}%
\providecommand \EOS [0]{\spacefactor3000\relax}%
\providecommand \BibitemShut  [1]{\csname bibitem#1\endcsname}%
\let\auto@bib@innerbib\@empty
\bibitem [{\citenamefont {Bogdanov}\ and\ \citenamefont
  {Yablonskii}(1989)}]{bogdanov1989ther}%
  \BibitemOpen
  \bibfield  {author} {\bibinfo {author} {\bibfnamefont {A.~N.}\ \bibnamefont
  {Bogdanov}}\ and\ \bibinfo {author} {\bibfnamefont {D.}~\bibnamefont
  {Yablonskii}},\ }\bibfield  {title} {\bibinfo {title} {Thermodynamically
  stable “vortices” in magnetically ordered crystals. the mixed state of
  magnets},\ }\href@noop {} {\bibfield  {journal} {\bibinfo  {journal} {Zh.
  Eksp. Teor. Fiz}\ }\textbf {\bibinfo {volume} {95}},\ \bibinfo {pages} {178}
  (\bibinfo {year} {1989})}\BibitemShut {NoStop}%
\bibitem [{\citenamefont {Bogdanov}\ and\ \citenamefont
  {Hubert}(1994)}]{bogdanov1994ther}%
  \BibitemOpen
  \bibfield  {author} {\bibinfo {author} {\bibfnamefont {A.}~\bibnamefont
  {Bogdanov}}\ and\ \bibinfo {author} {\bibfnamefont {A.}~\bibnamefont
  {Hubert}},\ }\bibfield  {title} {\bibinfo {title} {Thermodynamically stable
  magnetic vortex states in magnetic crystals},\ }\href
  {https://doi.org/10.1016/0304-8853(94)90046-9} {\bibfield  {journal}
  {\bibinfo  {journal} {Journal of magnetism and magnetic materials}\ }\textbf
  {\bibinfo {volume} {138}},\ \bibinfo {pages} {255} (\bibinfo {year}
  {1994})}\BibitemShut {NoStop}%
\bibitem [{\citenamefont {Roessler}\ \emph {et~al.}(2006)\citenamefont
  {Roessler}, \citenamefont {Bogdanov},\ and\ \citenamefont
  {Pfleiderer}}]{roessler2006spontaneous}%
  \BibitemOpen
  \bibfield  {author} {\bibinfo {author} {\bibfnamefont {U.~K.}\ \bibnamefont
  {Roessler}}, \bibinfo {author} {\bibfnamefont {A.}~\bibnamefont {Bogdanov}},\
  and\ \bibinfo {author} {\bibfnamefont {C.}~\bibnamefont {Pfleiderer}},\
  }\bibfield  {title} {\bibinfo {title} {Spontaneous skyrmion ground states in
  magnetic metals},\ }\href {https://doi.org/10.1038/nature05056} {\bibfield
  {journal} {\bibinfo  {journal} {Nature}\ }\textbf {\bibinfo {volume} {442}},\
  \bibinfo {pages} {797} (\bibinfo {year} {2006})}\BibitemShut {NoStop}%
\bibitem [{\citenamefont {M{\"u}hlbauer}\ \emph {et~al.}(2009)\citenamefont
  {M{\"u}hlbauer}, \citenamefont {Binz}, \citenamefont {Jonietz}, \citenamefont
  {Pfleiderer}, \citenamefont {Rosch}, \citenamefont {Neubauer}, \citenamefont
  {Georgii},\ and\ \citenamefont {B{\"o}ni}}]{muhlbauer2009sk}%
  \BibitemOpen
  \bibfield  {author} {\bibinfo {author} {\bibfnamefont {S.}~\bibnamefont
  {M{\"u}hlbauer}}, \bibinfo {author} {\bibfnamefont {B.}~\bibnamefont {Binz}},
  \bibinfo {author} {\bibfnamefont {F.}~\bibnamefont {Jonietz}}, \bibinfo
  {author} {\bibfnamefont {C.}~\bibnamefont {Pfleiderer}}, \bibinfo {author}
  {\bibfnamefont {A.}~\bibnamefont {Rosch}}, \bibinfo {author} {\bibfnamefont
  {A.}~\bibnamefont {Neubauer}}, \bibinfo {author} {\bibfnamefont
  {R.}~\bibnamefont {Georgii}},\ and\ \bibinfo {author} {\bibfnamefont
  {P.}~\bibnamefont {B{\"o}ni}},\ }\bibfield  {title} {\bibinfo {title}
  {Skyrmion lattice in a chiral magnet},\ }\href
  {https://doi.org/10.1126/science.1166767} {\bibfield  {journal} {\bibinfo
  {journal} {Science}\ }\textbf {\bibinfo {volume} {323}},\ \bibinfo {pages}
  {915} (\bibinfo {year} {2009})}\BibitemShut {NoStop}%
\bibitem [{\citenamefont {Yu}\ \emph {et~al.}(2010)\citenamefont {Yu},
  \citenamefont {Onose}, \citenamefont {Kanazawa}, \citenamefont {Park},
  \citenamefont {Han}, \citenamefont {Matsui}, \citenamefont {Nagaosa},\ and\
  \citenamefont {Tokura}}]{yu2010re}%
  \BibitemOpen
  \bibfield  {author} {\bibinfo {author} {\bibfnamefont {X.}~\bibnamefont
  {Yu}}, \bibinfo {author} {\bibfnamefont {Y.}~\bibnamefont {Onose}}, \bibinfo
  {author} {\bibfnamefont {N.}~\bibnamefont {Kanazawa}}, \bibinfo {author}
  {\bibfnamefont {J.}~\bibnamefont {Park}}, \bibinfo {author} {\bibfnamefont
  {J.}~\bibnamefont {Han}}, \bibinfo {author} {\bibfnamefont {Y.}~\bibnamefont
  {Matsui}}, \bibinfo {author} {\bibfnamefont {N.}~\bibnamefont {Nagaosa}},\
  and\ \bibinfo {author} {\bibfnamefont {Y.}~\bibnamefont {Tokura}},\
  }\bibfield  {title} {\bibinfo {title} {Real-space observation of a
  two-dimensional skyrmion crystal},\ }\href
  {https://doi.org/10.1038/nature09124} {\bibfield  {journal} {\bibinfo
  {journal} {Nature}\ }\textbf {\bibinfo {volume} {465}},\ \bibinfo {pages}
  {901} (\bibinfo {year} {2010})}\BibitemShut {NoStop}%
\bibitem [{\citenamefont {Nagaosa}\ and\ \citenamefont
  {Tokura}(2013)}]{nagaosa2013topo}%
  \BibitemOpen
  \bibfield  {author} {\bibinfo {author} {\bibfnamefont {N.}~\bibnamefont
  {Nagaosa}}\ and\ \bibinfo {author} {\bibfnamefont {Y.}~\bibnamefont
  {Tokura}},\ }\bibfield  {title} {\bibinfo {title} {Topological properties and
  dynamics of magnetic skyrmions},\ }\href
  {https://doi.org/10.1038/nnano.2013.243} {\bibfield  {journal} {\bibinfo
  {journal} {Nature Nanotechnology}\ }\textbf {\bibinfo {volume} {8}},\
  \bibinfo {pages} {899} (\bibinfo {year} {2013})}\BibitemShut {NoStop}%
\bibitem [{\citenamefont {Fert}\ \emph {et~al.}(2017)\citenamefont {Fert},
  \citenamefont {Reyren},\ and\ \citenamefont {Cros}}]{fert2017magnetic}%
  \BibitemOpen
  \bibfield  {author} {\bibinfo {author} {\bibfnamefont {A.}~\bibnamefont
  {Fert}}, \bibinfo {author} {\bibfnamefont {N.}~\bibnamefont {Reyren}},\ and\
  \bibinfo {author} {\bibfnamefont {V.}~\bibnamefont {Cros}},\ }\bibfield
  {title} {\bibinfo {title} {Magnetic skyrmions: advances in physics and
  potential applications},\ }\href {https://doi.org/10.1038/natrevmats.2017.31}
  {\bibfield  {journal} {\bibinfo  {journal} {Nature Reviews Materials}\
  }\textbf {\bibinfo {volume} {2}},\ \bibinfo {pages} {1} (\bibinfo {year}
  {2017})}\BibitemShut {NoStop}%
\bibitem [{\citenamefont {Everschor-Sitte}\ \emph {et~al.}(2018)\citenamefont
  {Everschor-Sitte}, \citenamefont {Masell}, \citenamefont {Reeve},\ and\
  \citenamefont {Kl{\"a}ui}}]{everschor2018perspective}%
  \BibitemOpen
  \bibfield  {author} {\bibinfo {author} {\bibfnamefont {K.}~\bibnamefont
  {Everschor-Sitte}}, \bibinfo {author} {\bibfnamefont {J.}~\bibnamefont
  {Masell}}, \bibinfo {author} {\bibfnamefont {R.~M.}\ \bibnamefont {Reeve}},\
  and\ \bibinfo {author} {\bibfnamefont {M.}~\bibnamefont {Kl{\"a}ui}},\
  }\bibfield  {title} {\bibinfo {title} {Perspective: Magnetic
  skyrmions—overview of recent progress in an active research field},\ }\href
  {https://doi.org/10.1063/1.5048972} {\bibfield  {journal} {\bibinfo
  {journal} {Journal of Applied Physics}\ }\textbf {\bibinfo {volume} {124}},\
  \bibinfo {pages} {240901} (\bibinfo {year} {2018})}\BibitemShut {NoStop}%
\bibitem [{\citenamefont {Dzyaloshinsky}(1958)}]{DM1}%
  \BibitemOpen
  \bibfield  {author} {\bibinfo {author} {\bibfnamefont {I.}~\bibnamefont
  {Dzyaloshinsky}},\ }\bibfield  {title} {\bibinfo {title} {A thermodynamic
  theory of “weak” ferromagnetism of antiferromagnetics},\ }\href
  {https://doi.org/10.1016/0022-3697(58)90076-3} {\bibfield  {journal}
  {\bibinfo  {journal} {Journal of physics and chemistry of solids}\ }\textbf
  {\bibinfo {volume} {4}},\ \bibinfo {pages} {241} (\bibinfo {year}
  {1958})}\BibitemShut {NoStop}%
\bibitem [{\citenamefont {Moriya}(1960)}]{DM2}%
  \BibitemOpen
  \bibfield  {author} {\bibinfo {author} {\bibfnamefont {T.}~\bibnamefont
  {Moriya}},\ }\bibfield  {title} {\bibinfo {title} {New mechanism of
  anisotropic superexchange interaction},\ }\href
  {https://doi.org/10.1103/PhysRevLett.4.228} {\bibfield  {journal} {\bibinfo
  {journal} {Physical Review Letters}\ }\textbf {\bibinfo {volume} {4}},\
  \bibinfo {pages} {228} (\bibinfo {year} {1960})}\BibitemShut {NoStop}%
\bibitem [{\citenamefont {Okubo}\ \emph {et~al.}(2012)\citenamefont {Okubo},
  \citenamefont {Chung},\ and\ \citenamefont {Kawamura}}]{Okubo2012}%
  \BibitemOpen
  \bibfield  {author} {\bibinfo {author} {\bibfnamefont {T.}~\bibnamefont
  {Okubo}}, \bibinfo {author} {\bibfnamefont {S.}~\bibnamefont {Chung}},\ and\
  \bibinfo {author} {\bibfnamefont {H.}~\bibnamefont {Kawamura}},\ }\bibfield
  {title} {\bibinfo {title} {Multiple-q states and the skyrmion lattice of the
  triangular-lattice heisenberg antiferromagnet under magnetic fields},\ }\href
  {https://doi.org/10.1103/PhysRevLett.108.017206} {\bibfield  {journal}
  {\bibinfo  {journal} {Physical Review Letters}\ }\textbf {\bibinfo {volume}
  {108}},\ \bibinfo {pages} {017206} (\bibinfo {year} {2012})}\BibitemShut
  {NoStop}%
\bibitem [{\citenamefont {Mohylna}\ \emph {et~al.}(2022)\citenamefont
  {Mohylna}, \citenamefont {Albarrac{\'\i}n}, \citenamefont
  {{\v{Z}}ukovi{\v{c}}},\ and\ \citenamefont
  {Rosales}}]{mohylna2022spontaneous}%
  \BibitemOpen
  \bibfield  {author} {\bibinfo {author} {\bibfnamefont {M.}~\bibnamefont
  {Mohylna}}, \bibinfo {author} {\bibfnamefont {F.~G.}\ \bibnamefont
  {Albarrac{\'\i}n}}, \bibinfo {author} {\bibfnamefont {M.}~\bibnamefont
  {{\v{Z}}ukovi{\v{c}}}},\ and\ \bibinfo {author} {\bibfnamefont
  {H.}~\bibnamefont {Rosales}},\ }\bibfield  {title} {\bibinfo {title}
  {Spontaneous antiferromagnetic skyrmion/antiskyrmion lattice and spiral
  spin-liquid states in the frustrated triangular lattice},\ }\href
  {https://doi.org/10.1103/PhysRevB.106.224406} {\bibfield  {journal} {\bibinfo
   {journal} {Physical Review B}\ }\textbf {\bibinfo {volume} {106}},\ \bibinfo
  {pages} {224406} (\bibinfo {year} {2022})}\BibitemShut {NoStop}%
\bibitem [{\citenamefont {Gao}\ \emph {et~al.}(2020)\citenamefont {Gao},
  \citenamefont {Rosales}, \citenamefont {G{\'o}mez~Albarrac{\'\i}n},
  \citenamefont {Tsurkan}, \citenamefont {Kaur}, \citenamefont {Fennell},
  \citenamefont {Steffens}, \citenamefont {Boehm}, \citenamefont
  {{\v{C}}erm{\'a}k}, \citenamefont {Schneidewind} \emph
  {et~al.}}]{Gao2020Nat}%
  \BibitemOpen
  \bibfield  {author} {\bibinfo {author} {\bibfnamefont {S.}~\bibnamefont
  {Gao}}, \bibinfo {author} {\bibfnamefont {H.~D.}\ \bibnamefont {Rosales}},
  \bibinfo {author} {\bibfnamefont {F.~A.}\ \bibnamefont
  {G{\'o}mez~Albarrac{\'\i}n}}, \bibinfo {author} {\bibfnamefont
  {V.}~\bibnamefont {Tsurkan}}, \bibinfo {author} {\bibfnamefont
  {G.}~\bibnamefont {Kaur}}, \bibinfo {author} {\bibfnamefont {T.}~\bibnamefont
  {Fennell}}, \bibinfo {author} {\bibfnamefont {P.}~\bibnamefont {Steffens}},
  \bibinfo {author} {\bibfnamefont {M.}~\bibnamefont {Boehm}}, \bibinfo
  {author} {\bibfnamefont {P.}~\bibnamefont {{\v{C}}erm{\'a}k}}, \bibinfo
  {author} {\bibfnamefont {A.}~\bibnamefont {Schneidewind}}, \emph {et~al.},\
  }\bibfield  {title} {\bibinfo {title} {Fractional antiferromagnetic skyrmion
  lattice induced by anisotropic couplings},\ }\href
  {https://doi.org/10.1038/s41586-020-2716-8} {\bibfield  {journal} {\bibinfo
  {journal} {Nature}\ }\textbf {\bibinfo {volume} {586}},\ \bibinfo {pages}
  {37} (\bibinfo {year} {2020})}\BibitemShut {NoStop}%
\bibitem [{\citenamefont {Amoroso}\ \emph {et~al.}(2020)\citenamefont
  {Amoroso}, \citenamefont {Barone},\ and\ \citenamefont
  {Picozzi}}]{Amoroso2020}%
  \BibitemOpen
  \bibfield  {author} {\bibinfo {author} {\bibfnamefont {D.}~\bibnamefont
  {Amoroso}}, \bibinfo {author} {\bibfnamefont {P.}~\bibnamefont {Barone}},\
  and\ \bibinfo {author} {\bibfnamefont {S.}~\bibnamefont {Picozzi}},\
  }\bibfield  {title} {\bibinfo {title} {Spontaneous skyrmionic lattice from
  anisotropic symmetric exchange in a ni-halide monolayer},\ }\href
  {https://doi.org/10.1038/s41467-020-19535-w} {\bibfield  {journal} {\bibinfo
  {journal} {Nature Communications}\ }\textbf {\bibinfo {volume} {11}},\
  \bibinfo {pages} {1} (\bibinfo {year} {2020})}\BibitemShut {NoStop}%
\bibitem [{\citenamefont {Wang}\ \emph {et~al.}(2021)\citenamefont {Wang},
  \citenamefont {Su}, \citenamefont {Lin},\ and\ \citenamefont
  {Batista}}]{Wang2021}%
  \BibitemOpen
  \bibfield  {author} {\bibinfo {author} {\bibfnamefont {Z.}~\bibnamefont
  {Wang}}, \bibinfo {author} {\bibfnamefont {Y.}~\bibnamefont {Su}}, \bibinfo
  {author} {\bibfnamefont {S.-Z.}\ \bibnamefont {Lin}},\ and\ \bibinfo {author}
  {\bibfnamefont {C.~D.}\ \bibnamefont {Batista}},\ }\bibfield  {title}
  {\bibinfo {title} {Meron, skyrmion, and vortex crystals in centrosymmetric
  tetragonal magnets},\ }\href {https://doi.org/10.1103/PhysRevB.103.104408}
  {\bibfield  {journal} {\bibinfo  {journal} {Physical Review B}\ }\textbf
  {\bibinfo {volume} {103}},\ \bibinfo {pages} {104408} (\bibinfo {year}
  {2021})}\BibitemShut {NoStop}%
\bibitem [{\citenamefont {Hayami}\ and\ \citenamefont
  {Motome}(2021{\natexlab{a}})}]{Hayami2021-a}%
  \BibitemOpen
  \bibfield  {author} {\bibinfo {author} {\bibfnamefont {S.}~\bibnamefont
  {Hayami}}\ and\ \bibinfo {author} {\bibfnamefont {Y.}~\bibnamefont
  {Motome}},\ }\bibfield  {title} {\bibinfo {title} {Square skyrmion crystal in
  centrosymmetric itinerant magnets},\ }\href
  {https://doi.org/10.1103/PhysRevB.103.024439} {\bibfield  {journal} {\bibinfo
   {journal} {Physical Review B}\ }\textbf {\bibinfo {volume} {103}},\ \bibinfo
  {pages} {024439} (\bibinfo {year} {2021}{\natexlab{a}})}\BibitemShut
  {NoStop}%
\bibitem [{\citenamefont {Utesov}(2021)}]{Utesov2021}%
  \BibitemOpen
  \bibfield  {author} {\bibinfo {author} {\bibfnamefont {O.~I.}\ \bibnamefont
  {Utesov}},\ }\bibfield  {title} {\bibinfo {title} {Thermodynamically stable
  skyrmion lattice in a tetragonal frustrated antiferromagnet with dipolar
  interaction},\ }\href {https://doi.org/10.1103/PhysRevB.103.064414}
  {\bibfield  {journal} {\bibinfo  {journal} {Physical Review B}\ }\textbf
  {\bibinfo {volume} {103}},\ \bibinfo {pages} {064414} (\bibinfo {year}
  {2021})}\BibitemShut {NoStop}%
\bibitem [{\citenamefont {Yambe}\ and\ \citenamefont
  {Hayami}(2021)}]{Yambe2021}%
  \BibitemOpen
  \bibfield  {author} {\bibinfo {author} {\bibfnamefont {R.}~\bibnamefont
  {Yambe}}\ and\ \bibinfo {author} {\bibfnamefont {S.}~\bibnamefont {Hayami}},\
  }\bibfield  {title} {\bibinfo {title} {Skyrmion crystals in centrosymmetric
  itinerant magnets without horizontal mirror plane},\ }\href
  {https://doi.org/10.1038/s41598-021-90308-1} {\bibfield  {journal} {\bibinfo
  {journal} {Scientific Reports}\ }\textbf {\bibinfo {volume} {11}},\ \bibinfo
  {pages} {1} (\bibinfo {year} {2021})}\BibitemShut {NoStop}%
\bibitem [{\citenamefont {Amoroso}\ \emph {et~al.}(2021)\citenamefont
  {Amoroso}, \citenamefont {Barone},\ and\ \citenamefont
  {Picozzi}}]{Amoroso2021}%
  \BibitemOpen
  \bibfield  {author} {\bibinfo {author} {\bibfnamefont {D.}~\bibnamefont
  {Amoroso}}, \bibinfo {author} {\bibfnamefont {P.}~\bibnamefont {Barone}},\
  and\ \bibinfo {author} {\bibfnamefont {S.}~\bibnamefont {Picozzi}},\
  }\bibfield  {title} {\bibinfo {title} {Interplay between single-ion and
  two-ion anisotropies in frustrated 2d semiconductors and tuning of magnetic
  structures topology},\ }\href {https://doi.org/10.3390/nano11081873}
  {\bibfield  {journal} {\bibinfo  {journal} {Nanomaterials}\ }\textbf
  {\bibinfo {volume} {11}},\ \bibinfo {pages} {1873} (\bibinfo {year}
  {2021})}\BibitemShut {NoStop}%
\bibitem [{\citenamefont {Rosales}\ \emph {et~al.}(2022)\citenamefont
  {Rosales}, \citenamefont {G\'omez~Albarrac\'{\i}n}, \citenamefont
  {Guratinder}, \citenamefont {Tsurkan}, \citenamefont {Prodan}, \citenamefont
  {Ressouche},\ and\ \citenamefont {Zaharko}}]{Rosales22}%
  \BibitemOpen
  \bibfield  {author} {\bibinfo {author} {\bibfnamefont {H.~D.}\ \bibnamefont
  {Rosales}}, \bibinfo {author} {\bibfnamefont {F.~A.}\ \bibnamefont
  {G\'omez~Albarrac\'{\i}n}}, \bibinfo {author} {\bibfnamefont
  {K.}~\bibnamefont {Guratinder}}, \bibinfo {author} {\bibfnamefont
  {V.}~\bibnamefont {Tsurkan}}, \bibinfo {author} {\bibfnamefont
  {L.}~\bibnamefont {Prodan}}, \bibinfo {author} {\bibfnamefont
  {E.}~\bibnamefont {Ressouche}},\ and\ \bibinfo {author} {\bibfnamefont
  {O.}~\bibnamefont {Zaharko}},\ }\bibfield  {title} {\bibinfo {title}
  {Anisotropy-driven response of the fractional antiferromagnetic skyrmion
  lattice in ${\mathrm{mnsc}}_{2}{\mathrm{s}}_{4}$ to applied magnetic
  fields},\ }\href {https://doi.org/10.1103/PhysRevB.105.224402} {\bibfield
  {journal} {\bibinfo  {journal} {Phys. Rev. B}\ }\textbf {\bibinfo {volume}
  {105}},\ \bibinfo {pages} {224402} (\bibinfo {year} {2022})}\BibitemShut
  {NoStop}%
\bibitem [{\citenamefont {Hayami}(2022)}]{Hayami2022-a}%
  \BibitemOpen
  \bibfield  {author} {\bibinfo {author} {\bibfnamefont {S.}~\bibnamefont
  {Hayami}},\ }\bibfield  {title} {\bibinfo {title} {Multiple skyrmion crystal
  phases by itinerant frustration in centrosymmetric tetragonal magnets},\
  }\href {https://doi.org/10.7566/JPSJ.91.023705} {\bibfield  {journal}
  {\bibinfo  {journal} {Journal of the Physical Society of Japan}\ }\textbf
  {\bibinfo {volume} {91}},\ \bibinfo {pages} {023705} (\bibinfo {year}
  {2022})}\BibitemShut {NoStop}%
\bibitem [{\citenamefont {Akagi}\ \emph {et~al.}(2012)\citenamefont {Akagi},
  \citenamefont {Udagawa},\ and\ \citenamefont
  {Motome}}]{PhysRevLett.108.096401}%
  \BibitemOpen
  \bibfield  {author} {\bibinfo {author} {\bibfnamefont {Y.}~\bibnamefont
  {Akagi}}, \bibinfo {author} {\bibfnamefont {M.}~\bibnamefont {Udagawa}},\
  and\ \bibinfo {author} {\bibfnamefont {Y.}~\bibnamefont {Motome}},\
  }\bibfield  {title} {\bibinfo {title} {Hidden multiple-spin interactions as
  an origin of spin scalar chiral order in frustrated kondo lattice models},\
  }\href {https://doi.org/10.1103/PhysRevLett.108.096401} {\bibfield  {journal}
  {\bibinfo  {journal} {Phys. Rev. Lett.}\ }\textbf {\bibinfo {volume} {108}},\
  \bibinfo {pages} {096401} (\bibinfo {year} {2012})}\BibitemShut {NoStop}%
\bibitem [{\citenamefont {Wang}\ \emph {et~al.}(2020)\citenamefont {Wang},
  \citenamefont {Su}, \citenamefont {Lin},\ and\ \citenamefont
  {Batista}}]{WangRKKY2020}%
  \BibitemOpen
  \bibfield  {author} {\bibinfo {author} {\bibfnamefont {Z.}~\bibnamefont
  {Wang}}, \bibinfo {author} {\bibfnamefont {Y.}~\bibnamefont {Su}}, \bibinfo
  {author} {\bibfnamefont {S.-Z.}\ \bibnamefont {Lin}},\ and\ \bibinfo {author}
  {\bibfnamefont {C.~D.}\ \bibnamefont {Batista}},\ }\bibfield  {title}
  {\bibinfo {title} {Skyrmion crystal from rkky interaction mediated by 2d
  electron gas},\ }\href {https://doi.org/10.1103/PhysRevLett.124.207201}
  {\bibfield  {journal} {\bibinfo  {journal} {Physical Review Letters}\
  }\textbf {\bibinfo {volume} {124}},\ \bibinfo {pages} {207201} (\bibinfo
  {year} {2020})}\BibitemShut {NoStop}%
\bibitem [{\citenamefont {Hayami}\ and\ \citenamefont
  {Motome}(2021{\natexlab{b}})}]{hayami21}%
  \BibitemOpen
  \bibfield  {author} {\bibinfo {author} {\bibfnamefont {S.}~\bibnamefont
  {Hayami}}\ and\ \bibinfo {author} {\bibfnamefont {Y.}~\bibnamefont
  {Motome}},\ }\bibfield  {title} {\bibinfo {title} {Topological spin crystals
  by itinerant frustration},\ }\href {https://doi.org/10.1088/1361-648X/ac1a30}
  {\bibfield  {journal} {\bibinfo  {journal} {Journal of Physics: Condensed
  Matter}\ }\textbf {\bibinfo {volume} {33}},\ \bibinfo {pages} {443001}
  (\bibinfo {year} {2021}{\natexlab{b}})}\BibitemShut {NoStop}%
\bibitem [{\citenamefont {Wang}\ and\ \citenamefont
  {Batista}(2023)}]{10.21468/SciPostPhys.15.4.161}%
  \BibitemOpen
  \bibfield  {author} {\bibinfo {author} {\bibfnamefont {Z.}~\bibnamefont
  {Wang}}\ and\ \bibinfo {author} {\bibfnamefont {C.~D.}\ \bibnamefont
  {Batista}},\ }\bibfield  {title} {\bibinfo {title} {{Skyrmion crystals in the
  triangular Kondo lattice model}},\ }\href
  {https://doi.org/10.21468/SciPostPhys.15.4.161} {\bibfield  {journal}
  {\bibinfo  {journal} {SciPost Phys.}\ }\textbf {\bibinfo {volume} {15}},\
  \bibinfo {pages} {161} (\bibinfo {year} {2023})}\BibitemShut {NoStop}%
\bibitem [{\citenamefont {Paul}\ \emph {et~al.}(2020)\citenamefont {Paul},
  \citenamefont {Haldar}, \citenamefont {von Malottki},\ and\ \citenamefont
  {Heinze}}]{Paul2020}%
  \BibitemOpen
  \bibfield  {author} {\bibinfo {author} {\bibfnamefont {S.}~\bibnamefont
  {Paul}}, \bibinfo {author} {\bibfnamefont {S.}~\bibnamefont {Haldar}},
  \bibinfo {author} {\bibfnamefont {S.}~\bibnamefont {von Malottki}},\ and\
  \bibinfo {author} {\bibfnamefont {S.}~\bibnamefont {Heinze}},\ }\bibfield
  {title} {\bibinfo {title} {Role of higher-order exchange interactions for
  skyrmion stability},\ }\href {https://doi.org/10.1038/s41467-020-18473-x}
  {\bibfield  {journal} {\bibinfo  {journal} {Nature Communications}\ }\textbf
  {\bibinfo {volume} {11}},\ \bibinfo {pages} {1} (\bibinfo {year}
  {2020})}\BibitemShut {NoStop}%
\bibitem [{\citenamefont {Ezawa}(2011)}]{ezawa2011c}%
  \BibitemOpen
  \bibfield  {author} {\bibinfo {author} {\bibfnamefont {M.}~\bibnamefont
  {Ezawa}},\ }\bibfield  {title} {\bibinfo {title} {Compact merons and
  skyrmions in thin chiral magnetic films},\ }\href
  {https://doi.org/10.1103/PhysRevB.83.100408} {\bibfield  {journal} {\bibinfo
  {journal} {Phys. Rev. B}\ }\textbf {\bibinfo {volume} {83}},\ \bibinfo
  {pages} {100408} (\bibinfo {year} {2011})}\BibitemShut {NoStop}%
\bibitem [{\citenamefont {Huang}\ \emph {et~al.}(2020)\citenamefont {Huang},
  \citenamefont {Sch{\"o}nenberger}, \citenamefont {Cantoni}, \citenamefont
  {Heinen}, \citenamefont {Magrez}, \citenamefont {Rosch}, \citenamefont
  {Carbone},\ and\ \citenamefont {R{\o}nnow}}]{huang2020m}%
  \BibitemOpen
  \bibfield  {author} {\bibinfo {author} {\bibfnamefont {P.}~\bibnamefont
  {Huang}}, \bibinfo {author} {\bibfnamefont {T.}~\bibnamefont
  {Sch{\"o}nenberger}}, \bibinfo {author} {\bibfnamefont {M.}~\bibnamefont
  {Cantoni}}, \bibinfo {author} {\bibfnamefont {L.}~\bibnamefont {Heinen}},
  \bibinfo {author} {\bibfnamefont {A.}~\bibnamefont {Magrez}}, \bibinfo
  {author} {\bibfnamefont {A.}~\bibnamefont {Rosch}}, \bibinfo {author}
  {\bibfnamefont {F.}~\bibnamefont {Carbone}},\ and\ \bibinfo {author}
  {\bibfnamefont {H.~M.}\ \bibnamefont {R{\o}nnow}},\ }\bibfield  {title}
  {\bibinfo {title} {Melting of a skyrmion lattice to a skyrmion liquid via a
  hexatic phase},\ }\href {https://doi.org/10.1038/s41565-020-0716-3}
  {\bibfield  {journal} {\bibinfo  {journal} {Nature Nanotechnology}\ }\textbf
  {\bibinfo {volume} {15}},\ \bibinfo {pages} {761} (\bibinfo {year}
  {2020})}\BibitemShut {NoStop}%
\bibitem [{\citenamefont {Bal{\'a}{\v{z}}}\ \emph {et~al.}(2021)\citenamefont
  {Bal{\'a}{\v{z}}}, \citenamefont {Pa{\'s}ciak},\ and\ \citenamefont
  {Hlinka}}]{balavz2021m}%
  \BibitemOpen
  \bibfield  {author} {\bibinfo {author} {\bibfnamefont {P.}~\bibnamefont
  {Bal{\'a}{\v{z}}}}, \bibinfo {author} {\bibfnamefont {M.}~\bibnamefont
  {Pa{\'s}ciak}},\ and\ \bibinfo {author} {\bibfnamefont {J.}~\bibnamefont
  {Hlinka}},\ }\bibfield  {title} {\bibinfo {title} {Melting of n{\'e}el
  skyrmion lattice},\ }\href {https://doi.org/10.1103/PhysRevB.103.174411}
  {\bibfield  {journal} {\bibinfo  {journal} {Phys. Rev. B}\ }\textbf {\bibinfo
  {volume} {103}},\ \bibinfo {pages} {174411} (\bibinfo {year}
  {2021})}\BibitemShut {NoStop}%
\bibitem [{\citenamefont {Nishikawa}\ \emph {et~al.}(2019)\citenamefont
  {Nishikawa}, \citenamefont {Hukushima},\ and\ \citenamefont
  {Krauth}}]{nishikawa2019s}%
  \BibitemOpen
  \bibfield  {author} {\bibinfo {author} {\bibfnamefont {Y.}~\bibnamefont
  {Nishikawa}}, \bibinfo {author} {\bibfnamefont {K.}~\bibnamefont
  {Hukushima}},\ and\ \bibinfo {author} {\bibfnamefont {W.}~\bibnamefont
  {Krauth}},\ }\bibfield  {title} {\bibinfo {title} {Solid-liquid transition of
  skyrmions in a two-dimensional chiral magnet},\ }\href
  {https://doi.org/10.1103/PhysRevB.99.064435} {\bibfield  {journal} {\bibinfo
  {journal} {Phys. Rev. B}\ }\textbf {\bibinfo {volume} {99}},\ \bibinfo
  {pages} {064435} (\bibinfo {year} {2019})}\BibitemShut {NoStop}%
\bibitem [{\citenamefont {Rosales}\ \emph {et~al.}(2023)\citenamefont
  {Rosales}, \citenamefont {Albarrac{\'\i}n}, \citenamefont {Pujol},\ and\
  \citenamefont {Jaubert}}]{rosales2023skyrmion}%
  \BibitemOpen
  \bibfield  {author} {\bibinfo {author} {\bibfnamefont {H.~D.}\ \bibnamefont
  {Rosales}}, \bibinfo {author} {\bibfnamefont {F.~A.~G.}\ \bibnamefont
  {Albarrac{\'\i}n}}, \bibinfo {author} {\bibfnamefont {P.}~\bibnamefont
  {Pujol}},\ and\ \bibinfo {author} {\bibfnamefont {L.~D.}\ \bibnamefont
  {Jaubert}},\ }\bibfield  {title} {\bibinfo {title} {Skyrmion fluid and
  bimeron glass protected by a chiral spin liquid on a kagome lattice},\ }\href
  {https://doi.org/10.1103/PhysRevLett.130.106703} {\bibfield  {journal}
  {\bibinfo  {journal} {Physical Review Letters}\ }\textbf {\bibinfo {volume}
  {130}},\ \bibinfo {pages} {106703} (\bibinfo {year} {2023})}\BibitemShut
  {NoStop}%
\bibitem [{\citenamefont {Yi}\ \emph {et~al.}(2009)\citenamefont {Yi},
  \citenamefont {Onoda}, \citenamefont {Nagaosa},\ and\ \citenamefont
  {Han}}]{yi09_skyrm_anomal_hall_effec_dzyal}%
  \BibitemOpen
  \bibfield  {author} {\bibinfo {author} {\bibfnamefont {S.~D.}\ \bibnamefont
  {Yi}}, \bibinfo {author} {\bibfnamefont {S.}~\bibnamefont {Onoda}}, \bibinfo
  {author} {\bibfnamefont {N.}~\bibnamefont {Nagaosa}},\ and\ \bibinfo {author}
  {\bibfnamefont {J.~H.}\ \bibnamefont {Han}},\ }\bibfield  {title} {\bibinfo
  {title} {Skyrmions and anomalous hall effect in a dzyaloshinskii-moriya
  spiral magnet},\ }\href {https://doi.org/10.1103/physrevb.80.054416}
  {\bibfield  {journal} {\bibinfo  {journal} {Phys. Rev. B}\ }\textbf {\bibinfo
  {volume} {80}},\ \bibinfo {pages} {054416} (\bibinfo {year}
  {2009})}\BibitemShut {NoStop}%
\bibitem [{\citenamefont {Hamamoto}\ \emph {et~al.}(2015)\citenamefont
  {Hamamoto}, \citenamefont {Ezawa},\ and\ \citenamefont
  {Nagaosa}}]{hamamoto2015quantized}%
  \BibitemOpen
  \bibfield  {author} {\bibinfo {author} {\bibfnamefont {K.}~\bibnamefont
  {Hamamoto}}, \bibinfo {author} {\bibfnamefont {M.}~\bibnamefont {Ezawa}},\
  and\ \bibinfo {author} {\bibfnamefont {N.}~\bibnamefont {Nagaosa}},\
  }\bibfield  {title} {\bibinfo {title} {Quantized topological hall effect in
  skyrmion crystal},\ }\href {https://doi.org/10.1103/PhysRevB.92.115417}
  {\bibfield  {journal} {\bibinfo  {journal} {Phys. Rev. B}\ }\textbf {\bibinfo
  {volume} {92}},\ \bibinfo {pages} {115417} (\bibinfo {year}
  {2015})}\BibitemShut {NoStop}%
\bibitem [{\citenamefont {Essafi}\ \emph {et~al.}(2016)\citenamefont {Essafi},
  \citenamefont {Benton},\ and\ \citenamefont {Jaubert}}]{essafi2016k}%
  \BibitemOpen
  \bibfield  {author} {\bibinfo {author} {\bibfnamefont {K.}~\bibnamefont
  {Essafi}}, \bibinfo {author} {\bibfnamefont {O.}~\bibnamefont {Benton}},\
  and\ \bibinfo {author} {\bibfnamefont {L.~D.}\ \bibnamefont {Jaubert}},\
  }\bibfield  {title} {\bibinfo {title} {A kagome map of spin liquids from xxz
  to dzyaloshinskii--moriya ferromagnet},\ }\href
  {https://doi.org/10.1038/ncomms10297} {\bibfield  {journal} {\bibinfo
  {journal} {Nature Communications}\ }\textbf {\bibinfo {volume} {7}},\
  \bibinfo {pages} {1} (\bibinfo {year} {2016})}\BibitemShut {NoStop}%
\bibitem [{\citenamefont {Luttinger}\ and\ \citenamefont
  {Tisza}(1946)}]{Luttinger1946}%
  \BibitemOpen
  \bibfield  {author} {\bibinfo {author} {\bibfnamefont {J.}~\bibnamefont
  {Luttinger}}\ and\ \bibinfo {author} {\bibfnamefont {L.}~\bibnamefont
  {Tisza}},\ }\bibfield  {title} {\bibinfo {title} {Theory of dipole
  interaction in crystals},\ }\href {https://doi.org/10.1103/PhysRev.70.954}
  {\bibfield  {journal} {\bibinfo  {journal} {Physical Review}\ }\textbf
  {\bibinfo {volume} {70}},\ \bibinfo {pages} {954} (\bibinfo {year}
  {1946})}\BibitemShut {NoStop}%
\bibitem [{\citenamefont {Luttinger}(1951)}]{Luttinger1951}%
  \BibitemOpen
  \bibfield  {author} {\bibinfo {author} {\bibfnamefont {J.}~\bibnamefont
  {Luttinger}},\ }\bibfield  {title} {\bibinfo {title} {A note on the ground
  state in antiferromagnetics},\ }\href
  {https://doi.org/10.1103/PhysRev.81.1015} {\bibfield  {journal} {\bibinfo
  {journal} {Physical Review}\ }\textbf {\bibinfo {volume} {81}},\ \bibinfo
  {pages} {1015} (\bibinfo {year} {1951})}\BibitemShut {NoStop}%
\bibitem [{\citenamefont {Rosales}\ \emph {et~al.}(2015)\citenamefont
  {Rosales}, \citenamefont {Cabra},\ and\ \citenamefont
  {Pujol}}]{rosales2015three}%
  \BibitemOpen
  \bibfield  {author} {\bibinfo {author} {\bibfnamefont {H.~D.}\ \bibnamefont
  {Rosales}}, \bibinfo {author} {\bibfnamefont {D.~C.}\ \bibnamefont {Cabra}},\
  and\ \bibinfo {author} {\bibfnamefont {P.}~\bibnamefont {Pujol}},\ }\bibfield
   {title} {\bibinfo {title} {Three-sublattice skyrmion crystal in the
  antiferromagnetic triangular lattice},\ }\href
  {https://doi.org/10.1103/PhysRevB.92.214439} {\bibfield  {journal} {\bibinfo
  {journal} {Phys. Rev. B}\ }\textbf {\bibinfo {volume} {92}},\ \bibinfo
  {pages} {214439} (\bibinfo {year} {2015})}\BibitemShut {NoStop}%
\bibitem [{\citenamefont {Henley}(2010)}]{henley10a}%
  \BibitemOpen
  \bibfield  {author} {\bibinfo {author} {\bibfnamefont {C.~L.}\ \bibnamefont
  {Henley}},\ }\bibfield  {title} {\bibinfo {title} {The "coulomb phase" in
  frustrated systems},\ }\href {https://doi.org/DOI
  10.1146/annurev-conmatphys-070909-104138} {\bibfield  {journal} {\bibinfo
  {journal} {Annu. Rev. Condens. Matter Phys.}\ }\textbf {\bibinfo {volume}
  {1}},\ \bibinfo {pages} {179} (\bibinfo {year} {2010})}\BibitemShut {NoStop}%
\bibitem [{\citenamefont {Chalker}\ \emph {et~al.}(1992)\citenamefont
  {Chalker}, \citenamefont {Holdsworth},\ and\ \citenamefont
  {Shender}}]{Chalker92a}%
  \BibitemOpen
  \bibfield  {author} {\bibinfo {author} {\bibfnamefont {J.~T.}\ \bibnamefont
  {Chalker}}, \bibinfo {author} {\bibfnamefont {P.~C.~W.}\ \bibnamefont
  {Holdsworth}},\ and\ \bibinfo {author} {\bibfnamefont {E.~F.}\ \bibnamefont
  {Shender}},\ }\bibfield  {title} {\bibinfo {title} {Hidden order in a
  frustrated system: Properties of the heisenberg kagom\'e antiferromagnet},\
  }\href {https://doi.org/10.1103/PhysRevLett.68.855} {\bibfield  {journal}
  {\bibinfo  {journal} {Phys. Rev. Lett.}\ }\textbf {\bibinfo {volume} {68}},\
  \bibinfo {pages} {855} (\bibinfo {year} {1992})}\BibitemShut {NoStop}%
\bibitem [{\citenamefont {Davier}\ \emph {et~al.}(2023)\citenamefont {Davier},
  \citenamefont {G\'omez~Albarrac\'{\i}n}, \citenamefont {Rosales},\ and\
  \citenamefont {Pujol}}]{davier2023combined}%
  \BibitemOpen
  \bibfield  {author} {\bibinfo {author} {\bibfnamefont {N.}~\bibnamefont
  {Davier}}, \bibinfo {author} {\bibfnamefont {F.~A.}\ \bibnamefont
  {G\'omez~Albarrac\'{\i}n}}, \bibinfo {author} {\bibfnamefont {H.~D.}\
  \bibnamefont {Rosales}},\ and\ \bibinfo {author} {\bibfnamefont
  {P.}~\bibnamefont {Pujol}},\ }\bibfield  {title} {\bibinfo {title} {Combined
  approach to analyze and classify families of classical spin liquids},\ }\href
  {https://doi.org/10.1103/PhysRevB.108.054408} {\bibfield  {journal} {\bibinfo
   {journal} {Phys. Rev. B}\ }\textbf {\bibinfo {volume} {108}},\ \bibinfo
  {pages} {054408} (\bibinfo {year} {2023})}\BibitemShut {NoStop}%
\bibitem [{\citenamefont {Yan}\ \emph {et~al.}(2023)\citenamefont {Yan},
  \citenamefont {Benton}, \citenamefont {Moessner},\ and\ \citenamefont
  {Nevidomskyy}}]{yan2023classification}%
  \BibitemOpen
  \bibfield  {author} {\bibinfo {author} {\bibfnamefont {H.}~\bibnamefont
  {Yan}}, \bibinfo {author} {\bibfnamefont {O.}~\bibnamefont {Benton}},
  \bibinfo {author} {\bibfnamefont {R.}~\bibnamefont {Moessner}},\ and\
  \bibinfo {author} {\bibfnamefont {A.~H.}\ \bibnamefont {Nevidomskyy}},\
  }\bibfield  {title} {\bibinfo {title} {Classification of classical spin
  liquids: Typology and resulting landscape},\ }\href
  {https://doi.org/10.48550/arXiv.2305.00155} {\bibfield  {journal} {\bibinfo
  {journal} {arXiv preprint arXiv:2305.00155}\ } (\bibinfo {year}
  {2023})}\BibitemShut {NoStop}%
\bibitem [{\citenamefont {Essafi}\ \emph {et~al.}(2017)\citenamefont {Essafi},
  \citenamefont {Benton},\ and\ \citenamefont {Jaubert}}]{Essafi2017}%
  \BibitemOpen
  \bibfield  {author} {\bibinfo {author} {\bibfnamefont {K.}~\bibnamefont
  {Essafi}}, \bibinfo {author} {\bibfnamefont {O.}~\bibnamefont {Benton}},\
  and\ \bibinfo {author} {\bibfnamefont {L.~D.~C.}\ \bibnamefont {Jaubert}},\
  }\bibfield  {title} {\bibinfo {title} {Generic nearest-neighbor kagome model:
  Xyz and dzyaloshinskii-moriya couplings with comparison to the
  pyrochlore-lattice case},\ }\href
  {https://doi.org/10.1103/PhysRevB.96.205126} {\bibfield  {journal} {\bibinfo
  {journal} {Phys. Rev. B}\ }\textbf {\bibinfo {volume} {96}},\ \bibinfo
  {pages} {205126} (\bibinfo {year} {2017})}\BibitemShut {NoStop}%
\bibitem [{\citenamefont {Garanin}\ \emph {et~al.}(2020)\citenamefont
  {Garanin}, \citenamefont {Chudnovsky}, \citenamefont {Zhang},\ and\
  \citenamefont {Zhang}}]{GARANIN2020165724}%
  \BibitemOpen
  \bibfield  {author} {\bibinfo {author} {\bibfnamefont {D.~A.}\ \bibnamefont
  {Garanin}}, \bibinfo {author} {\bibfnamefont {E.~M.}\ \bibnamefont
  {Chudnovsky}}, \bibinfo {author} {\bibfnamefont {S.}~\bibnamefont {Zhang}},\
  and\ \bibinfo {author} {\bibfnamefont {X.}~\bibnamefont {Zhang}},\ }\bibfield
   {title} {\bibinfo {title} {Thermal creation of skyrmions in ferromagnetic
  films with perpendicular anisotropy and dzyaloshinskii-moriya interaction},\
  }\href {https://www.sciencedirect.com/science/article/pii/S0304885319316713}
  {\bibfield  {journal} {\bibinfo  {journal} {Journal of Magnetism and Magnetic
  Materials}\ }\textbf {\bibinfo {volume} {493}},\ \bibinfo {pages} {165724}
  (\bibinfo {year} {2020})}\BibitemShut {NoStop}%
\bibitem [{\citenamefont {Ohara}\ \emph {et~al.}(2022)\citenamefont {Ohara},
  \citenamefont {Zhang}, \citenamefont {Chen}, \citenamefont {Kato},
  \citenamefont {Xia}, \citenamefont {Ezawa}, \citenamefont {Tretiakov},
  \citenamefont {Hou}, \citenamefont {Zhou}, \citenamefont {Zhao},
  \citenamefont {Yang},\ and\ \citenamefont {Liu}}]{OharaNANO}%
  \BibitemOpen
  \bibfield  {author} {\bibinfo {author} {\bibfnamefont {K.}~\bibnamefont
  {Ohara}}, \bibinfo {author} {\bibfnamefont {X.}~\bibnamefont {Zhang}},
  \bibinfo {author} {\bibfnamefont {Y.}~\bibnamefont {Chen}}, \bibinfo {author}
  {\bibfnamefont {S.}~\bibnamefont {Kato}}, \bibinfo {author} {\bibfnamefont
  {J.}~\bibnamefont {Xia}}, \bibinfo {author} {\bibfnamefont {M.}~\bibnamefont
  {Ezawa}}, \bibinfo {author} {\bibfnamefont {O.~A.}\ \bibnamefont
  {Tretiakov}}, \bibinfo {author} {\bibfnamefont {Z.}~\bibnamefont {Hou}},
  \bibinfo {author} {\bibfnamefont {Y.}~\bibnamefont {Zhou}}, \bibinfo {author}
  {\bibfnamefont {G.}~\bibnamefont {Zhao}}, \bibinfo {author} {\bibfnamefont
  {J.}~\bibnamefont {Yang}},\ and\ \bibinfo {author} {\bibfnamefont
  {X.}~\bibnamefont {Liu}},\ }\bibfield  {title} {\bibinfo {title} {Reversible
  transformation between isolated skyrmions and bimerons},\ }\href
  {https://doi.org/10.1021/acs.nanolett.2c03106} {\bibfield  {journal}
  {\bibinfo  {journal} {Nano Letters}\ }\textbf {\bibinfo {volume} {22}},\
  \bibinfo {pages} {8559} (\bibinfo {year} {2022})}\BibitemShut {NoStop}%
\bibitem [{\citenamefont {Wang}\ \emph {et~al.}(2022)\citenamefont {Wang},
  \citenamefont {Sun},\ and\ \citenamefont {Li}}]{Wang-Wenbin}%
  \BibitemOpen
  \bibfield  {author} {\bibinfo {author} {\bibfnamefont {W.}~\bibnamefont
  {Wang}}, \bibinfo {author} {\bibfnamefont {J.}~\bibnamefont {Sun}},\ and\
  \bibinfo {author} {\bibfnamefont {H.}~\bibnamefont {Li}},\ }\bibfield
  {title} {\bibinfo {title} {Stabilization of skyrmions in two-dimensional
  systems with next-nearest-neighbor exchange interactions},\ }\href
  {https://www.frontiersin.org/articles/10.3389/fphy.2022.995902} {\bibfield
  {journal} {\bibinfo  {journal} {Frontiers in Physics}\ }\textbf {\bibinfo
  {volume} {10}},\ \bibinfo {pages} {995902} (\bibinfo {year}
  {2022})}\BibitemShut {NoStop}%
\bibitem [{\citenamefont {Du}\ \emph {et~al.}(2022)\citenamefont {Du},
  \citenamefont {Dou}, \citenamefont {He}, \citenamefont {Dai}, \citenamefont
  {Huang},\ and\ \citenamefont {Ma}}]{Du-Wenhui}%
  \BibitemOpen
  \bibfield  {author} {\bibinfo {author} {\bibfnamefont {W.}~\bibnamefont
  {Du}}, \bibinfo {author} {\bibfnamefont {K.}~\bibnamefont {Dou}}, \bibinfo
  {author} {\bibfnamefont {Z.}~\bibnamefont {He}}, \bibinfo {author}
  {\bibfnamefont {Y.}~\bibnamefont {Dai}}, \bibinfo {author} {\bibfnamefont
  {B.}~\bibnamefont {Huang}},\ and\ \bibinfo {author} {\bibfnamefont
  {Y.}~\bibnamefont {Ma}},\ }\bibfield  {title} {\bibinfo {title} {Spontaneous
  magnetic skyrmions in single-layer crinx3 (x = te, se)},\ }\href
  {https://doi.org/10.1021/acs.nanolett.2c00836} {\bibfield  {journal}
  {\bibinfo  {journal} {Nano Letters}\ }\textbf {\bibinfo {volume} {22}},\
  \bibinfo {pages} {3440} (\bibinfo {year} {2022})}\BibitemShut {NoStop}%
\bibitem [{\citenamefont {Han}\ \emph {et~al.}(2010)\citenamefont {Han},
  \citenamefont {Zang}, \citenamefont {Yang}, \citenamefont {Park},\ and\
  \citenamefont {Nagaosa}}]{han2010}%
  \BibitemOpen
  \bibfield  {author} {\bibinfo {author} {\bibfnamefont {J.~H.}\ \bibnamefont
  {Han}}, \bibinfo {author} {\bibfnamefont {J.}~\bibnamefont {Zang}}, \bibinfo
  {author} {\bibfnamefont {Z.}~\bibnamefont {Yang}}, \bibinfo {author}
  {\bibfnamefont {J.-H.}\ \bibnamefont {Park}},\ and\ \bibinfo {author}
  {\bibfnamefont {N.}~\bibnamefont {Nagaosa}},\ }\bibfield  {title} {\bibinfo
  {title} {Skyrmion lattice in a two-dimensional chiral magnet},\ }\href
  {https://doi.org/10.1103/PhysRevB.82.094429} {\bibfield  {journal} {\bibinfo
  {journal} {Phys. Rev. B}\ }\textbf {\bibinfo {volume} {82}},\ \bibinfo
  {pages} {094429} (\bibinfo {year} {2010})}\BibitemShut {NoStop}%
\bibitem [{\citenamefont {Chern}\ and\ \citenamefont
  {Tchernyshyov}(2012)}]{Tchernyshyov2012}%
  \BibitemOpen
  \bibfield  {author} {\bibinfo {author} {\bibfnamefont {G.-W.}\ \bibnamefont
  {Chern}}\ and\ \bibinfo {author} {\bibfnamefont {O.}~\bibnamefont
  {Tchernyshyov}},\ }\bibfield  {title} {\bibinfo {title} {Magnetic charge and
  ordering in kagome spin ice},\ }\href
  {https://doi.org/https://doi.org/10.1098/rsta.2011.0388} {\bibfield
  {journal} {\bibinfo  {journal} {Phyl. Trans. Roy. Soc. A}\ }\textbf {\bibinfo
  {volume} {370}},\ \bibinfo {pages} {5718} (\bibinfo {year}
  {2012})}\BibitemShut {NoStop}%
\bibitem [{\citenamefont {Mohanta}\ \emph {et~al.}(2020)\citenamefont
  {Mohanta}, \citenamefont {Christianson}, \citenamefont {Okamoto},\ and\
  \citenamefont {Dagotto}}]{mohanta2020}%
  \BibitemOpen
  \bibfield  {author} {\bibinfo {author} {\bibfnamefont {N.}~\bibnamefont
  {Mohanta}}, \bibinfo {author} {\bibfnamefont {A.~D.}\ \bibnamefont
  {Christianson}}, \bibinfo {author} {\bibfnamefont {S.}~\bibnamefont
  {Okamoto}},\ and\ \bibinfo {author} {\bibfnamefont {E.}~\bibnamefont
  {Dagotto}},\ }\bibfield  {title} {\bibinfo {title} {Signatures of a
  liquid$-$crystal transition in spin$-$wave excitations of skyrmions},\ }\href
  {https://doi.org/10.1038/s42005-020-00489-w} {\bibfield  {journal} {\bibinfo
  {journal} {Communications Physics}\ }\textbf {\bibinfo {volume} {3}},\
  \bibinfo {pages} {1} (\bibinfo {year} {2020})}\BibitemShut {NoStop}%
\bibitem [{\citenamefont {Robert}\ \emph {et~al.}(2008)\citenamefont {Robert},
  \citenamefont {Canals}, \citenamefont {Simonet},\ and\ \citenamefont
  {Ballou}}]{robert08a}%
  \BibitemOpen
  \bibfield  {author} {\bibinfo {author} {\bibfnamefont {J.}~\bibnamefont
  {Robert}}, \bibinfo {author} {\bibfnamefont {B.}~\bibnamefont {Canals}},
  \bibinfo {author} {\bibfnamefont {V.}~\bibnamefont {Simonet}},\ and\ \bibinfo
  {author} {\bibfnamefont {R.}~\bibnamefont {Ballou}},\ }\bibfield  {title}
  {\bibinfo {title} {Propagation and ghosts in the classical kagome
  antiferromagnet},\ }\href {https://doi.org/10.1103/PhysRevLett.101.117207}
  {\bibfield  {journal} {\bibinfo  {journal} {Phys. Rev. Lett.}\ }\textbf
  {\bibinfo {volume} {101}},\ \bibinfo {pages} {117207} (\bibinfo {year}
  {2008})}\BibitemShut {NoStop}%
\bibitem [{\citenamefont {Mizoguchi}\ \emph {et~al.}(2018)\citenamefont
  {Mizoguchi}, \citenamefont {Jaubert}, \citenamefont {Moessner},\ and\
  \citenamefont {Udagawa}}]{mizoguchi2018m}%
  \BibitemOpen
  \bibfield  {author} {\bibinfo {author} {\bibfnamefont {T.}~\bibnamefont
  {Mizoguchi}}, \bibinfo {author} {\bibfnamefont {L.~D.}\ \bibnamefont
  {Jaubert}}, \bibinfo {author} {\bibfnamefont {R.}~\bibnamefont {Moessner}},\
  and\ \bibinfo {author} {\bibfnamefont {M.}~\bibnamefont {Udagawa}},\
  }\bibfield  {title} {\bibinfo {title} {Magnetic clustering, half-moons, and
  shadow pinch points as signals of a proximate coulomb phase in frustrated
  heisenberg magnets},\ }\href {https://doi.org/10.1103/PhysRevB.98.144446}
  {\bibfield  {journal} {\bibinfo  {journal} {Phys. Rev. B}\ }\textbf {\bibinfo
  {volume} {98}},\ \bibinfo {pages} {144446} (\bibinfo {year}
  {2018})}\BibitemShut {NoStop}%
\bibitem [{\citenamefont {Yan}\ \emph {et~al.}(2018)\citenamefont {Yan},
  \citenamefont {Pohle},\ and\ \citenamefont {Shannon}}]{yan2018half}%
  \BibitemOpen
  \bibfield  {author} {\bibinfo {author} {\bibfnamefont {H.}~\bibnamefont
  {Yan}}, \bibinfo {author} {\bibfnamefont {R.}~\bibnamefont {Pohle}},\ and\
  \bibinfo {author} {\bibfnamefont {N.}~\bibnamefont {Shannon}},\ }\bibfield
  {title} {\bibinfo {title} {Half moons are pinch points with dispersion},\
  }\href {https://doi.org/10.1103/PhysRevB.98.140402} {\bibfield  {journal}
  {\bibinfo  {journal} {Phys. Rev. B}\ }\textbf {\bibinfo {volume} {98}},\
  \bibinfo {pages} {140402} (\bibinfo {year} {2018})}\BibitemShut {NoStop}%
\bibitem [{\citenamefont {Timm}\ \emph {et~al.}(1998)\citenamefont {Timm},
  \citenamefont {Girvin},\ and\ \citenamefont {Fertig}}]{timm1998sk}%
  \BibitemOpen
  \bibfield  {author} {\bibinfo {author} {\bibfnamefont {C.}~\bibnamefont
  {Timm}}, \bibinfo {author} {\bibfnamefont {S.}~\bibnamefont {Girvin}},\ and\
  \bibinfo {author} {\bibfnamefont {H.}~\bibnamefont {Fertig}},\ }\bibfield
  {title} {\bibinfo {title} {Skyrmion lattice melting in the quantum hall
  system},\ }\href {https://doi.org/10.1103/PhysRevB.58.10634} {\bibfield
  {journal} {\bibinfo  {journal} {Phys. Rev. B}\ }\textbf {\bibinfo {volume}
  {58}},\ \bibinfo {pages} {10634} (\bibinfo {year} {1998})}\BibitemShut
  {NoStop}%
\bibitem [{\citenamefont {Nelson}\ and\ \citenamefont
  {Halperin}(1979)}]{nelson1979d}%
  \BibitemOpen
  \bibfield  {author} {\bibinfo {author} {\bibfnamefont {D.~R.}\ \bibnamefont
  {Nelson}}\ and\ \bibinfo {author} {\bibfnamefont {B.}~\bibnamefont
  {Halperin}},\ }\bibfield  {title} {\bibinfo {title} {Dislocation-mediated
  melting in two dimensions},\ }\href
  {https://doi.org/10.1103/PhysRevB.19.2457} {\bibfield  {journal} {\bibinfo
  {journal} {Phys. Rev. B}\ }\textbf {\bibinfo {volume} {19}},\ \bibinfo
  {pages} {2457} (\bibinfo {year} {1979})}\BibitemShut {NoStop}%
\bibitem [{\citenamefont {Kosterlitz}\ and\ \citenamefont
  {Thouless}(1973)}]{Kosterlitz1973}%
  \BibitemOpen
  \bibfield  {author} {\bibinfo {author} {\bibfnamefont {J.~M.}\ \bibnamefont
  {Kosterlitz}}\ and\ \bibinfo {author} {\bibfnamefont {D.~J.}\ \bibnamefont
  {Thouless}},\ }\bibfield  {title} {\bibinfo {title} {Ordering, metastability
  and phase transitions in two-dimensional systems},\ }\href
  {https://doi.org/10.1088/0022-3719/6/7/010} {\bibfield  {journal} {\bibinfo
  {journal} {Journal of Physics C: Solid State Physics}\ }\textbf {\bibinfo
  {volume} {6}},\ \bibinfo {pages} {1181} (\bibinfo {year} {1973})}\BibitemShut
  {NoStop}%
\bibitem [{\citenamefont {Halperin}\ and\ \citenamefont
  {Nelson}(1978)}]{Halperin1978}%
  \BibitemOpen
  \bibfield  {author} {\bibinfo {author} {\bibfnamefont {B.~I.}\ \bibnamefont
  {Halperin}}\ and\ \bibinfo {author} {\bibfnamefont {D.~R.}\ \bibnamefont
  {Nelson}},\ }\bibfield  {title} {\bibinfo {title} {Theory of two-dimensional
  melting},\ }\href {https://doi.org/10.1103/PhysRevLett.41.121} {\bibfield
  {journal} {\bibinfo  {journal} {Phys. Rev. Lett.}\ }\textbf {\bibinfo
  {volume} {41}},\ \bibinfo {pages} {121} (\bibinfo {year} {1978})}\BibitemShut
  {NoStop}%
\bibitem [{\citenamefont {Young}(1979)}]{Young1979}%
  \BibitemOpen
  \bibfield  {author} {\bibinfo {author} {\bibfnamefont {A.~P.}\ \bibnamefont
  {Young}},\ }\bibfield  {title} {\bibinfo {title} {Melting and the vector
  coulomb gas in two dimensions},\ }\href
  {https://doi.org/10.1103/PhysRevB.19.1855} {\bibfield  {journal} {\bibinfo
  {journal} {Phys. Rev. B}\ }\textbf {\bibinfo {volume} {19}},\ \bibinfo
  {pages} {1855} (\bibinfo {year} {1979})}\BibitemShut {NoStop}%
\bibitem [{\citenamefont {Ganguli}\ \emph {et~al.}(2015)\citenamefont
  {Ganguli}, \citenamefont {Singh}, \citenamefont {Saraswat}, \citenamefont
  {Ganguly}, \citenamefont {Bagwe}, \citenamefont {Shirage}, \citenamefont
  {Thamizhavel},\ and\ \citenamefont {Raychaudhuri}}]{ganguli2015d}%
  \BibitemOpen
  \bibfield  {author} {\bibinfo {author} {\bibfnamefont {S.~C.}\ \bibnamefont
  {Ganguli}}, \bibinfo {author} {\bibfnamefont {H.}~\bibnamefont {Singh}},
  \bibinfo {author} {\bibfnamefont {G.}~\bibnamefont {Saraswat}}, \bibinfo
  {author} {\bibfnamefont {R.}~\bibnamefont {Ganguly}}, \bibinfo {author}
  {\bibfnamefont {V.}~\bibnamefont {Bagwe}}, \bibinfo {author} {\bibfnamefont
  {P.}~\bibnamefont {Shirage}}, \bibinfo {author} {\bibfnamefont
  {A.}~\bibnamefont {Thamizhavel}},\ and\ \bibinfo {author} {\bibfnamefont
  {P.}~\bibnamefont {Raychaudhuri}},\ }\bibfield  {title} {\bibinfo {title}
  {Disordering of the vortex lattice through successive destruction of
  positional and orientational order in a weakly pinned co 0.0075 nbse 2 single
  crystal},\ }\href {https://doi.org/10.1038/srep10613} {\bibfield  {journal}
  {\bibinfo  {journal} {Scientific reports}\ }\textbf {\bibinfo {volume} {5}},\
  \bibinfo {pages} {1} (\bibinfo {year} {2015})}\BibitemShut {NoStop}%
\bibitem [{\citenamefont {Batista}\ \emph {et~al.}(2016)\citenamefont
  {Batista}, \citenamefont {Lin}, \citenamefont {Hayami},\ and\ \citenamefont
  {Kamiya}}]{batista16a}%
  \BibitemOpen
  \bibfield  {author} {\bibinfo {author} {\bibfnamefont {C.~D.}\ \bibnamefont
  {Batista}}, \bibinfo {author} {\bibfnamefont {S.-Z.}\ \bibnamefont {Lin}},
  \bibinfo {author} {\bibfnamefont {S.}~\bibnamefont {Hayami}},\ and\ \bibinfo
  {author} {\bibfnamefont {Y.}~\bibnamefont {Kamiya}},\ }\bibfield  {title}
  {\bibinfo {title} {Frustration and chiral orderings in correlated electron
  systems},\ }\href {https://doi.org/10.1088/0034-4885/79/8/084504} {\bibfield
  {journal} {\bibinfo  {journal} {Reports on Progress in Physics}\ }\textbf
  {\bibinfo {volume} {79}},\ \bibinfo {pages} {084504} (\bibinfo {year}
  {2016})}\BibitemShut {NoStop}%
\bibitem [{\citenamefont {Taguchi}\ \emph {et~al.}(2001)\citenamefont
  {Taguchi}, \citenamefont {Oohara}, \citenamefont {Yoshizawa}, \citenamefont
  {Nagaosa},\ and\ \citenamefont {Tokura}}]{doi:10.1126/science.1058161}%
  \BibitemOpen
  \bibfield  {author} {\bibinfo {author} {\bibfnamefont {Y.}~\bibnamefont
  {Taguchi}}, \bibinfo {author} {\bibfnamefont {Y.}~\bibnamefont {Oohara}},
  \bibinfo {author} {\bibfnamefont {H.}~\bibnamefont {Yoshizawa}}, \bibinfo
  {author} {\bibfnamefont {N.}~\bibnamefont {Nagaosa}},\ and\ \bibinfo {author}
  {\bibfnamefont {Y.}~\bibnamefont {Tokura}},\ }\bibfield  {title} {\bibinfo
  {title} {Spin chirality, berry phase, and anomalous hall effect in a
  frustrated ferromagnet},\ }\href {https://doi.org/10.1126/science.1058161}
  {\bibfield  {journal} {\bibinfo  {journal} {Science}\ }\textbf {\bibinfo
  {volume} {291}},\ \bibinfo {pages} {2573} (\bibinfo {year}
  {2001})}\BibitemShut {NoStop}%
\bibitem [{\citenamefont {Machida}\ \emph {et~al.}(2007)\citenamefont
  {Machida}, \citenamefont {Nakatsuji}, \citenamefont {Maeno}, \citenamefont
  {Tayama}, \citenamefont {Sakakibara},\ and\ \citenamefont
  {Onoda}}]{PhysRevLett.98.057203}%
  \BibitemOpen
  \bibfield  {author} {\bibinfo {author} {\bibfnamefont {Y.}~\bibnamefont
  {Machida}}, \bibinfo {author} {\bibfnamefont {S.}~\bibnamefont {Nakatsuji}},
  \bibinfo {author} {\bibfnamefont {Y.}~\bibnamefont {Maeno}}, \bibinfo
  {author} {\bibfnamefont {T.}~\bibnamefont {Tayama}}, \bibinfo {author}
  {\bibfnamefont {T.}~\bibnamefont {Sakakibara}},\ and\ \bibinfo {author}
  {\bibfnamefont {S.}~\bibnamefont {Onoda}},\ }\bibfield  {title} {\bibinfo
  {title} {Unconventional anomalous hall effect enhanced by a noncoplanar spin
  texture in the frustrated kondo lattice
  ${\mathrm{pr}}_{2}{\mathrm{ir}}_{2}{\mathrm{o}}_{7}$},\ }\href
  {https://doi.org/10.1103/PhysRevLett.98.057203} {\bibfield  {journal}
  {\bibinfo  {journal} {Phys. Rev. Lett.}\ }\textbf {\bibinfo {volume} {98}},\
  \bibinfo {pages} {057203} (\bibinfo {year} {2007})}\BibitemShut {NoStop}%
\bibitem [{\citenamefont {Uehara}\ \emph {et~al.}(2022)\citenamefont {Uehara},
  \citenamefont {Ohtsuki}, \citenamefont {Udagawa}, \citenamefont {Nakatsuji},\
  and\ \citenamefont {Machida}}]{Uehara22}%
  \BibitemOpen
  \bibfield  {author} {\bibinfo {author} {\bibfnamefont {T.}~\bibnamefont
  {Uehara}}, \bibinfo {author} {\bibfnamefont {T.}~\bibnamefont {Ohtsuki}},
  \bibinfo {author} {\bibfnamefont {M.}~\bibnamefont {Udagawa}}, \bibinfo
  {author} {\bibfnamefont {S.}~\bibnamefont {Nakatsuji}},\ and\ \bibinfo
  {author} {\bibfnamefont {Y.}~\bibnamefont {Machida}},\ }\bibfield  {title}
  {\bibinfo {title} {Phonon thermal hall effect in a metallic spin ice},\
  }\href {https://doi.org/10.1038/s41467-022-32375-0} {\bibfield  {journal}
  {\bibinfo  {journal} {Nature Communications}\ }\textbf {\bibinfo {volume}
  {13}},\ \bibinfo {pages} {4604} (\bibinfo {year} {2022})}\BibitemShut
  {NoStop}%
\bibitem [{\citenamefont {Ohgushi}\ \emph {et~al.}(2000)\citenamefont
  {Ohgushi}, \citenamefont {Murakami},\ and\ \citenamefont
  {Nagaosa}}]{ohgushi2000spin}%
  \BibitemOpen
  \bibfield  {author} {\bibinfo {author} {\bibfnamefont {K.}~\bibnamefont
  {Ohgushi}}, \bibinfo {author} {\bibfnamefont {S.}~\bibnamefont {Murakami}},\
  and\ \bibinfo {author} {\bibfnamefont {N.}~\bibnamefont {Nagaosa}},\
  }\bibfield  {title} {\bibinfo {title} {Spin anisotropy and quantum hall
  effect in the kagom{\'e} lattice: Chiral spin state based on a ferromagnet},\
  }\href {https://doi.org/10.1103/PhysRevB.62.R6065} {\bibfield  {journal}
  {\bibinfo  {journal} {Phys. Rev. B}\ }\textbf {\bibinfo {volume} {62}},\
  \bibinfo {pages} {R6065} (\bibinfo {year} {2000})}\BibitemShut {NoStop}%
\bibitem [{\citenamefont {Taillefumier}\ \emph {et~al.}(2006)\citenamefont
  {Taillefumier}, \citenamefont {Canals}, \citenamefont {Lacroix},
  \citenamefont {Dugaev},\ and\ \citenamefont {Bruno}}]{Taillefumier06a}%
  \BibitemOpen
  \bibfield  {author} {\bibinfo {author} {\bibfnamefont {M.}~\bibnamefont
  {Taillefumier}}, \bibinfo {author} {\bibfnamefont {B.}~\bibnamefont
  {Canals}}, \bibinfo {author} {\bibfnamefont {C.}~\bibnamefont {Lacroix}},
  \bibinfo {author} {\bibfnamefont {V.~K.}\ \bibnamefont {Dugaev}},\ and\
  \bibinfo {author} {\bibfnamefont {P.}~\bibnamefont {Bruno}},\ }\bibfield
  {title} {\bibinfo {title} {Anomalous hall effect due to spin chirality in the
  kagom\'e lattice},\ }\href {https://doi.org/10.1103/PhysRevB.74.085105}
  {\bibfield  {journal} {\bibinfo  {journal} {Phys. Rev. B}\ }\textbf {\bibinfo
  {volume} {74}},\ \bibinfo {pages} {085105} (\bibinfo {year}
  {2006})}\BibitemShut {NoStop}%
\bibitem [{\citenamefont {Xu}\ \emph {et~al.}(2015)\citenamefont {Xu},
  \citenamefont {Lian},\ and\ \citenamefont {Zhang}}]{xu15a}%
  \BibitemOpen
  \bibfield  {author} {\bibinfo {author} {\bibfnamefont {G.}~\bibnamefont
  {Xu}}, \bibinfo {author} {\bibfnamefont {B.}~\bibnamefont {Lian}},\ and\
  \bibinfo {author} {\bibfnamefont {S.-C.}\ \bibnamefont {Zhang}},\ }\bibfield
  {title} {\bibinfo {title} {Intrinsic quantum anomalous hall effect in the
  kagome lattice ${\mathrm{cs}}_{2}{\mathrm{limn}}_{3}{\mathrm{f}}_{12}$},\
  }\href {https://doi.org/10.1103/PhysRevLett.115.186802} {\bibfield  {journal}
  {\bibinfo  {journal} {Phys. Rev. Lett.}\ }\textbf {\bibinfo {volume} {115}},\
  \bibinfo {pages} {186802} (\bibinfo {year} {2015})}\BibitemShut {NoStop}%
\bibitem [{\citenamefont {Rosales}\ \emph {et~al.}(2019)\citenamefont
  {Rosales}, \citenamefont {G{\'o}mez~Albarrac{\'\i}n},\ and\ \citenamefont
  {Pujol}}]{rosales2019frustrated}%
  \BibitemOpen
  \bibfield  {author} {\bibinfo {author} {\bibfnamefont {H.}~\bibnamefont
  {Rosales}}, \bibinfo {author} {\bibfnamefont {F.}~\bibnamefont
  {G{\'o}mez~Albarrac{\'\i}n}},\ and\ \bibinfo {author} {\bibfnamefont
  {P.}~\bibnamefont {Pujol}},\ }\bibfield  {title} {\bibinfo {title} {From
  frustrated magnetism to spontaneous chern insulators},\ }\href
  {https://doi.org/10.1103/PhysRevB.99.035163} {\bibfield  {journal} {\bibinfo
  {journal} {Phys. Rev. B}\ }\textbf {\bibinfo {volume} {99}},\ \bibinfo
  {pages} {035163} (\bibinfo {year} {2019})}\BibitemShut {NoStop}%
\bibitem [{\citenamefont {Busch}\ \emph {et~al.}(2020)\citenamefont {Busch},
  \citenamefont {G\"obel},\ and\ \citenamefont {Mertig}}]{Mertig2020}%
  \BibitemOpen
  \bibfield  {author} {\bibinfo {author} {\bibfnamefont {O.}~\bibnamefont
  {Busch}}, \bibinfo {author} {\bibfnamefont {B.}~\bibnamefont {G\"obel}},\
  and\ \bibinfo {author} {\bibfnamefont {I.}~\bibnamefont {Mertig}},\
  }\bibfield  {title} {\bibinfo {title} {Microscopic origin of the anomalous
  hall effect in noncollinear kagome magnets},\ }\href
  {https://doi.org/10.1103/PhysRevResearch.2.033112} {\bibfield  {journal}
  {\bibinfo  {journal} {Phys. Rev. Res.}\ }\textbf {\bibinfo {volume} {2}},\
  \bibinfo {pages} {033112} (\bibinfo {year} {2020})}\BibitemShut {NoStop}%
\bibitem [{\citenamefont {Chen}\ \emph {et~al.}(2014)\citenamefont {Chen},
  \citenamefont {Niu},\ and\ \citenamefont {MacDonald}}]{Chen14a}%
  \BibitemOpen
  \bibfield  {author} {\bibinfo {author} {\bibfnamefont {H.}~\bibnamefont
  {Chen}}, \bibinfo {author} {\bibfnamefont {Q.}~\bibnamefont {Niu}},\ and\
  \bibinfo {author} {\bibfnamefont {A.~H.}\ \bibnamefont {MacDonald}},\
  }\bibfield  {title} {\bibinfo {title} {Anomalous hall effect arising from
  noncollinear antiferromagnetism},\ }\href
  {https://doi.org/10.1103/PhysRevLett.112.017205} {\bibfield  {journal}
  {\bibinfo  {journal} {Phys. Rev. Lett.}\ }\textbf {\bibinfo {volume} {112}},\
  \bibinfo {pages} {017205} (\bibinfo {year} {2014})}\BibitemShut {NoStop}%
\bibitem [{\citenamefont {Chen}(2022)}]{Chen2022}%
  \BibitemOpen
  \bibfield  {author} {\bibinfo {author} {\bibfnamefont {H.}~\bibnamefont
  {Chen}},\ }\bibfield  {title} {\bibinfo {title} {Electronic chiralization as
  an indicator of the anomalous hall effect in unconventional magnetic
  systems},\ }\href {https://doi.org/10.1103/PhysRevB.106.024421} {\bibfield
  {journal} {\bibinfo  {journal} {Phys. Rev. B}\ }\textbf {\bibinfo {volume}
  {106}},\ \bibinfo {pages} {024421} (\bibinfo {year} {2022})}\BibitemShut
  {NoStop}%
\bibitem [{\citenamefont {Udagawa}\ and\ \citenamefont
  {Moessner}(2013)}]{Udagawa13a}%
  \BibitemOpen
  \bibfield  {author} {\bibinfo {author} {\bibfnamefont {M.}~\bibnamefont
  {Udagawa}}\ and\ \bibinfo {author} {\bibfnamefont {R.}~\bibnamefont
  {Moessner}},\ }\bibfield  {title} {\bibinfo {title} {Anomalous hall effect
  from frustration-tuned scalar chirality distribution in
  ${\mathrm{pr}}_{2}{\mathrm{ir}}_{2}{\mathbf{o}}_{7}$},\ }\href
  {https://doi.org/10.1103/PhysRevLett.111.036602} {\bibfield  {journal}
  {\bibinfo  {journal} {Phys. Rev. Lett.}\ }\textbf {\bibinfo {volume} {111}},\
  \bibinfo {pages} {036602} (\bibinfo {year} {2013})}\BibitemShut {NoStop}%
\bibitem [{\citenamefont {Ishizuka}\ and\ \citenamefont
  {Motome}(2013)}]{Ishizuka2013}%
  \BibitemOpen
  \bibfield  {author} {\bibinfo {author} {\bibfnamefont {H.}~\bibnamefont
  {Ishizuka}}\ and\ \bibinfo {author} {\bibfnamefont {Y.}~\bibnamefont
  {Motome}},\ }\bibfield  {title} {\bibinfo {title} {Quantum anomalous hall
  effect in kagome ice},\ }\href {https://doi.org/10.1103/PhysRevB.87.081105}
  {\bibfield  {journal} {\bibinfo  {journal} {Phys. Rev. B}\ }\textbf {\bibinfo
  {volume} {87}},\ \bibinfo {pages} {081105} (\bibinfo {year}
  {2013})}\BibitemShut {NoStop}%
\bibitem [{\citenamefont {Chern}\ \emph {et~al.}(2014)\citenamefont {Chern},
  \citenamefont {Rahmani}, \citenamefont {Martin},\ and\ \citenamefont
  {Batista}}]{Chern2014}%
  \BibitemOpen
  \bibfield  {author} {\bibinfo {author} {\bibfnamefont {G.-W.}\ \bibnamefont
  {Chern}}, \bibinfo {author} {\bibfnamefont {A.}~\bibnamefont {Rahmani}},
  \bibinfo {author} {\bibfnamefont {I.}~\bibnamefont {Martin}},\ and\ \bibinfo
  {author} {\bibfnamefont {C.~D.}\ \bibnamefont {Batista}},\ }\bibfield
  {title} {\bibinfo {title} {Quantum hall ice},\ }\href
  {https://doi.org/10.1103/PhysRevB.90.241102} {\bibfield  {journal} {\bibinfo
  {journal} {Phys. Rev. B}\ }\textbf {\bibinfo {volume} {90}},\ \bibinfo
  {pages} {241102} (\bibinfo {year} {2014})}\BibitemShut {NoStop}%
\bibitem [{\citenamefont {Ye}\ \emph {et~al.}(2018)\citenamefont {Ye},
  \citenamefont {Kang}, \citenamefont {Liu}, \citenamefont {von Cube},
  \citenamefont {Wicker}, \citenamefont {Suzuki}, \citenamefont {Jozwiak},
  \citenamefont {Bostwick}, \citenamefont {Rotenberg}, \citenamefont {Bell},
  \citenamefont {Fu}, \citenamefont {Comin},\ and\ \citenamefont
  {Checkelsky}}]{ye2018}%
  \BibitemOpen
  \bibfield  {author} {\bibinfo {author} {\bibfnamefont {L.}~\bibnamefont
  {Ye}}, \bibinfo {author} {\bibfnamefont {M.}~\bibnamefont {Kang}}, \bibinfo
  {author} {\bibfnamefont {J.}~\bibnamefont {Liu}}, \bibinfo {author}
  {\bibfnamefont {F.}~\bibnamefont {von Cube}}, \bibinfo {author}
  {\bibfnamefont {C.~R.}\ \bibnamefont {Wicker}}, \bibinfo {author}
  {\bibfnamefont {T.}~\bibnamefont {Suzuki}}, \bibinfo {author} {\bibfnamefont
  {C.}~\bibnamefont {Jozwiak}}, \bibinfo {author} {\bibfnamefont
  {A.}~\bibnamefont {Bostwick}}, \bibinfo {author} {\bibfnamefont
  {E.}~\bibnamefont {Rotenberg}}, \bibinfo {author} {\bibfnamefont {D.~C.}\
  \bibnamefont {Bell}}, \bibinfo {author} {\bibfnamefont {L.}~\bibnamefont
  {Fu}}, \bibinfo {author} {\bibfnamefont {R.}~\bibnamefont {Comin}},\ and\
  \bibinfo {author} {\bibfnamefont {J.~G.}\ \bibnamefont {Checkelsky}},\
  }\bibfield  {title} {\bibinfo {title} {Massive dirac fermions in a
  ferromagnetic kagome metal},\ }\href {https://doi.org/10.1038/nature25987}
  {\bibfield  {journal} {\bibinfo  {journal} {Nature}\ }\textbf {\bibinfo
  {volume} {555}},\ \bibinfo {pages} {638} (\bibinfo {year}
  {2018})}\BibitemShut {NoStop}%
\bibitem [{\citenamefont {Onoda}\ \emph {et~al.}(2004)\citenamefont {Onoda},
  \citenamefont {Tatara},\ and\ \citenamefont
  {Nagaosa}}]{doi:10.1143/JPSJ.73.2624}%
  \BibitemOpen
  \bibfield  {author} {\bibinfo {author} {\bibfnamefont {M.}~\bibnamefont
  {Onoda}}, \bibinfo {author} {\bibfnamefont {G.}~\bibnamefont {Tatara}},\ and\
  \bibinfo {author} {\bibfnamefont {N.}~\bibnamefont {Nagaosa}},\ }\bibfield
  {title} {\bibinfo {title} {Anomalous hall effect and skyrmion number in real
  and momentum spaces},\ }\href {https://doi.org/10.1143/JPSJ.73.2624}
  {\bibfield  {journal} {\bibinfo  {journal} {Journal of the Physical Society
  of Japan}\ }\textbf {\bibinfo {volume} {73}},\ \bibinfo {pages} {2624}
  (\bibinfo {year} {2004})}\BibitemShut {NoStop}%
\bibitem [{\citenamefont {G{\"o}bel}\ \emph
  {et~al.}(2017{\natexlab{a}})\citenamefont {G{\"o}bel}, \citenamefont {Mook},
  \citenamefont {Henk},\ and\ \citenamefont
  {Mertig}}]{gobel2017unconventional}%
  \BibitemOpen
  \bibfield  {author} {\bibinfo {author} {\bibfnamefont {B.}~\bibnamefont
  {G{\"o}bel}}, \bibinfo {author} {\bibfnamefont {A.}~\bibnamefont {Mook}},
  \bibinfo {author} {\bibfnamefont {J.}~\bibnamefont {Henk}},\ and\ \bibinfo
  {author} {\bibfnamefont {I.}~\bibnamefont {Mertig}},\ }\bibfield  {title}
  {\bibinfo {title} {Unconventional topological hall effect in skyrmion
  crystals caused by the topology of the lattice},\ }\href
  {https://doi.org/10.1103/PhysRevB.95.094413} {\bibfield  {journal} {\bibinfo
  {journal} {Phys. Rev. B}\ }\textbf {\bibinfo {volume} {95}},\ \bibinfo
  {pages} {094413} (\bibinfo {year} {2017}{\natexlab{a}})}\BibitemShut
  {NoStop}%
\bibitem [{\citenamefont {G{\"o}bel}\ \emph
  {et~al.}(2017{\natexlab{b}})\citenamefont {G{\"o}bel}, \citenamefont {Mook},
  \citenamefont {Henk},\ and\ \citenamefont {Mertig}}]{gobel2017signatures}%
  \BibitemOpen
  \bibfield  {author} {\bibinfo {author} {\bibfnamefont {B.}~\bibnamefont
  {G{\"o}bel}}, \bibinfo {author} {\bibfnamefont {A.}~\bibnamefont {Mook}},
  \bibinfo {author} {\bibfnamefont {J.}~\bibnamefont {Henk}},\ and\ \bibinfo
  {author} {\bibfnamefont {I.}~\bibnamefont {Mertig}},\ }\bibfield  {title}
  {\bibinfo {title} {Signatures of lattice geometry in quantum and topological
  hall effect},\ }\href {https://dx.doi.org/10.1088/1367-2630/aa709b}
  {\bibfield  {journal} {\bibinfo  {journal} {New Journal of Physics}\ }\textbf
  {\bibinfo {volume} {19}},\ \bibinfo {pages} {063042} (\bibinfo {year}
  {2017}{\natexlab{b}})}\BibitemShut {NoStop}%
\bibitem [{\citenamefont {G{\"o}bel}\ \emph {et~al.}(2018)\citenamefont
  {G{\"o}bel}, \citenamefont {Mook}, \citenamefont {Henk},\ and\ \citenamefont
  {Mertig}}]{gobel2018family}%
  \BibitemOpen
  \bibfield  {author} {\bibinfo {author} {\bibfnamefont {B.}~\bibnamefont
  {G{\"o}bel}}, \bibinfo {author} {\bibfnamefont {A.}~\bibnamefont {Mook}},
  \bibinfo {author} {\bibfnamefont {J.}~\bibnamefont {Henk}},\ and\ \bibinfo
  {author} {\bibfnamefont {I.}~\bibnamefont {Mertig}},\ }\bibfield  {title}
  {\bibinfo {title} {The family of topological hall effects for electrons in
  skyrmion crystals},\ }\href {https://doi.org/10.1140/epjb/e2018-90090-0}
  {\bibfield  {journal} {\bibinfo  {journal} {The European Physical Journal B}\
  }\textbf {\bibinfo {volume} {91}},\ \bibinfo {pages} {179} (\bibinfo {year}
  {2018})}\BibitemShut {NoStop}%
\bibitem [{\citenamefont {Tom{\'e}}\ and\ \citenamefont
  {Rosales}(2021)}]{tome2021topo}%
  \BibitemOpen
  \bibfield  {author} {\bibinfo {author} {\bibfnamefont {M.}~\bibnamefont
  {Tom{\'e}}}\ and\ \bibinfo {author} {\bibfnamefont {H.}~\bibnamefont
  {Rosales}},\ }\bibfield  {title} {\bibinfo {title} {Topological phase
  transition driven by magnetic field and topological hall effect in an
  antiferromagnetic skyrmion lattice},\ }\href
  {https://doi.org/10.1103/PhysRevB.103.L020403} {\bibfield  {journal}
  {\bibinfo  {journal} {Phys. Rev. B}\ }\textbf {\bibinfo {volume} {103}},\
  \bibinfo {pages} {L020403} (\bibinfo {year} {2021})}\BibitemShut {NoStop}%
\bibitem [{\citenamefont {Kolincio}\ \emph {et~al.}(2023)\citenamefont
  {Kolincio}, \citenamefont {Hirschberger}, \citenamefont {Masell},
  \citenamefont {Arima}, \citenamefont {Nagaosa},\ and\ \citenamefont
  {Tokura}}]{Kolincio23}%
  \BibitemOpen
  \bibfield  {author} {\bibinfo {author} {\bibfnamefont {K.~K.}\ \bibnamefont
  {Kolincio}}, \bibinfo {author} {\bibfnamefont {M.}~\bibnamefont
  {Hirschberger}}, \bibinfo {author} {\bibfnamefont {J.}~\bibnamefont
  {Masell}}, \bibinfo {author} {\bibfnamefont {T.-h.}\ \bibnamefont {Arima}},
  \bibinfo {author} {\bibfnamefont {N.}~\bibnamefont {Nagaosa}},\ and\ \bibinfo
  {author} {\bibfnamefont {Y.}~\bibnamefont {Tokura}},\ }\bibfield  {title}
  {\bibinfo {title} {Kagome lattice promotes chiral spin fluctuations},\ }\href
  {https://doi.org/10.1103/PhysRevLett.130.136701} {\bibfield  {journal}
  {\bibinfo  {journal} {Phys. Rev. Lett.}\ }\textbf {\bibinfo {volume} {130}},\
  \bibinfo {pages} {136701} (\bibinfo {year} {2023})}\BibitemShut {NoStop}%
\bibitem [{\citenamefont {Kathyat}\ \emph {et~al.}(2021)\citenamefont
  {Kathyat}, \citenamefont {Mukherjee},\ and\ \citenamefont
  {Kumar}}]{Kathyat2021}%
  \BibitemOpen
  \bibfield  {author} {\bibinfo {author} {\bibfnamefont {D.~S.}\ \bibnamefont
  {Kathyat}}, \bibinfo {author} {\bibfnamefont {A.}~\bibnamefont {Mukherjee}},\
  and\ \bibinfo {author} {\bibfnamefont {S.}~\bibnamefont {Kumar}},\ }\bibfield
   {title} {\bibinfo {title} {Electronic mechanism for nanoscale skyrmions and
  topological metals},\ }\href {https://doi.org/10.1103/PhysRevB.103.035111}
  {\bibfield  {journal} {\bibinfo  {journal} {Phys. Rev. B}\ }\textbf {\bibinfo
  {volume} {103}},\ \bibinfo {pages} {035111} (\bibinfo {year}
  {2021})}\BibitemShut {NoStop}%
\bibitem [{\citenamefont {Mukherjee}\ \emph {et~al.}(2023)\citenamefont
  {Mukherjee}, \citenamefont {Sanyal},\ and\ \citenamefont
  {Dagotto}}]{Mukherjee2023}%
  \BibitemOpen
  \bibfield  {author} {\bibinfo {author} {\bibfnamefont {A.}~\bibnamefont
  {Mukherjee}}, \bibinfo {author} {\bibfnamefont {A.~B.}\ \bibnamefont
  {Sanyal}},\ and\ \bibinfo {author} {\bibfnamefont {E.}~\bibnamefont
  {Dagotto}},\ }\bibfield  {title} {\bibinfo {title} {Unconventional skyrmions
  in an interfacial frustrated triangular lattice},\ }\href
  {https://doi.org/10.1103/PhysRevB.108.014408} {\bibfield  {journal} {\bibinfo
   {journal} {Phys. Rev. B}\ }\textbf {\bibinfo {volume} {108}},\ \bibinfo
  {pages} {014408} (\bibinfo {year} {2023})}\BibitemShut {NoStop}%
\bibitem [{\citenamefont {Terasawa}\ and\ \citenamefont
  {Ishizuka}(2023)}]{terasawa2023anomalous}%
  \BibitemOpen
  \bibfield  {author} {\bibinfo {author} {\bibfnamefont {R.}~\bibnamefont
  {Terasawa}}\ and\ \bibinfo {author} {\bibfnamefont {H.}~\bibnamefont
  {Ishizuka}},\ }\bibfield  {title} {\bibinfo {title} {Anomalous hall effect by
  chiral spin textures in two-dimensional luttinger model},\ }\href@noop {}
  {\bibfield  {journal} {\bibinfo  {journal}
  {arXiv:cond-mat.mes-hall/2310.03576}\ } (\bibinfo {year} {2023})}\BibitemShut
  {NoStop}%
\bibitem [{\citenamefont {Furukawa}(1994)}]{furukawa1994transport}%
  \BibitemOpen
  \bibfield  {author} {\bibinfo {author} {\bibfnamefont {N.}~\bibnamefont
  {Furukawa}},\ }\bibfield  {title} {\bibinfo {title} {Transport properties of
  the kondo lattice model in the limit s$=\infty$ and d=$=\infty$},\ }\href
  {https://doi.org/10.1143/JPSJ.63.3214} {\bibfield  {journal} {\bibinfo
  {journal} {Journal of the Physical Society of Japan}\ }\textbf {\bibinfo
  {volume} {63}},\ \bibinfo {pages} {3214} (\bibinfo {year}
  {1994})}\BibitemShut {NoStop}%
\bibitem [{\citenamefont {Udagawa}\ \emph {et~al.}(2012)\citenamefont
  {Udagawa}, \citenamefont {Ishizuka},\ and\ \citenamefont
  {Motome}}]{Udagawa2012}%
  \BibitemOpen
  \bibfield  {author} {\bibinfo {author} {\bibfnamefont {M.}~\bibnamefont
  {Udagawa}}, \bibinfo {author} {\bibfnamefont {H.}~\bibnamefont {Ishizuka}},\
  and\ \bibinfo {author} {\bibfnamefont {Y.}~\bibnamefont {Motome}},\
  }\bibfield  {title} {\bibinfo {title} {Non-kondo mechanism for resistivity
  minimum in spin ice conduction systems},\ }\href
  {https://doi.org/10.1103/PhysRevLett.108.066406} {\bibfield  {journal}
  {\bibinfo  {journal} {Phys. Rev. Lett.}\ }\textbf {\bibinfo {volume} {108}},\
  \bibinfo {pages} {066406} (\bibinfo {year} {2012})}\BibitemShut {NoStop}%
\bibitem [{\citenamefont {Udagawa}(2015)}]{Udagawa2015}%
  \BibitemOpen
  \bibfield  {author} {\bibinfo {author} {\bibfnamefont {M.}~\bibnamefont
  {Udagawa}},\ }\bibfield  {title} {\bibinfo {title} {Magnetic response of
  itinerant spin ice},\ }\href {https://doi.org/10.1142/S2010324715400044}
  {\bibfield  {journal} {\bibinfo  {journal} {SPIN}\ }\textbf {\bibinfo
  {volume} {05}},\ \bibinfo {pages} {1540004} (\bibinfo {year}
  {2015})}\BibitemShut {NoStop}%
\bibitem [{\citenamefont {Wang}\ \emph {et~al.}(2016)\citenamefont {Wang},
  \citenamefont {Barros}, \citenamefont {Chern}, \citenamefont {Maslov},\ and\
  \citenamefont {Batista}}]{Wang2016a}%
  \BibitemOpen
  \bibfield  {author} {\bibinfo {author} {\bibfnamefont {Z.}~\bibnamefont
  {Wang}}, \bibinfo {author} {\bibfnamefont {K.}~\bibnamefont {Barros}},
  \bibinfo {author} {\bibfnamefont {G.-W.}\ \bibnamefont {Chern}}, \bibinfo
  {author} {\bibfnamefont {D.~L.}\ \bibnamefont {Maslov}},\ and\ \bibinfo
  {author} {\bibfnamefont {C.~D.}\ \bibnamefont {Batista}},\ }\bibfield
  {title} {\bibinfo {title} {Resistivity minimum in highly frustrated itinerant
  magnets},\ }\href {https://doi.org/10.1103/PhysRevLett.117.206601} {\bibfield
   {journal} {\bibinfo  {journal} {Phys. Rev. Lett.}\ }\textbf {\bibinfo
  {volume} {117}},\ \bibinfo {pages} {206601} (\bibinfo {year}
  {2016})}\BibitemShut {NoStop}%
\bibitem [{\citenamefont {Chern}\ \emph {et~al.}(2013)\citenamefont {Chern},
  \citenamefont {Maiti}, \citenamefont {Fernandes},\ and\ \citenamefont
  {W\"olfle}}]{Chern2013a}%
  \BibitemOpen
  \bibfield  {author} {\bibinfo {author} {\bibfnamefont {G.-W.}\ \bibnamefont
  {Chern}}, \bibinfo {author} {\bibfnamefont {S.}~\bibnamefont {Maiti}},
  \bibinfo {author} {\bibfnamefont {R.~M.}\ \bibnamefont {Fernandes}},\ and\
  \bibinfo {author} {\bibfnamefont {P.}~\bibnamefont {W\"olfle}},\ }\bibfield
  {title} {\bibinfo {title} {Electronic transport in the coulomb phase of the
  pyrochlore spin ice},\ }\href
  {https://doi.org/10.1103/PhysRevLett.110.146602} {\bibfield  {journal}
  {\bibinfo  {journal} {Phys. Rev. Lett.}\ }\textbf {\bibinfo {volume} {110}},\
  \bibinfo {pages} {146602} (\bibinfo {year} {2013})}\BibitemShut {NoStop}%
\end{thebibliography}

%

\end{document}